\documentclass[aps,pre,amsmath,amssymb,reprint,showkeys,showpacs,dvipdfmx]{revtex4-1}
\newcommand{\del}{\partial}
\newcommand{\nn}{\nonumber}
\newcommand{\ra}{\rightarrow}
\newcommand{\la}{\leftarrow}

\newcommand{\ep}{\epsilon}

\newcommand{\ve}[1]{{\bm #1}}

\newcommand{\h}{\hat}

\newcommand{\dgamma}{\dot{\gamma}}

\newcommand{\vp}{\ve{p}}
\newcommand{\vF}{\ve{F}}
\newcommand{\vq}{\ve{q}}
\newcommand{\vk}{\ve{k}}
\newcommand{\vka}{\ve{\kappa}}

\newcommand{\vj}{\ve{j}}
\newcommand{\vH}{\ve{H}}

\newcommand{\Liouv}{\mathcal{L}}
\newcommand{\tLiouv}{\tilde{\Liouv}}
\newcommand{\tOmega}{\tilde{\Omega}}
\newcommand{\tit}{\tilde{t}}
\newcommand{\tvr}{\tilde{\ve{r}}}

\newcommand{\eav}[1]{\left\langle \left. #1 \right. \right\rangle}
\newcommand{\vg}{\ve{\Gamma}}
\newcommand{\mP}{\mathcal{P}}
\newcommand{\mQ}{\mathcal{Q}}
\newcommand{\bq}{\bar{q}}
\newcommand{\bk}{\bar{k}}
\newcommand{\bp}{\bar{p}}

\usepackage{graphicx}
\usepackage{dcolumn}
\usepackage{bm}

\usepackage{version}
\begin{document}


\title{Nonequilibrium mode-coupling theory for uniformly sheared underdamped systems}

\author{Koshiro Suzuki}
\email{\vspace{-2.0em}suzuki.koshiro@canon.co.jp}
\affiliation{
Analysis Technology Development Center,
Canon Inc.,
30-2 Shimomaruko 3-chome,
Ohta-ku,
Tokyo 146-8501,
Japan
}
\author{Hisao Hayakawa}
\affiliation{
Yukawa Institute for Theoretical Physics,
Kyoto University,
Kitashirakawa Oiwake-cho,
Kyoto 606-8502,
Japan
}

\date{\today}
\begin{abstract}
Nonequilibrium mode-coupling theory (MCT) for uniformly sheared
underdamped systems is developed, starting from the microscopic
thermostatted SLLOD equation, and the corresponding Liouville equation.
Special attention is paid to the translational invariance in the
sheared frame, which requires an appropriate definition of the transient
time-correlators.
The derived MCT equation satisfies the alignment of the wavevectors, and
is manifestly translationally invariant.
Isothermal condition is implemented by the introduction of the current
fluctuation in the dissipative coupling to the thermostat.
This current fluctation grows in the $\alpha$-relaxation regime, which
generates a deviation of the yield stress in the glassy phase from the
overdamped case.
Response to a perturbation of the shear rate demonstrates an inertia
effect which is not observed in the overdamped case.
Our theory turns out to be a non-trivial extension of the theory by
Fuchs and Cates (J.~Rheol.~{\bf 53(4)}, 2009) to underdamped systems.
Since our starting point is identical to that by Chong and Kim
(Phys.~Rev.~E~{\bf 79}, 2009), the contradictions between Fuchs-Cates and
Chong-Kim are resolved.
\end{abstract}
\maketitle


\section{Introduction}

Liquids under shear are attracting continuous interest, not only for
the importance of their application to industry, but also for their
significance to the study of nonequilibrium statistical physics as
ideal but experimentally accessible systems.
Among them, special attention has been paid to dense supercooled
liquids in the vicinity of the glass transition, i.e. glassy materials,
due to their difficulty of understanding and their peculiarity compared
with conventional systems which exhibit thermodynamic phase transitions.

Although not yet perfect, the mode-coupling theory (MCT) presents
remarkable success in its application to glassy materials such as
colloidal suspensions \cite{GS1992, G, Miyazaki2007}.
Probably the most striking feature of MCT is that it predicts a two-step
relaxation phenomenon (the $\beta$-relaxation followed by the
$\alpha$-relaxation) \cite{2step1, 2step2} of the intermediate
scattering function (i.e. density time-correlator)
in the vicinity of a critical packing fraction $\varphi_c$, which is
referred to as the ``MCT transition point''.
This two-step relaxation is thoroughly investigated, and it is
established that MCT is able to explain, e.g. the following properties:
(i) square-root cusp anomaly of the temperature dependence of the
plateau height of the density time-correlator (called the non-ergodic
parameter, NEP) \cite{NEP},
(ii) power-law time-dependence with von Schweidler exponents of the
density time-correlator around the $\beta$-relaxation \cite{beta_exp1,
beta_exp2, beta_sim},
(iii) time-temperature superposition principle realized by the
Kohlrausch-Williams-Watts-type behavior of the density time-correlator
at the $\alpha$-relaxation \cite{TTS}.
On the other hand, MCT is marred with the problem that the NEP can
survive for temperatures below the critical temperature $T_c$, while it
decays to zero in actual glassy states \cite{2step1, KA1994}.
Moreover, $\varphi_c$ is about 10\% smaller than the experimental value.
To overcome these difficulties, successive studies beyond the
conventional MCT have been carried out eagerly as well, e.g.
(i) field-theoretic formulations \cite{ABL2006, KK2007, KK2008, NH2008,
JW2011},
(ii) inclusion of higher-order correlations \cite{Szamel2003, Szamel2012, GMCT},
(iii) estimation of the dynamic correlation length \cite{IMCT}.

The framework of MCT itself is generic, and extensions to other
materials such as dense granular materials \cite{HO2008, KSZ2010,
SH2012p2} have also been attempted.
In particular, MCT is able to incorporate the effect of shear and its
resulting demolition of the ``cage effect''.

The introduction of shear into MCT has been worked out in the pioneering
papers by Fuchs and Cates (FC) \cite{FC2002} and Miyazaki and Reichman
\cite{MR2002}, both of which have been formulated for sheared systems
immersed in solvents, such as colloidal suspensions.
The weak point of the approach of Ref.~\cite{MR2002}, which was followed
by Ref.~\cite{HO2008}, has been pointed out by Chong and Kim (CK)
\cite{CK2009}.
They argued that the approach of Refs.~\cite{MR2002, HO2008}, where the
{\it steady-state} structure factor (or, equivalently, the radial
distribution function) is plugged in as an input, and then the
steady-state shear stress is calculated as an output, is inconsistent,
since they should be treated on the same footing.
Rather, the {\it equilibrium} structure factor should be plugged in as
an input, which is the case for Ref.~\cite{FC2002}.
(This scheme \cite{EM} is referred to as the
``integration-through-transient (ITT)`` in Ref.~\cite{FC2005}.)

In this spirit, CK \cite{CK2009} have constructed MCT for a sheared
system associated with a thermostat, where the SLLOD equation \cite{EM,
SLLOD} is chosen as a microscopic starting point.
In contrast to the theory of FC \cite{FC2002, FC2005, FC2009}, where a
Brownian system governed by the Smoluchowski operator is the starting
point, the SLLOD equation is governed by the Liouville operator
(Liouvillian), and the momentum variables are left unintegrated, i.e. it
is an {\it underdamped} system.

Besides this issue, apparent discrepancies between the resulting
equations of CK \cite{CK2009} and FC \cite{FC2002, FC2005, FC2009} have
been noticed.
For instance, the ``initial decay rate'' of the Debye relaxation, which
exhibits the conventional Taylor dispersion, is time-dependent in FC
\cite{FC2002, FC2005, FC2009}, while it is not in CK \cite{CK2009}.
Probably the most notable difference is that there appears an additional
memory kernel in CK \cite{CK2009}, in addition to the conventional one
which also appears commonly in other MCTs.
As is well known, and is also argued in CK \cite{CK2009}, these
discrepancies cannot be the result of the difference of underdamped and
overdamped systems, which exhibit quite similar behavior for long-time
dynamics \cite{SL1991, GKB1998, SF2004}.
These contradictions were apparent, at least at the formal level, but as
far as we know, no reconciliation has been proposed so far.

In this paper, we reformulate the MCT for uniformly sheared {\it
underdamped} systems, and resolve the contradictions mentioned above.
We point out that the definition of the time-correlators by CK
\cite{CK2009} should be modified to satisfy the translational invariance
in the sheared frame and fulfill the requirement of physical
correlations between the Fourier modes of the fluctuations.
This modification leads to complexities, which are absent in the
formulation of CK \cite{CK2009}, due to the non-commutativity and the
non-Hermiticity of the Liouvillians.
The path which handles these complexities closely follows that of FC
\cite{FC2009}, which we believe to be the most sensible so far
\footnote{There is a subtle problem in defining the ``random force'' in
the present formulation, as is also pointed out in
Ref.~\cite{FC2009}. This issue will be discussed in section
\ref{sec:MCA} }.

The crucial difference of our formulation to the CK's is the
introduction of the isothermal condition, which is implemented at the
level of the Mori-type equations.
This is realized by incorporating current fluctuations into the
dissipative coupling to the thermostat, accompanied by the introduction
of a multiplier.
The growth of this fluctuation in the $\alpha$-relaxation regime
generates deviation of the yield stress in the glassy phase from that of
the overdamped case.
The resulting equation formally corresponds to that of FC \cite{FC2009}
in the overdamped limit, but the effect of the inertia or the
fluctuating coupling to the thermostat renders this correspondence to be
highly non-trivial.
Hence, our framework appears to be a non-trivial extension of FC
\cite{FC2009} to underdamped systems.
The significance of the underdamped formulation is also demonstrated by
its application to the calculation of the response to a perturbation of
the shear rate, where the inertia effect is clearly observed.

The paper is organized as follows.
In section \ref{sec:MSP}, we briefly review the microscopic starting
points.
In section \ref{sec:TransInv}, special attention is paid to the
translational invariance in the sheared frame and the physically
sensible definition of the time-correlators.
Adjoint Liouvillians are introduced, which turns out to be natural in
the treatment of sheared systems.
In section \ref{sec:MoriEq}, Mori-type equations are derived by the
application of the projection operator formalism, first without the
isothermal condition.
After that, the isothermal condition is formulated, and the Mori-type
equation where this condition is implemented is derived.
In section \ref{sec:MCA}, the mode-coupling approximation is worked out,
and a closed equation for the time-correlators (MCT equation) is
derived.
In section \ref{sec:SSP}, a general formula of time-correlators for the
steady-state quantities is presented, and the specific forms of the the
dissipative coupling and the multiplier are discussed.
A formula for the steady-state shear stress in this formulation is
derived.
In section \ref{sec:Num}, the result of the numerical calculation is
shown for the density time-correlator and the steady-state shear stress,
where the current fluctuation in the coupling to the thermostat plays an
important role in the $\alpha$-relaxation.
The results of the CK theory~\cite{CK2009} and the overdamped limit are
also shown and are compared to the above result.
In addition, response to a perturbation of the shear rate is shown.
Section~\ref{sec:Disc} is provided for discussion.
Here, we first compare our work with the major preceding works.
We hope that issues which were previously confusing are
clarified.
Then, the physical significance of the present formulation and the
possibility of its extension are discussed.
Finally, in section~\ref{sec:Sum}, we summarize our results and conclude
the paper.
An application to the response theory (Appendix~\ref{App:sec:response}),
comparison of the Mori-type equations in the underdamped and overdamped
cases (Appendix~\ref{App:sec:Mori}),
and technical details (Appendices~\ref{App:sec:Details} and
\ref{App:sec:Num}) are collected in the Appendices.

\section{Microscopic starting points}
\label{sec:MSP}

Here we briefly summarize our microscopic starting points, i.e. the
SLLOD equation, the Liouville equation, and the steady-state formula.
Our treatment closely follows that of CK \cite{CK2009}, so the details
in common are to be referred to Ref.~\cite{CK2009}.
The crucial difference, however, is that our formulation is intended to
satisfy the isothermal condition, which will be presented in section
\ref{subsec:IsoTh}.
Attentions related to this issue will be paid.

\subsection{SLLOD equation}
\label{subsec:SLLOD}

We deal with an assembly of $N$ equivalent spheres with diameter $d$ and
mass $m$, interacting with themselves and a thermostat in a volume $V$.
The interaction between the spheres is assumed to be a two-body
soft-core potential force.
Uniform shear is imposed on this system, where the shear velocity is
given by $\ve{v}_{\mathrm{sh}} = \ve{\kappa}\cdot\ve{r}$.
Here, $\ve{\kappa}$ is the time-independent shear-rate tensor, which is
assumed to be of the form $\ve{\kappa}_{\mu \nu} = \dgamma \delta_{\mu
x} \delta_{\nu y}$ in this paper, where the shear rate is denoted as
$\dgamma$.
$\ve{r}$ is the spatial coordinate, and Greek indices $\lambda, \mu,
\nu, \cdots$ denote spatial components $\{ x, y, z\}$ in the remainder.
We introduce the distance of the two shear boundaries, $L$, for later
convenience.
Then, the shear velocity at the boundaries are $v_0^{(\pm)} \equiv \pm
\dot{\gamma}L/2$ for $y=\pm L/2$, respectively.
The Newtonian equation of motion for the $i$-th sphere
($i=1,2,\cdots,N$) is given by the following SLLOD equation
\cite{EM, SLLOD},
\begin{eqnarray}
\dot{\ve{r}}_i(t)
&=&
\frac{\ve{p}_i(t)}{m} + \ve{\kappa}\cdot \ve{r}_i(t),
\label{rdot}
\\
\dot{\ve{p}}_i(t)
&=&
\ve{F}_i(t) - \ve{\kappa}\cdot \ve{p}_i(t)
- \alpha \left( \vg \right) \ve{p}_i(t).
\label{pdot}
\end{eqnarray}
Here, $\{\ve{r}_i(t), \ve{p}_i(t) \}$ is the position and the momentum
of the $i$-th sphere at time $t$, and $\ve{\Gamma}(t) \equiv \{
\ve{r}_i(t), \ve{p}_i(t) \}_{i=1}^N$ is the phase-space coordinate.
In Eq.~(\ref{rdot}), $\ve{p}_i(t)$ is defined as a relative momentum in
the sheared frame, and is referred to as the {\it peculiar} or the {\it
thermal} momentum.
$\ve{F}_i(t) \equiv - \del U/ \del \ve{r}_i(t) = - \sum_j \del u \left(
r_{ij}(t) \right) / \del \ve{r}_{ij}(t)$ is the conservative force
acting on the $i$-th sphere from other spheres, $U$ is the total
potential, $u \left( r_{ij}(t) \right)$ is the two-body potential,
$\ve{r}_{ij}(t) \equiv \ve{r}_i(t) - \ve{r}_j(t)$ is the relative
position and $r_{ij}(t) \equiv | \ve{r}_{ij}(t)|$ is the relative
distance, between the $i$-th and $j$-th spheres, respectively.
The parameter $\alpha \left( \vg \right)$ represents the strength of the
coupling of spheres to the thermostat, which in general implicitly
depends on time through its dependence on $\vg(t)$.
Note that $\alpha(\vg)$ has been assumed to be a constant in
Ref.~\cite{CK2009}.
It is possible to prevent the system from heating-up and control the
kinetic temperature to be a constant by tuning the coupling
$\alpha(\vg)$, in which case $\alpha(\vg)$ should be regarded as a
multiplier, rather than an independent physical parameter.
A well-known example can be found in molecular dynamics, which is the
Gaussian isokinetic (GIK) thermostat \cite{EM}.
This thermostat is distinguished by its generality and its importance,
which has the explicit time dependence as
\begin{eqnarray}
\alpha(t)
&=&
\frac{
\sum_i \left[ \ve{F}_i(t) \cdot \ve{p}_i(t) -
\dot{\gamma} p_i^x(t) p_i^y(t) \right]
}
{
\sum_j \ve{p}_j(t) \cdot
\ve{p}_j(t)
},
\label{Eq:alpha_GIK}
\end{eqnarray}
which can be derived from the condition $\frac{d}{dt}
\sum_i \ve{p}_i(t)^2 = 0$.
The specific form of $\alpha(\vg)$ in MCT with isothermal condition
will be discussed later in section \ref{sec:IsoTh_MCT}.
Note that Eqs.~\eqref{rdot} and (\ref{pdot}) reduce to Newtonian
equations if we introduce the viscous force $\vF_i^{(\mathrm{vis})}(t)
\equiv -\alpha(\vg) m \left(\dot{\ve{r}}_i(t) -\ve{\kappa} \cdot
\ve{r}_i(t) \right)$, except for the instant onset of the shear.
%

\subsection{Liouville equation}
\label{subsec:LiouvilleEq}

The Liouville equation for a phase-space variable $A(\ve{\Gamma}(t))$
reads
\begin{eqnarray}
\frac{d}{dt} A\left( \ve{\Gamma}(t) \right)
&=&
\dot{\ve{\Gamma}}(\ve{\Gamma}(t)) \cdot
\left.
\frac{\del A(\ve{\Gamma})}{\del \ve{\Gamma}}
\right|_{\ve{\Gamma} = \ve{\Gamma}(t)}
\nn \\
&=&
e^{i\Liouv t}
i\Liouv
A\left(\ve{\Gamma}(0) \right)
=
i\Liouv A\left(\ve{\Gamma}(t) \right),
\label{LiouvilleEq}
\end{eqnarray}
where the action of the Liouvillian $i\Liouv$ is defined as
\begin{eqnarray}
i\Liouv
A\left(\ve{\Gamma}(0) \right)
&\equiv&
\dot{\ve{\Gamma}}(\ve{\Gamma}) \cdot
\left.
\frac{\del A(\ve{\Gamma}')}{\del \ve{\Gamma}'}
\right|_{\ve{\Gamma}' = \ve{\Gamma}}.
\label{Eq:Liouvillian}
\end{eqnarray}
Here, and in the remainder, the abbreviated notation $\ve{\Gamma} \equiv
\ve{\Gamma}(0)$ is adopted.
The formal solution of Eq.~\eqref{LiouvilleEq} is given by $A\left(
\ve{\Gamma}(t)\right) = e^{i\Liouv t} A\left( \ve{\Gamma}\right)$.
Note that, in Eqs.~(\ref{LiouvilleEq}) and (\ref{Eq:Liouvillian}), the
Liouvillian is assumed not to bear explicit time-dependence.
This is compatible with a time-independent shear, which we consider in
this paper.

On the other hand, the Liouville equation for the nonequilibrium
distribution function $\rho\left( \ve{\Gamma}, t\right)$ reads
\begin{eqnarray}
\hspace{-1.5em}
\frac{\del \rho(\ve{\Gamma}, t)}{\del t}
&=&
-
\left[
\dot{\ve{\Gamma}}\cdot \frac{\del}{\del \ve{\Gamma}} + \Lambda(\ve{\Gamma})
\right]
\rho(\ve{\Gamma}, t)
=
-i \Liouv^{\dagger} \rho(\ve{\Gamma}, t),
\label{eom-rho}
\end{eqnarray}
where
\begin{eqnarray}
i \Liouv^{\dag}
&\equiv &
\dot{\ve{\Gamma}}\cdot \frac{\del}{\del \ve{\Gamma}}
+ \Lambda(\ve{\Gamma})
=
i\Liouv + \Lambda(\ve{\Gamma})
\label{Ldag}
\end{eqnarray}
is the adjoint Liouvillian, and
\begin{eqnarray}
\Lambda(\ve{\Gamma})
&\equiv &
\frac{\del}{\del \ve{\Gamma}} \cdot \dot{\ve{\Gamma}}
\end{eqnarray}
is the volume contraction factor of the phase space.
The formal solution of Eq.~(\ref{eom-rho}) is $\rho\left( \ve{\Gamma},
t\right) = e^{-i\Liouv^\dagger t} \rho\left( \ve{\Gamma}, 0\right)$.
In general, $\Lambda(\ve{\Gamma}) \neq 0$ for nonequilibrium systems, and
hence the Liouvillian is non-Hermitian.
In our case, $\Lambda(\ve{\Gamma})$ is
\begin{eqnarray}
\Lambda(\ve{\Gamma})
&=&
\sum_i
\left(
\frac{\del}{\del \ve{r}_i} \cdot \dot{\ve{r}_i}
+
\frac{\del}{\del \ve{p}_i} \cdot \dot{\ve{p}_i}
\right)
\nn \\
&=&
- 3N \alpha(\vg)
- \sum_i \frac{\del \alpha(\vg)}{\del \ve{p}_i} \cdot \ve{p}_i
\nn \\
&\simeq&
- 3N \alpha(\vg),
\label{Lambda}
\end{eqnarray}
where the last approximate equality holds in the thermodynamic limit.
This will be verified for our specific choice of $\alpha(\vg)$ in
section \ref{sec:IsoTh_MCT}.

Let us decompose the Liouvillian as follows for convenience:
\begin{eqnarray}
i \Liouv
&=&
i \Liouv_0 + i \Liouv_{\dgamma} + i \Liouv_\alpha,
\label{Liouv}
\\
i \Liouv_0
&\equiv &
\sum_i
\left(
\frac{\ve{p}_i}{m} \cdot \frac{\del}{\del \ve{r}_i}
+
\ve{F}_i \cdot \frac{\del}{\del \ve{p}_i}
\right),
\label{Liouv_0}
\\
i \Liouv_{\dgamma}
&\equiv&
\sum_i
\left(
\ve{r}_i \cdot \ve{\kappa}^T \cdot \frac{\del}{\del \ve{r}_i}
-
\ve{p}_i \cdot \ve{\kappa}^T \cdot \frac{\del}{\del \ve{p}_i}
\right),
\label{Liouv_dgamma}
\\
i \Liouv_\alpha
&\equiv &
-
\alpha(\vg)
\sum_i
\ve{p}_i \cdot \frac{\del}{\del \ve{p}_i}.
\label{Liouv_alpha}
\end{eqnarray}
As can be easily seen, $i\Liouv_0$, $i\Liouv_{\dgamma}$, and
$i\Liouv_\alpha$ are the time-evolution generators of the
conservative force, shearing, and the thermostat, respectively.
We consider a situation where the system is initially in equilibrium
with temperature $T$, and at time $t=0$ a uniform shear with shear rate
$\dgamma$ and a thermostat with strength $\alpha(\vg)$ is turned on.

The Liouvillian obeys the following adjoint relation inside the integral
with respect to the phase-space coordinates,
\begin{eqnarray}
\hspace{-1.5em}
\int d\vg \left[ i\Liouv A(\vg)\right] B(\vg)
&=& - \int d\vg A(\vg) \left[ i\Liouv^\dagger B(\vg) \right],
\label{adj1}
\end{eqnarray}
whose repeated use leads to
\begin{eqnarray}
\hspace{-1.5em}
\int d\vg \left[ e^{i\Liouv t} A(\vg)\right] B(\vg)
&=& \int d\vg A(\vg) \left[ e^{-i\Liouv^\dagger t} B(\vg) \right].
\label{adj2}
\end{eqnarray}
Equation~(\ref{adj1}) can be easily shown by partial integration.
The adjoint relation Eq.~(\ref{adj1}) and the
definition of the adjoint Liouvillian Eq.~(\ref{Ldag}) leads to the
following  important relation which indicates the action of the
Liouvillian {\it inside the time-correlators},
\begin{eqnarray}
&&
\hspace{-2em}
\eav{\left[i\Liouv A(\vg(t))\right]B(\vg)^*}
\nn \\
&=&
-
\eav{A(\vg(t))\left[ i\Liouv B(\vg)\right]^*}
+
\eav{A(\vg(t))B(\vg)^* \Omega(\vg)},
\hspace{1em}
\label{adj3}
\end{eqnarray}
where $\Omega(\vg)$ is the work function,
\begin{eqnarray}
\Omega(\vg)
&\equiv&
- \beta \dgamma \sigma_{xy}(\vg)
- 2 \beta \alpha(\vg) \delta K(\vg).
\label{Eq:workfunction}
\end{eqnarray}
Here, $\sigma_{xy}(\vg) \equiv \sum_i \left( p_i^x p_i^y/m + y_i F_i^x
\right)$ and $\delta K(\vg) \equiv K(\vg) - 3 N k_B T / 2$ are the
zero-wavevector limit of the shear stress and the fluctuation of the
kinetic energy $K(\vg) \equiv \sum_i \ve{p}_i^2 / (2m)$, respectively,
and $\beta \equiv 1/\left( k_B T \right)$.
The derivation of Eq.~(\ref{Eq:workfunction}) is shown in Appendix
\ref{App:subsec:SS}.
The ensemble average $\eav{\cdots}$ in Eq.~(\ref{adj3}) is defined for a
phase-space variable $A(\vg(t))$ as
\begin{eqnarray}
\eav{A(\vg(t))}
&\equiv&
\int d\vg
\rho_{\mathrm{ini}}(\vg)
A(\vg(t)),
\label{eavA}
\end{eqnarray}
where $\rho_{\mathrm{ini}}(\vg)$ is the initial Maxwell-Boltzmann
distribution, and the definition of Eq.~\eqref{eavA} corresponds to the
``Heisenberg picture'' \cite{EM}, which we adopt in this work.

Finally, we state here a steady-state formula for a phase-space
variable $A(\vg(t))$,
\begin{eqnarray}
\eav{ A}_{\mathrm{SS}}
&=&
\eav{ A(\vg)}
-
\beta \dgamma
\int_0^\infty dt
\eav{ A(\vg(t)) \sigma_{xy}(\vg) }
\nn \\
&&
-
2 \beta
\int_0^\infty dt
\eav{ A(\vg(t)) \alpha(\vg) \delta K(\vg) },
\label{Ass}
\end{eqnarray}
where
\begin{eqnarray}
\eav{ A}_{\mathrm{SS}}
&\equiv&
\lim_{t \ra \infty}
\eav{ A(\vg(t))}
\label{Eq:Ass_def}
\end{eqnarray}
is the ensemble average in a nonequilibrium steady-state.
Refer to Appendix \ref{App:subsec:SS} for the derivation of
Eq.~(\ref{Ass}).

\section{Translational invariance in the sheared frame}
\label{sec:TransInv}

So far we have repeated the formulation by CK \cite{CK2009}, aside from
the introduction of the dependence on the phase-space variables $\vg$ of
the dissipative coupling to the thermostat, $\alpha(\vg)$.
In this section, a detailed discussion is devoted to the translational
invariance in the sheared frame, and to the definition of the
time-correlators.
This discussion clearly indicates a crucial mistake in the Fourier
transform of the phase-space variables in CK.

\subsection{Fourier transform}
\label{subsec:FT}

A crucial feature of the sheared system is that the translational
invariance is preserved only in the sheared frame.
Hence, in order to examine the implications of the translational
invariance, we must move on to this frame.
As derived in Appendix \ref{App:subsec:FT}, Eq.~(\ref{A_tilq}), the
wavevector of the Fourier transform in the sheared frame is
Affine-deformed from that of the experimental frame (we will refer to it
as the ``advected wavevector'' in the following):
\begin{eqnarray}
A_{\ve{q}(-t)} (\tit)
&=&
\int d^3 \tvr
A(\tvr, \tit)
e^{i\ve{q}(-t)\cdot\tvr},
\label{FT}
\\
\ve{q}(t)
&\equiv&
\ve{q} - \ve{q} \cdot \ve{\kappa} t.
\label{advk}
\end{eqnarray}
Here, $\tit \equiv t$ and $\tvr \equiv \ve{r} - \left( \ve{\kappa} \cdot
\ve{r} \right)t$ are the temporal and spatial coordinates in the sheared
frame, respectively.
We adopt the definition of the advected wavevector of FC
\cite{FC2009}, which differs from the conventional one, $\ve{q}(t)
\equiv \ve{q} + \ve{q} \cdot \ve{\kappa} t$ \cite{CK2009, MR2002,
HO2008}.
The reason of this choice will be explained later.
The values of the phase-space variables are equivalent, irrespective of
the frames, so the following equality holds:
\begin{eqnarray}
A_{\ve{q}(-t)} (\tit)
&=&
A_\ve{q}(t).
\label{equality}
\end{eqnarray}
The time-evolution of the Fourier transform in the sheared frame is
generated by $i\Liouv_0$, $i\Liouv_\alpha$, and $i\Liouv_{\dgamma_p}$,
where $i\Liouv_{\dgamma_p}$ is the momentum part of $i\Liouv_{\dgamma}$:
\begin{eqnarray}
\frac{\del}{\del \tit}
A_{\ve{q}(-t)} (\tit)
&=&
i\tLiouv
A_{\ve{q}(-t)}(\tit),
\label{eomq}
\\
i\tLiouv
&\equiv&
i\Liouv_0
+
i\Liouv_\alpha
+
i\Liouv_{\dgamma_p},
\label{tLiouv}
\\
i\Liouv_{\dgamma_p}
&\equiv&
-
\sum_i
\ve{p}_i \cdot \ve{\kappa}^T \cdot \frac{\del}{\del \ve{p}_i}.
\label{Liouv_dgammap}
\end{eqnarray}
The derivation is shown in Appendix \ref{App:subsec:FT}. 
%
Note that the action of the coordinate part of $i\Liouv_{\dgamma}$,
which we denote
\begin{eqnarray}
i\Liouv_{\dgamma_r}
&\equiv&
\sum_i
\ve{r}_i \cdot \ve{\kappa}^T \cdot \frac{\del}{\del \ve{r}_i},
\label{Liouv_dgammar}
\end{eqnarray}
is already incorporated in the advected wavevector.
This fact already has been pointed out by Hayakawa and Otsuki (HO)
\cite{HO2008}.
We refer to $i\Liouv_{\dgamma_r}$ as the ``advection Liouvillian'',
since it generates an advection of the wavevectors of plane waves,
$e^{-i\Liouv_{\dgamma_r} t} e^{i\vq\cdot\ve{r}} =
e^{i\vq(t)\cdot\ve{r}}$.

\subsection{Adjoint Liouvillians inside the time-correlators}
\label{subsec:Adjoint}

We have seen above that the Fourier transform separates the Liouvillian
$i\Liouv$ into $i\tLiouv$ and the advection Liouvillian
$i\Liouv_{\dgamma_r}$.
Hence it is convenient to decompose the adjoint relation Eq.~(\ref{adj3})
into those for $i\tLiouv$ and $i\Liouv_{\dgamma_r}$, and define the
corresponding adjoint Liouvillians, $i\tLiouv^\dagger$ and
$i\Liouv_{\dgamma_r}^\dagger$.
As for $i\tLiouv$, it is
\begin{eqnarray}
\eav{\left[i\tLiouv A(\vg(t))\right]B(\vg)^*}
&=&
-
\eav{A(\vg(t))\left[ i\tLiouv^\dagger B(\vg)\right]^*},
\hspace{1.5em}
\label{adj4}
\\
i\tLiouv^\dagger
&\equiv&
i\tLiouv - \tOmega,
\label{tLiouv_dagger}
\end{eqnarray}
where
\begin{eqnarray}
\tOmega(\vg)
&\equiv&
-
\beta \dot{\gamma}
\sigma_{xy}^{(\mathrm{kin})}(\vg)
-
2 \beta \alpha(\vg)
\delta K(\vg)
\label{tOmega}
\end{eqnarray}
is the modified work function, which includes only the kinetic part of
the shear stress,
\begin{eqnarray}
\sigma_{xy}^{(\mathrm{kin})}
&\equiv&
\sum_i
\frac{p_i^x p_i^y}{m}.
\label{sigma_kin}
\end{eqnarray}
As for $i\Liouv_{\dgamma_r}$, it is
\begin{eqnarray}
\eav{\left[i\Liouv_{\dgamma_r} A(\vg(t))\right]B(\vg)^*}
&=&
-
\eav{A(\vg(t))\left[ i\Liouv_{\dgamma_r}^\dagger B(\vg)\right]^*},
\hspace{1.5em}
\label{adj6}
\\
i\Liouv_{\dgamma_r}^\dagger
&\equiv&
i\Liouv_{\dgamma_r}
+
\beta \dgamma
\sigma_{xy}^{(\mathrm{pot})},
\label{Ldgammar_dag}
\end{eqnarray}
where the repeated use of Eq.~(\ref{adj6}) results in
\begin{eqnarray}
&&
\hspace{-6em}
\eav{\left[ e^{i\Liouv_{\dgamma_r} t} A(\vg(t))\right]B(\vg)^*}
\nn \\
&=&
\eav{A(\vg(t))\left[ e^{-i\Liouv_{\dgamma_r}^\dagger t} B(\vg)\right]^*}.
\label{adj7}
\end{eqnarray}
Here, $\sigma_{xy}^{(\mathrm{pot})}$ is the potential part of the shear
stress,
\begin{eqnarray}
\sigma_{xy}^{(\mathrm{pot})}
&\equiv&
\sum_i
y_i F_i^x.
\label{sigma_pot}
\end{eqnarray}
Note that the adjoint Liouvillians $i\tLiouv^\dagger$ and
$i\Liouv_{\dgamma_r}^\dagger$ are well-defined {\it only inside the
time-correlators}, and are not to be confused with the adjoint
Liouvillian $i\Liouv^\dagger$, which is defined in Eq.~(\ref{Ldag}) as
an independent operator.

\subsection{General time-correlators}
\label{subsec:TC}

In nonequilibrium statistical mechanics, time-correlators play an
essential role.
Hence, we figure out the implications of the translational invariance on
the time-correlators.
An immediate consequence of the translational invariance in the sheared
frame is
\begin{eqnarray}
A(\tvr, \tit)
&=&
A(\tvr+\ve{a}(t), \tit),
\\
A_{\vk} (\tit)
&=&
e^{i\vk\cdot\ve{a}(t)}
A_{\vk} (\tit),
\label{Eq:Ak}
\end{eqnarray}
where $\ve{a}(t) \equiv \left( \dgamma t L, L, 0 \right)$.
Here, $L$ is the distance between the two shear boundaries, which is
introduced in section \ref{subsec:SLLOD}.
Equation~(\ref{Eq:Ak}) leads to the following ``selection rule'' of the
wavevectors of the one-point and two-point functions in the sheared
frame:
\begin{eqnarray}
\eav{A_{\vk} (\tit)}
&=&
\eav{A_{\vk = 0} (\tit)}
\delta_{\vk, 0},
\\
\eav{A_{\vk} (\tit) B_\ve{q}^* (0)}
&=&
\eav{A_{\ve{q}}(\tit) B_{\ve{q}}^* (0)}
\delta_{\vk, \vq}.
\label{TPF-sf}
\end{eqnarray}

Application of the equivalence of the Fourier transforms in the
experimental and the sheared frames, i.e. Eq.~(\ref{equality}), to
Eq.~(\ref{TPF-sf}) results in the following ``selection rule'' in the
experimental frame:
\begin{eqnarray}
\eav{A_{\vk(t)}(t) B_\vq^* (0)}
&=&
\eav{A_{\vq(t)}(t) B_\vq^* (0)}
\delta_{\vk,\vq}.
\label{TPF-ef}
\end{eqnarray}
Here, the Fourier transform with an advected wavevector $A_{\vq(t)}(t)$
is explicitly written in terms of the Liouvillians as
\begin{eqnarray}
A_{\vq(t)}(t)
&=&
\sum_i
e^{i\Liouv t}
A_i \left( \vg(0) \right)
e^{i\vq(t)\cdot\ve{r}_i}
\nn \\
&=&
\sum_i
e^{i\Liouv t}
A_i \left( \vg(0) \right)
e^{-i\Liouv_{\dgamma_r} t}
e^{i\vq\cdot\ve{r}_i},
\label{Aqt}
\end{eqnarray}
where $A_i\left( \vg(t) \right)$ is a Fourier coefficient which is
defined in Appendix \ref{App:subsec:FT}, Eq.~(\ref{A_Art}).

One might think that the two-point function Eq.~(\ref{TPF-ef}) involves
an ambiguity; it might seem that another choice, e.g. $\eav{A_{\vq}(t)
B_{\vq(-t)}^* (0)}$, is equally valid.
Actually, this has been the choice made in CK \cite{CK2009} and FC
\cite{FC2002, FC2005}.
However, the two expressions $\eav{A_{\vq(t)} (t) B_{\vq}^* (0)}$ and
$\eav{A_{\vq} (t) B_{\vq(-t)}^* (0)}$ are in fact inequivalent, due to
the non-commutativity and the non-Hermiticity of the Liouvillians.
We will prove this statement below.
First, with the use of Eq.~(\ref{Aqt}), $\eav{A_{\vq(t)} (t) B_{\vq}^*
(0)}$ can be written in the following form:
\begin{eqnarray}
&&
\hspace{-1.5em}
\eav{A_{\vq(t)}(t) B_{\vq}^* (0)}
\nn \\
&=&
\eav{
\left[
\sum_i
e^{i\Liouv t}
A_i(\vg(0))
e^{-i\Liouv_{\dgamma_r} t}
e^{i\vq\cdot\ve{r}_i}
\right]
\sum_j
B_j(\vg(0))^*
e^{-i\vq\cdot\ve{r}_j}
}.
\label{Cor-SH}
\nn \\
\end{eqnarray}
Similarly, $\eav{A_{\vq} (t) B_{\vq(-t)}^* (0)}$ can be written as
follows:
\begin{eqnarray}
&&
\hspace{-1.5em}
\eav{A_{\vq}(t) B_{\vq(-t)}^* (0)}
\nn \\
&=&
\eav{
\left[
\sum_i
e^{i\Liouv t}
A_i(\vg(0))
e^{i\vq\cdot\ve{r}_i}
\right]
\sum_j
B_j(\vg(0))^*
e^{i\Liouv_{\dgamma_r} t}
e^{-i\vq\cdot\ve{r}_j}
}.
\label{Cor-CK}
\nn \\
\end{eqnarray}
Even when the ``advection Liouvillian'' $i\Liouv_{\dgamma_r}$ commutes
with $A_i(\vg(0))$ and $B_j(\vg(0))$, which is the case of interest in
MCT where these variables are the density and the current-density
fluctuations, Eqs.~(\ref{Cor-SH}) and (\ref{Cor-CK}) are not equivalent.
This can be seen by the use of the adjoint relation of the Liouvillians
Eq.~(\ref{adj7}) and the relations $\left[ i\Liouv_{\dgamma_r}, i\Liouv
\right] \neq 0$ and $i\Liouv_{\dgamma_r} \neq i\Liouv_{\dgamma_r}^\dagger$.
It can also be foreseen from Eqs.~(\ref{Cor-SH}) and (\ref{Cor-CK}) that
different definitions of the two-point functions lead to different
physical consequences.

We assert that the specific definition Eq.~(\ref{TPF-ef}) is the
physically sensible choice.
It states that a fluctuation at time $t=0$ with a wavevector $\vq =
\vq(0)$ is correlated at time $t$ with a fluctuation with a wavevector
$\vq(t)$, exclusively.
Note that the definition of the advected wavevector Eq.~(\ref{advk}) is
chosen for the compatibility to the intuitive picture described above.
This definition is essentially coincident with the one adopted in the
previous studies \cite{HO2008, MR2002, FC2009}, although it appears to
be $\eav{A_{\vq(-t)}(t) B_\vq^* (0)}$ in Refs.~\cite{HO2008} and
\cite{MR2002}.
This superficial discrepancy with Eq.~(\ref{TPF-ef}) is due to the
different definition of the advected wavevector, $\vq(t) \equiv \vq +
\vq \cdot \ve{\kappa} t$.

\section{Mori-type equations}
\label{sec:MoriEq}

In liquid theory, slowly-varying (i.e. long-wavelength and
low-frequency) conserved variables are of interest.
Conventionally these are the density fluctuation and the current-density
(momentum) fluctuation.
In MCT for sheared thermostatted systems, the formula for the
steady-state quantities is formulated in terms of a time-correlator for
density fluctuations (density time-correlator).
In this section, we introduce the time-correlators of interest and
derive the Mori-type equations \cite{HM} for them by applying the
projection operator formalism.
We first derive the Mori-type equations without isothermal condition.
Then, we formulate the isothermal condition by introducing a multiplier
for this constraint.
Finally, we derive corrections to the Mori-type equations, where the
multiplier is introduced.

\subsection{Time-correlators of interest}
\label{subsec:TC2}

Density and current-density fluctuations at equilibrium are denoted in
Fourier space as
\begin{eqnarray}
n_\vq
&\equiv&
\sum_i
e^{i\vq\cdot\ve{r}_i}
-
N \delta_{\vq, 0},
\\
j_\vq^\lambda
&\equiv&
\sum_i
\frac{p_i^\lambda}{m}
e^{i\vq\cdot\ve{r}_i}
\hspace{1em}
(\lambda = x,y,z),
\end{eqnarray}
respectively.
The spatial dimension is assumed to be three, in accordance with the
numerical calculation which will be carried out in section
\ref{sec:Num}.
The corresponding time-correlators of interest of the form
Eq.~(\ref{TPF-ef}) are as follows:
\begin{eqnarray}
\Phi_\vq(t)
&\equiv &
\frac{1}{N}
\eav{ n_{\vq(t)}(t) n_{\vq}(0)^*},
\label{Phi}
\\
H_\vq^\lambda(t)
&\equiv &
\frac{i}{N}
\eav{j_{\vq(t)}^\lambda(t) n_{\vq}(0)^*}.
\label{H}
\end{eqnarray}
Here, $\Phi_\vq(t)$ is the density time-correlator, and
$H_\vq^\lambda(t)$ is referred to as the density-current cross
time-correlator.
As already mentioned, the Fourier coefficient $A_i(\vg(0))$ in
Eq.~(\ref{Aqt}) is $A_i(\vg(0))=1$ for the density fluctuation and
$A_i(\vg(0))=p_i^\lambda/m$ for the current-density fluctuation; hence
they commute with $i\Liouv_{\dgamma_r}$, respectively.
This leads to the following expression for the density and
current-density fluctuations at time $t$,
\begin{eqnarray}
\xi_{\vq(t)}(t)
&=&
e^{i\Liouv t}
e^{-i\Liouv_{\dgamma_r}t}
\xi_\vq
=
e^{i\Liouv t}
\xi_{\vq(t)}
=
U(t) \xi_\vq,
\end{eqnarray}
where
\begin{eqnarray}
\xi_{\vq(t)}
&\equiv&
e^{-i\Liouv_{\dgamma_r}t}
\xi_\vq,
\\
U(t)
&\equiv&
e^{i\Liouv t}
e^{-i\Liouv_{\dgamma_r}t}.
\end{eqnarray}
Here, $\xi$ is either of the hydrodynamic variables, $n$ or $j$, and
$U(t)$ is the time-evolution operator for $\xi$ in
Fourier space.
Note that $U(t)$ is non-unitary since the Liouvillians are
non-Hermitian.
Then, Eqs.~(\ref{Phi}) and (\ref{H}) are expressed respectively as
\begin{eqnarray}
\Phi_\vq(t)
&=&
\frac{1}{N}
\eav{ \left[ U(t) n_{\vq} \right] n_{\vq}^* },
\label{Phi2}
\\
H_\vq^\lambda(t)
&=&
\frac{i}{N}
\eav{ \left[ U(t) j_{\vq}^\lambda \right] n_{\vq}^*}.
\label{H2}
\end{eqnarray}
%

\subsection{Continuity equations}
\label{subsec:ContEq}

Now we derive the equations of motion for Eqs.~(\ref{Phi2}) and
(\ref{H2}).
The time derivative of $U(t)$ is
\begin{eqnarray}
\hspace{-1.5em}
\frac{d}{dt}
U(t)
&=&
e^{i\Liouv t}
\left(
i\Liouv
-
i\Liouv_{\dgamma_r}
\right)
e^{-i\Liouv_{\dgamma_r}t}
=
e^{i\Liouv t}
i\tLiouv
e^{-i\Liouv_{\dgamma_r}t},
\label{dU}
\end{eqnarray}
where $i\tLiouv$ is defined in Eq.~(\ref{tLiouv}).
The action of $i\tLiouv$ on $n_\vq$ is obtained from
Eqs.~(\ref{Liouv_0}), (\ref{Liouv_alpha}), and (\ref{Liouv_dgammap}) as
\begin{eqnarray}
i\tLiouv n_\vq
&=&
i \vq \cdot \vj_\vq.
\label{tLiouv-n}
\end{eqnarray}
From Eqs.~(\ref{H2})--(\ref{tLiouv-n}) we obtain
\begin{eqnarray}
\hspace{-1.5em}
\frac{d}{dt}
\Phi_\vq(t)
&=&
\frac{1}{N}
\eav{
\left[
\frac{d}{dt} U(t)
n_{\vq}
\right]
n_{\vq}^* }
\nn \\
&=&
\frac{1}{N}
\eav{
\left[
e^{i\Liouv t} i\tLiouv
n_{\vq(t)}
\right]
n_{\vq}^* }
=
\vq(t) \cdot \vH_\vq(t).
\label{eom-Phi}
\end{eqnarray}
On the other hand, the action of $i\tLiouv$ on $\vj_\vq$ cannot be
written in a concise form as simply as Eq.~(\ref{tLiouv-n}):
\begin{eqnarray}
\frac{d}{dt}
H_\vq^\lambda(t)
&=&
\frac{i}{N}
\eav{
\left[
\frac{d}{dt} U(t)
j_{\vq}^\lambda
\right]
n_{\vq}^* }
\nn \\
&=&
\frac{i}{N}
\eav{
\left[
e^{i\Liouv t} i\tLiouv
j_{\vq(t)}^\lambda
\right]
n_{\vq}^* }.
\label{eom-H}
\end{eqnarray}
A conventional way to handle Eq.~(\ref{eom-H}) is to deform it into a
Mori-type equation \cite{HM}.
This task is conducted in the next subsection by the application of the
projection operator formalism.

\subsection{Projection operator formalism}
\label{subsec:PO}

We have introduced the time-correlators, Eqs.~(\ref{Phi}) and (\ref{H}),
in a physically sensible way that a fluctuation at time $t=0$ with a
wavevector $\vq$ is correlated at time $t$ with a fluctuation with a
wavevector $\vq(t)$.
We refer to this feature as the ``{\it alignment of the wavevectors}''
in this paper.
Even if the wavevectors of the time-correlators are aligned, this
feature is not necessarily preserved in their entire continuity
equations.
As for the density time-correlator $\Phi_\vq(t)$, this is positive as
can be seen from Eq.~(\ref{eom-Phi}).
In deriving a Mori-type equation for the cross time-correlator
$H_\vq^\lambda(t)$, we demand the alignment of the wavevectors as a
principle.

For this purpose, we introduce the following {\it time-dependent}
projection operator \cite{FC2009}
\begin{eqnarray}
\hspace{-1.5em}
\mP(t) X
&\equiv &
\sum_\vk
\frac{\eav{X n_{\vk(t)}^*} }{NS_{k(t)}} n_{\vk(t)}
+
\sum_\vk
\frac{\eav{X j_{\vk(t)}^{\lambda *}} }{N v_T^2} j_{\vk(t)}^\lambda
,
\label{P}
\end{eqnarray}
and its complementary operator
\begin{eqnarray}
\mQ(t)
&\equiv &
1 - \mP(t).
\label{Q}
\end{eqnarray}
Here, $X$ is an arbitrary phase-space variable in Fourier space and the
normalization factors are determined from the equal-time (equilibrium)
correlators, $\eav{n_\vq n_{\vq'}^*} = N S_q \delta_{\vq, \vq'}$ and
$\eav{j_\vq^\lambda j_{\vq'}^\mu} = N v_T^2 \delta^{\lambda \mu}
\delta_{\vq, \vq'}$, where $S_q$ is the static structure factor
\cite{HM} and $v_T \equiv \sqrt{k_B T/m}$ is the thermal velocity.
These operators preserve the desired properties, (i) idempotency:
$\mP(t)^2 = \mP(t)$, $\mQ(t)^2 = \mQ(t)$, (ii) orthogonality:
$\mP(t)\mQ(t) = \mQ(t)\mP(t) = 0$, and (iii) Hermiticity:
\begin{eqnarray}
\hspace{-1.5em}
\eav{\left[\mP(t) A_{\vq(t)}(t)\right]B_{\vq}(0)^*}
&=&
\eav{A_{\vq(t)}(t)\left[ \mP(t) B_{\vq}(0)\right]^*},
\\
\hspace{-1.5em}
\eav{\left[\mQ(t) A_{\vq(t)}(t)\right]B_{\vq}(0)^*}
&=&
\eav{A_{\vq(t)}(t)\left[ \mQ(t) B_{\vq}(0)\right]^*}.
\end{eqnarray}
In addition, we further introduce a ``rescaled static projection
operator'' \cite{FC2009},
\begin{eqnarray}
\bar{\mP}_t X
&=&
\sum_\vk
\frac{\eav{ X n_{\vk}^*} }{NS_{k(t)}}
n_{\vk}
+
\sum_\vk
\frac{\eav{ X j_{\vk}^{\lambda *} } }{N v_T^2}
j_{\vk}^\lambda,
\label{bP}
\end{eqnarray}
whose {\it raison d'\^{e}tre} will be explained later.
Although Eq.~\eqref{bP} is a projection operator onto the subspace
spanned by the static density and current-density fluctuations $\{
n_\vk, \vj_\vk \}$, it involves a dependence on time through the
advected index of $S_{k(t)}$, so we appended a subscript $t$.
One can easily verify, by the use of Eq.~(\ref{adj7}), the following
relation between $\mP(t)$ and $\bar{\mP}_t$,
\begin{eqnarray}
\mP(t)
&=&
e^{-i\Liouv_{\dgamma_r} t}
\bar{\mP}_t
e^{i\Liouv_{\dgamma_r}^\dagger t},
\end{eqnarray}
where the adjoint of the advection Liouvillian $i\Liouv_{\dgamma
r}^\dagger$ is defined in Eq.~(\ref{Ldgammar_dag}).

Now we derive a Mori-type equation for Eq.~(\ref{eom-H}) by inserting
the projection operators as follows:
\begin{eqnarray}
\frac{d}{dt}
U(t)
&=&
e^{i\Liouv t}
\left[
\mP(t)+ \mQ(t)
\right]
i\tLiouv
e^{-i\Liouv_{\dgamma_r}t}
\nn \\
&=&
e^{i\Liouv t}
\left[
e^{-i\Liouv_{\dgamma_r}t}
\bar{\mP}_t
e^{i\Liouv_{\dgamma_r}^\dagger t}
+
\mQ(t)
\right]
i\tLiouv
e^{-i\Liouv_{\dgamma_r}t}
\nn \\
&=&
e^{i\Liouv t}
e^{-i\Liouv_{\dgamma_r}t}
\left[
\bar{\mP}_t
e^{i\Liouv_{\dgamma_r}^\dagger t}
+
e^{i\Liouv_{\dgamma_r}t}
\mQ(t)
\right]
i\tLiouv
e^{-i\Liouv_{\dgamma_r}t}
\nn \\
&=&
U(t)
\left[
\bar{\mP}_t
e^{i\Liouv_{\dgamma_r}^\dagger t}
i\tLiouv
e^{-i\Liouv_{\dgamma_r}t}
+
e^{i\Liouv_{\dgamma_r}t}
\mQ(t)
i\tLiouv
e^{-i\Liouv_{\dgamma_r}t}
\right].
\label{dU2}
\nn \\
\end{eqnarray}
The formal solution of Eq.~(\ref{dU2}) is given as
\begin{eqnarray}
U(t)
&=&
U_0(t,0)
+
\int_0^t ds
U(s)
\bar{\mP}_s
e^{i\Liouv_{\dgamma_r}^\dagger s}
i\tLiouv
e^{-i\Liouv_{\dgamma_r}s}
U_0(t,s),
\label{U}
\nn \\
\end{eqnarray}
where $U_0(t,t')$ is the solution of the homogeneous equation.
$U_0(t,t')$ can be written in terms of the time-ordered exponential,
$\exp_\ra \left[ \int_{t'}^t ds X(s) \right] \equiv 1 +
\sum_{n=1}^\infty \int_{t'}^t ds_1 \cdots \int_{t'}^{s_{n-1}} ds_n
X(s_n) \cdots X(s_1)$, as follows:
\begin{eqnarray}
U_0(t,t')
&=&
\exp_\ra
\left[
\int_{t'}^t ds
e^{i\Liouv_{\dgamma_r}s}
\mQ(s)
i\tLiouv
e^{-i\Liouv_{\dgamma_r}s}
\right].
\label{U0}
\end{eqnarray}
From Eqs.~(\ref{dU2}) and (\ref{U}), the time derivative of the
current-density fluctuation, which appears in Eq.~(\ref{eom-H}), can be
decomposed into the ``correlated part'' and the ``uncorrelated part'':
\begin{eqnarray}
&&
\frac{d}{dt}
U(t)
j_{\vq}^\lambda
=
U(t)
\bar{\mP}_t
e^{i\Liouv_{\dgamma_r}^\dagger t}
i\tLiouv
j_{\vq(t)}^\lambda
+
U_0(t,0)
e^{i\Liouv_{\dgamma_r}t}
\mQ(t)
i\tLiouv
j_{\vq(t)}^\lambda
\nn \\
&&
+
\int_0^t ds
U(s)
\bar{\mP}_s
e^{i\Liouv_{\dgamma_r}^\dagger s}
i\tLiouv
e^{-i\Liouv_{\dgamma_r}s}
U_0(t,s)
\cdot
e^{i\Liouv_{\dgamma_r}t}
\mQ(t)
i\tLiouv
j_{\vq(t)}^\lambda.
\nn \\
\label{dUj}
\end{eqnarray}
Here, the first term on the right-hand side (r.h.s.) is the ``correlated
part'' and the second and the third terms are the ``uncorrelated part''.
Applying Eq.~(\ref{bP}) to Eq.~(\ref{dUj}), and then substituting
Eq.~(\ref{dUj}) to Eq.~(\ref{eom-H}), we obtain the Mori-type equation.
Following the derivation in Appendix \ref{App:subsec:MoriEq}, we obtain
\begin{eqnarray}
\frac{d}{dt}
H_\vq^\lambda(t)
&=&
-
v_T^2
\frac{q(t)^\lambda }{S_{q(t)}}
\Phi_\vq(t)
-
\alpha_0 H_{\vq}^\lambda(t)
-
\left[
\ve{\kappa} \cdot \vH_{\vq}(t)
\right]^\lambda
\nn \\
&&
+
\frac{i}{N}
\eav{
R_{\vq(t)}^\lambda(t)
n_\vq^*
}
-
\int_0^t ds
L_{\vq}^\lambda(t,s)
\Phi_\vq(s)
\nn \\
&&
-
\int_0^t ds
M_{\vq}^{\lambda \mu}(t,s)
H_\vq^\mu(s),
\label{eom-H2}
\end{eqnarray}
where we have introduced
\begin{eqnarray}
R_{\vq(t)}^\lambda(t)
&\equiv&
U_0(t,0)
e^{i\Liouv_{\dgamma_r}t}
R_{\vq(t)}^\lambda,
\label{Randt}
\\
R_{\vq(t)}^\lambda
&\equiv&
\mQ(t)
i\tilde{\Liouv}
j_{\vq(t)}^\lambda,
\label{Rand}
\\
iL_\vq^\lambda (t,s)
&\equiv&
\frac{1}{NS_{q(t)}}
\eav{
\left[
i\tLiouv
\tilde{U}_0(t,s)
R_{\vq(t)}^\lambda
\right]
n_{\vq(s)}^*
},
\label{L}
\\
M_\vq^{\lambda \mu}(t,s)
&\equiv&
-
\frac{1}{N v_T^2}
\eav{
\left[
i\tLiouv
\tilde{U}_0(t,s)
R_{\vq(t)}^\lambda
\right]
j_{\vq(s)}^{\mu *}
},
\label{M}
\\
\tilde{U}_0(t,s)
&\equiv&
e^{-i\Liouv_{\dgamma_r}s}
U_0(t,s)
e^{i\Liouv_{\dgamma_r}t}.
\label{tU0}
\end{eqnarray}
Here, $R_{\vq(t)}^\lambda(t)$ is the ``random force'', whose
time-evolution is given by the projected time-evolution operator,
$U_0(t,0)$.
There appear two types of memory kernels, $L_\vq^\lambda(t,s)$ and
$M_\vq^{\lambda \mu}(t,s)$, due to the projection onto the
current-density and the density fluctuations, respectively.
The correction of the friction coefficient $\alpha_0$ due to the
introduction of the isothermal condition will be discussed in section
\ref{subsec:IsoThMori}.

Note that the ``random force'' Eq.~(\ref{Randt}) is not orthogonal to
the density fluctuation at this stage.
An additional requirement leads to the orthogonality, which will be
discussed in Eqs.~(\ref{insertQ}) and (\ref{RandOrth}) in section
\ref{sec:MCA}.
Note also that the memory kernels possess two time arguments, in
contrast to those which appear in CK.
This issue will also be discussed in section \ref{sec:MCA}.
%

\subsection{Isothermal condition}
\label{subsec:IsoTh}

So far we have derived the Mori-type equations without isothermal
condition, which is a condition necessary to hold the kinetic temperture
unchanged.
In this subsection, we attempt to implement the isothermal condition.
The derivation of the resulting Mori-type equations will be performed in
section \ref{subsec:IsoThMori}.

At the level of the SLLOD equation, it is known that the isothermal
condition is satisfied by the specific choice of $\alpha(t)$ given
by Eq.~(\ref{Eq:alpha_GIK}), which is referred to as the GIK thermostat.
In principle, the Mori-type equations derived from the corresponding
Liouville equation, with $\alpha(\vg)$ given by
Eq.~(\ref{Eq:alpha_GIK}), no longer contain any multiplier, and the
invariance of the kinetic temperature is assured automatically.
However, this choice of $\alpha(\vg)$ leads to a difficulty in deriving
the Mori-type equations, since the integral of a rational function with
Gaussian weight is difficult to perform explicitly.

This motivates us to implement the isothermal condition {\it at the
level of the Mori-type equations}, which is attained by requiring the
time derivative of the {\it average kinetic temperature} to vanish.
Note that this scheme requires the time derivative of the kinetic
temperature, not necessarily itself but at least its average, to vanish.
In this scheme, we avoid to fully incorporate the fluctuations as in
Eq.~(\ref{Eq:alpha_GIK}), and attempt to incorporate them partially,
compensating it with a {\it multiplier at the level of the Mori-type
equations}.
We denote this multiplier, which corresponds to the multiplier of the
Liouville equation $\alpha(\vg)$, as $\lambda_{\alpha}(t)$ in the
remainder.

By differentiating the generalized Green-Kubo formula for the kinetic
temperature, we obtain
\begin{eqnarray}
\hspace{-2em}
\frac{d}{dt}
\eav{T(\vg(t))}
&=&
\frac{2}{3N k_B}
\eav{K(\vg(t)) \Omega(\vg)}
\nn \\
&=&
- \frac{2\beta}{3N k_B}
\left\{
\dgamma G_{K,\sigma}(t)
+ 2 G_{K, \alpha \delta K}(t)
\right\},
\label{Eq:dTdt}
\end{eqnarray}
where
\begin{eqnarray}
G_{K,\sigma}(t)
&\equiv&
 \eav{[U(t)K(\vg)]\sigma_{xy}(\vg)}
\end{eqnarray}
and
\begin{eqnarray}
G_{K, \alpha\delta K}(t)
&\equiv&
\eav{[U(t)K(\vg)]\alpha(\vg)\delta K(\vg)}.
\label{Eq:GKadK_formal}
\end{eqnarray}
Hence, the isothermal condition at the level of the Mori-type equations
reads
\begin{eqnarray}
\dgamma G_{K,\sigma}(t)
+ 2 G_{K, \alpha \delta K}(t)
= 0.
\label{Eq:isoth_bare}
\end{eqnarray}
In order to satisfy this constraint, we introduce $\lambda_\alpha(t)$ in
$G_{K, \alpha \delta K}(t)$, the result of which we denote $G_{K, \alpha
\delta K}^{(\lambda)}(t)$, and recast Eq.~(\ref{Eq:isoth_bare}) as
\begin{eqnarray}
\dgamma G_{K,\sigma}(t)
+ 2 G_{K, \alpha \delta K}^{(\lambda)}(t)
= 0.
\label{Eq:isoth}
\end{eqnarray}
However, from Eq.~(\ref{Eq:GKadK_formal}), we can see that the
definition of $G_{K, \alpha \delta K}^{(\lambda)}(t)$ is
non-trivial, since the effect of the time-evolution of the multiplier
appears in $U(t)$, rather than $\alpha(\vg)$.
This leads us to define a ``renormalized'' time-evolution operator
$U_R(t)$, where
\begin{eqnarray}
U_R(t)
&\equiv&
\lambda_\alpha(t) U(t),
\label{Eq:multiplier}
\end{eqnarray}
and define $G_{K, \alpha \delta K}^{(\lambda)}(t)$ as
\begin{eqnarray}
G_{K, \alpha\delta K}^{(\lambda)}(t)
&\equiv&
\eav{[U_R(t)K(\vg)]\alpha(\vg)\delta K(\vg)}
\nn \\
&=&
\lambda_\alpha(t)
G_{K, \alpha\delta K}(t) .
\label{Eq:GKadK_formal2}
\end{eqnarray}
Here, $\lambda_\alpha(t)$ factors out, since it is not a phase-space
variable.
Then, if $G_{K, \alpha\delta K}(t) \neq 0$, choosing the multiplier as
\begin{eqnarray}
\lambda_\alpha(t)
&=&
- \frac{\dgamma}{2}
\frac{G_{K, \sigma}(t)}{G_{K, \alpha\delta K}(t)}
\label{Eq:lambdat}
\end{eqnarray}
assures the isothermal condition Eq.~(\ref{Eq:isoth}) to be satisfied.
Note that, while $\alpha(\vg)$ is expressed in terms of the phase-space
variables $\vg$, $\lambda_\alpha(t)$ is expressed in terms of
time-correlators at time $t$, when the constraint Eq.~(\ref{Eq:isoth})
is satisfied.
Concrete examples of $\alpha(\vg)$ and $\lambda_\alpha(t)$ will be
discussed in section \ref{sec:IsoTh_MCT}.
%

\subsection{Isothermal Mori-type equations}
\label{subsec:IsoThMori}

Now we figure out how the multiplier $\lambda_\alpha(t)$ enters in the
Mori-type equations.
The identification of the complete correction is possible, but rather
lengthy.
Furthermore, most of the modifications vanish in the mode-coupling
approximation, which we will discuss in section \ref{sec:MCA}.
Hence, we only focus on the modification which survives the
mode-coupling approximation, which is for the friction term $\alpha_0
H_\vq^\lambda(t)$ in Eq.~(\ref{eom-H2}).
As shown in Appendix \ref{App:subsec:MoriEq}, this terms arises from the
``correlated part'' in Eq.~(\ref{dUj}), which is explicitly
\begin{eqnarray}
&&
\hspace{-5em}
\eav{
\left[
U(t)
\bar{\mP}_t
e^{i\Liouv_{\dgamma_r}^\dagger t}
i\Liouv_\alpha
j_{\vq(t)}^\lambda
\right]
n_\vq^*
}
\nn \\
&=&
\eav{
\left[
U(t)
\bar{\mP}_t
e^{i\Liouv_{\dgamma_r}^\dagger t}
\alpha(\vg)
i\Liouv_p
j_{\vq(t)}^\lambda
\right]
n_\vq^*
}.
\end{eqnarray}
Here, we extract $\alpha(\vg)$ and define the remaining part as
$i\Liouv_p \equiv - \sum_i \ve{p}_i \cdot \del / \del \ve{p}_i$.
We introduce $\lambda_\alpha(t)$ here as is done in
Eq.~(\ref{Eq:multiplier}), which leads to
\begin{eqnarray}
\eav{
\left[
U_R(t)
\bar{\mP}_t
e^{i\Liouv_{\dgamma_r}^\dagger t}
\alpha(\vg)
i\Liouv_p
j_{\vq(t)}^\lambda
\right]
n_\vq^*
}
=
\lambda_\alpha(t)
\alpha_0,
\end{eqnarray}
where the explicit form of $\alpha(\vg)$ which will be specified in
section \ref{sec:IsoTh_MCT}, Eq.~(\ref{Eq:alpha}), is utilized.
Hence, the Mori-type equation with isothermal condition is
\begin{eqnarray}
\frac{d}{dt}
H_\vq^\lambda(t)
&=&
-
v_T^2
\frac{q(t)^\lambda }{S_{q(t)}}
\Phi_\vq(t)
-
\lambda_\alpha(t) \alpha_0 H_{\vq}^\lambda(t)
-
\left[
\ve{\kappa} \cdot \vH_{\vq}(t)
\right]^\lambda
\nn \\
&&
+
\frac{i}{N}
\eav{
R_{\vq(t)}^\lambda(t)
n_\vq^*
}
-
\int_0^t ds
L_{\vq}^\lambda(t,s)
\Phi_\vq(s)
\nn \\
&&
-
\int_0^t ds
M_{\vq}^{\lambda \mu}(t,s)
H_\vq^\mu(s).
\label{eom-H3}
\end{eqnarray}
The fact that $\lambda_\alpha(t)$ does not appear in the memory kernels
in the mode-coupling approximation can be verified in section
\ref{sec:MCA}, where the resulting memory kernels turn out to be
independent of $\alpha_0$.
In addition, as shown in section \ref{sec:MCA}, the vanishing of $\eav{
R_{\vq(t)}^\lambda(t)
n_\vq^*
}$ is not affected by $\alpha(\vg)$, and hence $\lambda_\alpha(t)$ does
not appear here either.

In principle, the Mori-type equations which follow from the choice
Eq.~(\ref{Eq:alpha_GIK}) of $\alpha(\vg)$ contain no multiplier, and
is expected to satisfy the isothermal condition
Eq.~(\ref{Eq:isoth_bare}) automatically.
On the other hand, the Mori-type equations Eqs.~(\ref{eom-Phi}) and
(\ref{eom-H3}), which follow from $\alpha(\vg)$ of Eq.~(\ref{Eq:alpha}),
contain a multiplier $\lambda_\alpha(t)$, and the isothermal condition
Eq.~(\ref{Eq:isoth}) is explicitly satisfied if $\lambda_\alpha(t)$ is
chosen as Eq.~(\ref{Eq:lambda_MCT}), which will be derived in section
\ref{sec:IsoTh_MCT}.
We expect that the physical equivalence is achieved for these two
schemes.

The Mori-type equations of the underdamped and overdamped \cite{FC2009}
cases are compared and discussed in Appendix \ref{App:sec:Mori}.

\section{Mode-coupling approximation}
\label{sec:MCA}

We have derived an isothermal Mori-type equation for the cross
time-correlator $H_\vq^\lambda(t)$, Eq.~(\ref{eom-H3}), in the previous
section.
However, this is not a closed equation for the time-correlators
$\Phi_\vq(t)$ and $H_\vq^\lambda(t)$, unless the memory kernels are
expressed in terms of them.
For this purpose, we introduce a {\it time-dependent} second projection
operator \cite{FC2009}, which extracts the dynamics correlated with the
slowly-varying pair-density modes:
\begin{eqnarray}
\mP_2 (t) X
&\equiv &
\sum_{\vk > \vp}
\frac{\eav{X n_{\vk(t)}^* n_{\vp(t)}^*} }{N^2 S_{k(t)} S_{p(t)}}
n_{\vk(t)} n_{\vp(t)}.
\label{P2}
\end{eqnarray}
The normalization factor is determined by the factorization
approximation of the equal-time (equilibrium) four-point function of the
density fluctuations, $\eav{n_{\vk(t)} n_{\vp(t)} n_{\vk'(t)}^*
n_{\vp'(t)}^*} \simeq \eav{n_{\vk(t)} n_{\vk'(t)}^*} \eav{n_{\vp(t)}
n_{\vp'(t)}^*} = \delta_{\vk, \vk'}\delta_{\vp, \vp'} N^2 S_{k(t)}
S_{p(t)} \hspace{1em} (\vk > \vp, \vk' > \vp')$.
The second projection operator Eq.~\eqref{P2} is idempotent and
Hermitian, similar to the projection operator Eq.~(\ref{P}).

In underdamped MCT, there exists a potential next-to-leading second
projection operator, which projects the dynamics onto the
density-current modes \cite{CSOH2012}.
However, it can be shown that this projection is negligible as compared
to the projection onto the pair-density modes, at least
for time-reversible thermostatted systems  \cite{SH2012p1}.
Hence, the choice of Eq.~(\ref{P2}) is assured.
Note that this feature does not always hold, e.g. for granular systems,
where the projection onto the density-current modes plays an
essential role \cite{SH2012p2, CSOH2012}.

There is one subtle issue we should handle in order for the application
of Eq.~(\ref{P2}) to work
\footnote{
The problem discussed here for the ``uncorrelated part'' is avoided for
the ``correlated part'' by the introduction of the rescaled static
projection operator $\bar{\mP}_t$ defined in Eq.~\eqref{bP}.
}.
The operator which appears in the memory kernels defined in
Eqs.~(\ref{L}) and (\ref{M}), $i\tLiouv \tilde{U}_0(t,s)
R_{\vq(t)}^\lambda$, can be deformed as follows,
\begin{eqnarray}
&&
\hspace{-3em}
i\tLiouv
\tilde{U}_0(t,s)
R_{\vq(t)}^\lambda
\nn \\
&=&
i\tLiouv
\left[
1 + \Sigma(s)
\right]
e^{-i\Liouv_{\dgamma_r}^\dagger s}
U_0(t,s)
e^{i\Liouv_{\dgamma_r} t}
\mQ(t)
R_{\vq(t)}^\lambda,
\label{approx}
\end{eqnarray}
where Eq.~(\ref{tU0}), the idempotency of $\mQ(t)$, and the identity
\begin{eqnarray}
e^{-i\Liouv_{\dgamma_r} t}
&=&
\left[
1
+
\Sigma(t)
\right]
e^{-i\Liouv_{\dgamma_r}^\dagger t}
\label{id}
\end{eqnarray}
is applied.
Here,
\begin{eqnarray}
\Sigma(t)
&\equiv&
\beta \dgamma
\int_0^t ds
e^{i\Liouv_{\dgamma_r}^\dagger s}
\sigma_{xy}^{(\mathrm{pot})}
e^{-i\Liouv_{\dgamma_r} s}
\label{Sigmat}
\end{eqnarray}
is the accumulated elastic energy due to shear, where
$\sigma_{xy}^{(\mathrm{pot})}$ is the potential part of the shear stress
introduced in Eq.~(\ref{sigma_pot}).
As discussed in FC \cite{FC2009}, the shear-induced term $\Sigma(t)$ is
an obstacle for the application of the second projection operator and
the factorization approximation.
We assume here $\Sigma(t) \simeq 0$, whose validation is discussed in FC
\cite{FC2009}.
At least, this assumption is valid in the weak-shear regime, since
$\Sigma(t)$ is proportional to $\dgamma$.

The neglect of $\Sigma(t)$ leads to the following relation,
\begin{eqnarray}
e^{-i\Liouv_{\dgamma_r}^\dagger s}
U_0(t,s)
&=&
\mQ(s) e^{-i\Liouv_{\dgamma_r}^\dagger s}
U_0(t,s),
\label{insertQ}
\end{eqnarray}
whose proof is given in Appendix \ref{App:subsec:InsertionQ}.
The first implication of Eq.~(\ref{insertQ}) is the orthogonality of the
random force,
\begin{eqnarray}
\hspace{-2em}
\eav{ R_{\vq(t)}^\lambda(t) \xi_\vq^*}
&=&
\eav{
\left[
U_0(t,0) e^{i\Liouv_{\dgamma_r} t} R_{\vq(t)}^\lambda
\right]
\xi_\vq^*}
\nn \\
&=&
\eav{
\left[
\mQ(0) U_0(t,0) e^{i\Liouv_{\dgamma_r} t} R_{\vq(t)}^\lambda
\right]
\xi_\vq^*}
\nn \\
&=&
\eav{
\left[
U_0(t,0) e^{i\Liouv_{\dgamma_r} t} R_{\vq(t)}^\lambda
\right]
\mQ(0) \xi_\vq^*}
=
0,
\label{RandOrth}
\end{eqnarray}
where $\xi = n$ or $j$.
The second implication is the form of the aforementioned operator
described in Eq.~\eqref{approx},
\begin{eqnarray}
&&
\hspace{-3em}
i\tLiouv
\tilde{U}_0(t,s)
R_{\vq(t)}^\lambda
\nn \\
&\simeq&
i\tLiouv
\mQ(s)
e^{-i\Liouv_{\dgamma_r}^\dagger s}
U_0(t,s)
e^{i\Liouv_{\dgamma_r} t}
\mQ(t)
R_{\vq(t)}^\lambda,
\label{approx2}
\end{eqnarray}
which now has the desirable feature, i.e. the projection of $U_0(t,s)$
is now complete.
The memory kernels Eqs.~(\ref{L}) and (\ref{M}) can be rewritten, from
Eq.~(\ref{approx2}), as follows:
\begin{eqnarray}
iL_\vq^\lambda (t,s)
&=&
\frac{1}{NS_{q(t)}}
\eav{
\left[
\tilde{U}_0'(t,s)
R_{\vq(t)}^\lambda
\right]
\mQ(s)
\left[
n_{\vq(s)}^* \tOmega
\right]
},
\hspace{1.5em}
\label{L2}
\end{eqnarray}
%
%
\begin{eqnarray}
M_\vq^{\lambda \mu}(t,s)
&=&
\frac{1}{N v_T^2}
\eav{
\left[
\tilde{U}_0'(t,s)
R_{\vq(t)}^\lambda
\right]
\Delta
R_{\vq(s)}^{\mu *}
}.
\label{M2}
\end{eqnarray}
Here,
\begin{eqnarray}
\tilde{U}_0'(t,s)
&\equiv&
\mQ(s)
e^{-i\Liouv_{\dgamma_r}^\dagger s}
U_0(t,s)
e^{i\Liouv_{\dgamma_r}t}
\mQ(t)
\label{U0prime}
\end{eqnarray}
is the modified projected time-evolution operator,
\begin{eqnarray}
\Delta
R_{\vq(s)}^{\mu}
&\equiv&
R_{\vq(s)}^{\mu}
-
\mQ(s)
\left[
j_{\vq(s)}^{\mu} \tOmega
\right]
\label{dRand}
\end{eqnarray}
is the modified random force, and $\tOmega$ is the modified work
function defined in Eq.~(\ref{tOmega}).
The derivation of the above equations Eqs.~(\ref{L2})--(\ref{dRand}) are
shown in Appendix \ref{App:subsec:MemKer}.

Now we insert the second projection operator Eq.~(\ref{P2}) into
Eqs.~(\ref{L2}) and (\ref{M2}) as $\mQ(s) e^{-i\Liouv_{\dgamma
r}^\dagger s} U_0(t,s) e^{i\Liouv_{\dgamma_r} t} \mQ(t)$ $\simeq
\mP_2(s) \mQ(s) e^{-i\Liouv_{\dgamma_r}^\dagger s} U_0(t,s)
e^{i\Liouv_{\dgamma_r} t} \mQ(t) \mP_2(t)$, which results in the
following form, i.e. products of vertex functions at times $s$ and $t$,
bridged by a propagator from time $s$ to $t$:
\begin{widetext}
\begin{eqnarray}
iL_\vq^\lambda (t,s)
&\simeq&
\frac{1}{NS_{q(t)}}
\sum_{\vk'>\vp'}
\sum_{\vk>\vp}
\frac{
\eav{
R_{\vq(t)}^\lambda
n_{\vk(t)}^*
n_{\vp(t)}^*
}
}{N^2 S_{k(t)} S_{p(t)}}
\cdot
\eav{
\left[
\tilde{U}_0'(t,s)
n_{\vk(t)}
n_{\vp(t)}
\right]
n_{\vk'(s)}^*
n_{\vp'(s)}^*
}
\cdot
\frac{
\eav{
n_{\vk'(s)}
n_{\vp'(s)}
\mQ(s)
\left[
n_{\vq(s)}^*
\tOmega
\right]
}
}{N^2 S_{k'(s)} S_{p'(s)}},
\label{L3}
\nn \\
\\
M_\vq^{\lambda \mu}(t,s)
&\simeq&
\frac{1}{N v_T^2}
\sum_{\vk'>\vp'}
\sum_{\vk>\vp}
\frac{
\eav{
R_{\vq(t)}^\lambda
n_{\vk(t)}^*
n_{\vp(t)}^*
}
}{N^2 S_{k(t)} S_{p(t)}}
\cdot
\eav{
\left[
\tilde{U}_0'(t,s)
n_{\vk(t)}
n_{\vp(t)}
\right]
n_{\vk'(s)}^*
n_{\vp'(s)}^*
}
\cdot
\frac{
\eav{
n_{\vk'(s)}
n_{\vp'(s)}
\Delta
R_{\vq(s)}^{\mu *}
}
}{N^2 S_{k'(s)} S_{p'(s)}}.
\label{M3}
\end{eqnarray}
\end{widetext}
We can see from the above expressions that the derived forms of the
memory kernels are consistent with the principle of the ``alignment of
the wavevectors''; the vertex function at time $t$ includes as indices
only the advected wavevectors with argument $t$, e.g. $\vq(t)$, and a
similar feature also holds for the propagator.

The remaining tasks are the calculation of the vertex functions and the
approximation of the propagators .
As for the vertex functions, the convolution approximation \cite{HM} is
applied.
We only show the results below, since the derivation, which is shown in
Appendix \ref{App:subsec:VF}, is straightforward.
\begin{eqnarray}
\frac{
\eav{
R_{\vq(t)}^\lambda
n_{\vk(t)}^*
n_{\vp(t)}^*
}
}{N^2 S_{k(t)} S_{p(t)}}
&=&
-i \frac{n}{N}
\delta_{\vq,\vk+\vp}
V_{\vq(t),\vk(t),\vp(t)}^{\lambda},
\label{vf1}
\end{eqnarray}
%
%
\begin{eqnarray}
\frac{
\eav{n_{\vk(t)} n_{\vp(t)} \Delta R_{\vq(t)}^{\lambda *} }
}
{N^2 S_{k(t)} S_{p(t)}}
&=&
i
\frac{n}{N}
\delta_{\vq,\vk+\vp}
V_{\vq(t),\vk(t),\vp(t)}^{\lambda *},
\label{vf2}
\end{eqnarray}
%
%
\begin{eqnarray}
V_{\vq,\vk,\vp}^{\lambda}
&\equiv &
v_T^2
\left( k^\lambda c_k + p^\lambda c_p  \right),
\label{V}
\end{eqnarray}
%
%
\begin{eqnarray}
\frac{
\eav{
n_{\vk(t)} n_{\vp(t)} \mQ(t) \left[n_{\vq(t)}^* \tOmega \right] }
}
{N^2 S_{k(t)} S_{p(t)}}
&=&
0.
\label{vf3}
\end{eqnarray}
As for the propagator, we adopt the factorization approximation, which
replaces it with the product of the projection-free propagators:
\begin{eqnarray}
&&
\hspace{-5em}
\frac{1}{N^2}
\eav{
\left[
\tilde{U}_0'(t,s)
n_{\vk(t)}
n_{\vp(t)}
\right]
n_{\vk'(s)}^*
n_{\vp'(s)}^*
}
\nn \\
&\simeq&
\delta_{\vk', \vk}
\delta_{\vp', \vp}
\Phi_{\vk(s)}(t-s)
\Phi_{\vp(s)}(t-s).
\label{FA}
\end{eqnarray}
The derivation of Eq.~(\ref{FA}) is shown in Appendix
\ref{App:subsec:FA}.

From Eqs.~(\ref{L3})--(\ref{FA}), we arrive at the final expressions for
the memory kernels,
\begin{eqnarray}
iL_\vq^\lambda(t,s)
&=&
0,
\label{L4}
\\
M_\vq^{\lambda \mu}(t,s)
&=&
\frac{n}{2 v_T^2}
\int \frac{d^3 \vk}{(2\pi)^3}
V_{\vq(t),\vk(t),\vp(t)}^{\lambda}
V_{\vq(s),\vk(s),\vp(s)}^{\mu *}
\nn \\
&&
\times
\Phi_{\vk(s)}(t-s)
\Phi_{\vp(s)}(t-s),
\label{M4}
\end{eqnarray}
where the summation of the wavevectors is replaced by the integral, and
$\vp \equiv \vq - \vk$.
Note that the memory kernel $L_\vq^\lambda$ introduced by CK
\cite{CK2009} vanishes in our formulation.
This helps us to connect our formulation with the previous ones
\cite{MR2002, FC2002, FC2009}.
Note also that the memory kernel in Eq.~(\ref{M4}) cannot be expressed
solely in terms of the time difference $t-s$, even after the application
of the mode-coupling approximation.
This is in contrast to the cases of CK \cite{CK2009}, where the memory
kernels depend only on the time difference $t-s$ at the level of the
Mori-type equations.

\section{Steady-state properties}
\label{sec:SSP}

In the previous sections, we have derived a set of closed equations for
the time-correlators $\Phi_\vq(t)$ and $H_\vq^\lambda(t)$, i.e.
Eqs.~(\ref{eom-Phi}) and (\ref{eom-H3}), where the memory kernels are
given by Eqs.~(\ref{L4}) and (\ref{M4}).
According to the ITT scheme \cite{FC2002, FC2005, FC2009}, the
steady-state properties are written in terms of the time-correlators,
with equilibrium quantities (e.g. the static structure factor) as the
only inputs.
We follow this scheme and derive a closed formula for the steady-state
quantities.
In the course of this procedure, we will derive an explicit expression
of the multiplier $\lambda_\alpha(t)$ introduced in section
\ref{subsec:IsoTh}.
A specific formula for the steady-state shear stress is derived in
section \ref{section:stressformula}.

\subsection{Time-correlators of interest}

From the analogue of the Green-Kubo relation Eq.~(\ref{Ass}),
time-correlators of interest are of the following form,
\begin{eqnarray}
\hspace{-0.5em}
G_{A,B}(t)
&\equiv&
\eav{A(\vg(t))B(\vg)}
=
\eav{\left[U(t)A(\vg) \right]B(\vg)},
\hspace{1em}
\label{Eq:GAB}
\\
B(\vg)
&\equiv&
\sigma_{xy}(\vg)
\hspace{0.5em}
{\rm or}
\hspace{0.5em}
\alpha(\vg) \delta K(\vg).
\end{eqnarray}
In this paper, we concentrate on the steady-state kinetic temperature
and the shear stress; i.e. we consider the specific cases, $A(\vg) =
K(\vg)$ and $\sigma_{xy}(\vg)$.

Similarly to the case of static projection operators \cite{CK2009}, we
can show that the time-correlator Eq.~(\ref{Eq:GAB}) resides in the subspace
orthogonal to the density and the current-density fluctuations, i.e.
\begin{eqnarray}
G_{A,B}(t)
&=&
\eav{\left[U_0(t,0) A(\vg) \right]B(\vg)},
\label{GA2}
\end{eqnarray}
whose proof is given in Appendix \ref{App:subsec:PTC}.

Now we apply the second projection operator to Eq.~(\ref{GA2}).
Since $B(\vg) = \sigma_{xy}(\vg)$ and $\alpha(\vg) \delta K(\vg)$ are
zero-wavevector quantities, it is sufficient to project onto the
``zero-mode'' pair-density correlator,
\begin{eqnarray}
\mP_2^{0}(t) X
&\equiv&
\sum_{\vk >0}
\frac{\eav{X n_{\vk(t)}^* n_{\vk(t)}} }{N^2 S_{k(t)}^2}
n_{\vk(t)} n_{\vk(t)}^*,
\end{eqnarray}
which is a restricted form of Eq.~(\ref{P2}).
The mode-coupling approximation to Eq.~(\ref{GA2}) is then
\begin{eqnarray}
\hspace{-1em}
G_{A,B}(t)
&\simeq &
\eav{\left[
\mP_{2}^0 (0)
\tilde{U}_0(t,0)
\mP_{2}^0 (t)
A(\vg)\right]B(\vg)}
\nn \\
&=&
\eav{\left[
\tilde{U}_0(t,0)
\mP_{2}^0 (t)
A(\vg)\right]
\mP_{2}^0 (0) B(\vg)},
\label{Eq:GAB_mct}
\end{eqnarray}
where the Hermiticity of the projection operator $\mP_2^0$ is applied in
the last equality.
Note that $\tilde{U}_0(t,0)$, rather than $U_0(t,0)$, appears due to the
insertion of $\mP_2^0(t)$.

\subsection{Isothermal condition in MCT}
\label{sec:IsoTh_MCT}

To proceed further, it is necessary to calculate the explicit form of
the multiplier $\lambda_\alpha(t)$ given by Eq.~(\ref{Eq:lambdat}).
For this purpose, we specify the form of $\alpha(\vg)$ and explicitly
calculate $G_{K, \sigma}(t)$ and $G_{K, \alpha \delta K}(t)$, which are
given in the mode-coupling approximation as
\begin{eqnarray}
G_{K, \sigma}(t)
&\simeq&
\left\langle
\left[ \tilde{U}_0(t,0) \mP_{2}^0 (t)
K(\vg)\right]
\mP_{2}^0 (0) \sigma_{xy}(\vg) \right\rangle,
\hspace{2.5em}
\label{Eq:GKsigma}
\\
G_{K, \alpha\delta K}(t)
&\simeq&
\left\langle
\left[ \tilde{U}_0(t,0) \mP_{2}^0 (t)
K(\vg)\right]
\mP_{2}^0 (0) \alpha(\vg) \delta K(\vg) \right\rangle.
\label{Eq:GKadK}
\nn \\
\end{eqnarray}

In the following, we examine concrete examples of $\alpha(\vg)$ and
$\lambda_\alpha(t)$.
The simplest choice of $\alpha(\vg)$ is to fully neglect its
$\vg$-dependence.
In this case, no fluctuations are incorporated, and $\alpha(\vg)$ is a
constant, which we denote $\alpha_0$.
However, we illustrate here that this choice always causes a heating-up
of the kinetic temperature
\footnote{The resulting kinetic temperature is not divergent, since the
density time-correlator decays to zero around $t \sim \dgamma^{-1}$, but
is independent of $\alpha_0$, i.e. unaffected by the strength of
dissipation}.
For $\alpha(\vg) = \alpha_0$, Eq.~(\ref{Eq:GKadK}) reads
\begin{eqnarray}
G_{K, \alpha\delta K}(t)
&=&
\alpha_0
\left\langle
\left[ \tilde{U}_0(t,0) \mP_{2}^0 (t)
K(\vg)\right]
\mP_{2}^0 (0) \delta K(\vg) \right\rangle,
\hspace{2.5em}
\end{eqnarray}
which vanishes due to $\mP_{2}^0 (0) \delta K(\vg) = 0$.
This implies, together with the fact that $G_{K,\sigma}(t) \neq 0$ which
is shown below, that the isothermal condition Eq.~(\ref{Eq:isoth})
cannot be satisfied. 
Hence, the average kinetic temperature calculated this way is unphysical
and cannot be regarded as a steady-state temperature.
Note that this case is the one adopted in CK~\cite{CK2009}.

From the above argument, it is shown that, in order to retain $G_{K,
\alpha \delta K}(t)$ non-vanishing, it is at least necesssary to
introduce fluctuations into $\alpha(\vg)$.
The simplest choice for this is to incorporate current fluctuations as
\begin{eqnarray}
\alpha(\vg)
&=&
\frac{\alpha_0}{\frac{3}{2}N k_B T}
\sum_i
\frac{\ve{p}_i^2}{2m},
\label{Eq:alpha}
\end{eqnarray}
where $\alpha_0$ is a constant which gives the initial strength of the
dissipative coupling to the thermostat.
Now we calculate the projected properties $\mP_{2}^0 (0) [\alpha(\vg)
\delta K(\vg)] $, $\mP_{2}^0 (0) \sigma_{xy}(\vg)$, and $\mP_{2}^0 (t)
K(\vg)$ which appear in Eqs.~(\ref{Eq:GKsigma}) and (\ref{Eq:GKadK}),
with the choice of Eq.~(\ref{Eq:alpha}) for $\alpha(\vg)$.
Straightforward calculation leads to
\begin{eqnarray}
\hspace{-2em}
\mP_{2}^0 (0) [\alpha(\vg) \delta K(\vg)]
&=&
\alpha_0
\frac{k_B T}{N}
\sum_{\ve{k}>0}
\frac{n_{\ve{k}} n_{\ve{k}}^*}{S_k},
\label{Eq:P2alphadK}
\\
\mP_{2}^0 (t) K(\vg)
&=&
\frac{3}{2} k_B T
\sum_{\ve{k}>0}
\frac{n_{\ve{k}(t)} n_{\ve{k}(t)}^*}{S_{k(t)}},
\label{Eq:P2K}
\end{eqnarray}
and
\begin{eqnarray}
\hspace{-2em}
\mP_{2}^0 (t)
\sigma_{xy}(\vg)
&=&
-\frac{k_B T}{N}
\sum_{\vk >0}
\frac{W_{\vk(t)}}{S_{k(t)}}
n_{\vk(t)} n_{\vk(t)}^*,
\label{Psigma}
\\
W_{\vk}
&\equiv&
\frac{k_x k_y}{k}
\frac{1}{S_{k}}
\frac{\del S_{k}}{\del k},
\end{eqnarray}
is derived in Appendix \ref{App:subsec:PSS}.
From Eqs.~(\ref{Eq:GKsigma}), (\ref{Eq:GKadK}), and
(\ref{Eq:P2alphadK})--(\ref{Psigma}), $G_{K,\sigma}(t)$ and $G_{K,
\alpha \delta K}(t)$ are given by
\begin{eqnarray}
G_{K, \sigma}(t)
&=&
- \frac{3}{2}
\frac{\left( k_B T \right)^2}{N}
\sum_{\vk>0}
\sum_{\vk'>0}
\frac{1}{S_{k(t)}}
\frac{W_{\vk'}}{S_{k'}}
\nn \\
&&
\times
\eav{\left[
\tilde{U}_0(t,0)
n_{\vk(t)} n_{\vk(t)}^*
\right]
n_{\vk'} n_{\vk'}^*}
\label{Eq:GKs}
\end{eqnarray}
and
\begin{eqnarray}
\hspace{-1.5em}
G_{K, \alpha \delta K}(t)
&=&
\frac{3}{2}
\alpha_0
\frac{\left( k_B T \right)^2}{N}
\sum_{\vk>0}
\sum_{\vk'>0}
\frac{1}{S_{k(t)}}
\frac{1}{S_{k'}}
\nn \\
&&
\times
\eav{\left[
\tilde{U}_0(t,0)
n_{\vk(t)} n_{\vk(t)}^*
\right]
n_{\vk'} n_{\vk'}^*},
\hspace{1em}
\label{Eq:GKalphadK}
\end{eqnarray}
respectively.
Application of the factorization approximation to the four-point
function reads
\begin{eqnarray}
&&
\hspace{-6em}
\frac{1}{N^2}
\eav{\left[
\tilde{U}_0(t,0)
n_{\vk(t)} n_{\vk(t)}^*
\right]
n_{\vk'} n_{\vk'}^*
}
\nn \\
&\simeq &
\left[
\delta_{\vk',\vk}
+
\delta_{\vk',-\vk}
\right]
\Phi_\vk(t)^2,
\label{4point}
\end{eqnarray}
where $n_\vk^* = n_{-\vk}$, $\Phi_{-\vk}(t) = \Phi_\vk(t)$ are applied.
The derivation of Eq.~(\ref{4point}) is shown in Appendix
\ref{App:subsec:FA}.
From Eqs.~(\ref{Eq:lambdat}) and (\ref{Eq:GKs})--(\ref{4point}), the
functional form of the multiplier $\lambda_\alpha(t)$ should be given by
\begin{eqnarray}
\lambda_\alpha(t)
&=&
\frac{\dgamma}{2\alpha_0}
\frac{
\sum_{\ve{k}>0}
\frac{1}{S_{k(t)}}
\frac{W_{\ve{k}}}{S_k}
\Phi_{\ve{k}}(t)^2
}
{
\sum_{\ve{k}>0}
\frac{1}{S_{k(t)}S_k}
\Phi_{\ve{k}}(t)^2
}
\label{Eq:lambda_MCT}
\end{eqnarray}
in order to satisfy the constraint Eq.~(\ref{Eq:isoth}).
This can be regarded as an analogue for MCT of the GIK thermostat,
Eq.~(\ref{Eq:alpha_GIK}), for molecular dynamics.

We note here that, with the choice of Eq.~(\ref{Eq:alpha}), the second
term of $\Lambda(\vg)$ in Eq.~(\ref{Lambda}) reads $\sum_i [\del
\alpha(\vg) / \del \ve{p}_i] \cdot \ve{p}_i = 2 \alpha(\vg)$.
This is negligible compared to the first term of Eq.~(\ref{Lambda}),
which is proportional to $N$, in the thermodynamic limit.

\subsection{Steady-state stress formula}
\label{section:stressformula}

We define the steady-state shear stress by the steady-state stress
tensor as follows,
\begin{eqnarray}
\eav{\sigma}_{\mathrm{SS}}
&\equiv&
- \frac{1}{V}
\eav{\sigma_{xy}}_{\mathrm{SS}}.
\end{eqnarray}
From the steady-state formula Eqs.~(\ref{Ass}) and (\ref{Eq:Ass_def}),
and reminding that $\eav{\sigma_{xy}(\vg)}=0$, the steady-state shear
stress is given by the time-correlators as
\begin{eqnarray}
\eav{\sigma}_{\mathrm{SS}}
&=&
\frac{\beta \dgamma}{V}
\int_0^\infty dt
G_{\sigma,\sigma}(t)
+
\frac{2\beta}{V}
\int_0^\infty dt
G_{\sigma,\alpha\delta K}^{(\lambda)}(t),
\hspace{2em}
\label{Eq:sigmaSS}
\end{eqnarray}
where the time-correlators $G_{\sigma, \sigma}(t)$ and $G_{\sigma,
\alpha \delta K}^{(\lambda)}(t)$ are given by
\begin{eqnarray}
G_{\sigma, \sigma}(t)
&\equiv&
\eav{\left[
U(t)
\sigma_{xy}(\vg)
\right]
\sigma_{xy}(\vg)}
\hspace{2em}
\label{Eq:Gss}
\end{eqnarray}
and
\begin{eqnarray}
G_{\sigma, \alpha \delta K}^{(\lambda)}(t)
&\equiv&
\eav{\left[
U_R(t)
\sigma_{xy}(\vg)
\right]
\alpha(\vg) \delta K(\vg)}
\nn \\
&=&
\lambda_\alpha(t)
G_{\sigma, \alpha \delta K}(t).
\label{Eq:GsalphadK}
\end{eqnarray}
In Eq.~(\ref{Eq:GsalphadK}), we have introduced the multiplier
$\lambda_\alpha(t)$ by applying $U_R(t)$, Eq.~(\ref{Eq:multiplier}),
which is required from the arguments of section \ref{subsec:IsoTh}.
In the mode-coupling approximation, $G_{\sigma, \sigma}(t)$ and
$G_{\sigma, \alpha \delta K}^{(\lambda)}(t)$ are approximated as
\begin{eqnarray}
G_{\sigma, \sigma}(t)
&\simeq&
\eav{\left[
\tilde{U}_0(t,0)
\mP_{2}^0 (t)
\sigma_{xy}(\vg)
\right]
\mP_{2}^0 (0) \sigma_{xy}(\vg)}
\hspace{2em}
\label{Eq:Gss_MCT}
\end{eqnarray}
and
\begin{eqnarray}
G_{\sigma, \alpha \delta K}^{(\lambda)}(t)
&\simeq&
\lambda_\alpha(t)
\left\langle
\left[
\tilde{U}_0(t,0)
\mP_{2}^0 (t)
\sigma_{xy}(\vg)
\right]
\right.
\nn \\
&&
\times
\left.
\mP_{2}^0 (0)
\left[
\alpha(\vg) \delta K(\vg)
\right]
\right\rangle,
\label{Eq:GsalphadK_MCT}
\end{eqnarray}
respectively.
From Eqs.~(\ref{Eq:P2alphadK}), (\ref{Psigma}), (\ref{4point}),
(\ref{Eq:Gss_MCT}), and (\ref{Eq:GsalphadK_MCT}), Eq.~(\ref{Eq:sigmaSS})
reads
\begin{eqnarray}
\eav{\sigma}_{\mathrm{SS}}
&=&
\frac{k_B T}{2}
\int_0^\infty dt
\nn \\
&&
\times
\int \frac{d^3 \vk}{(2\pi)^3}
\frac{W_{\vk(t)}}{S_{k(t)}}
\frac{\dgamma W_{\vk} - 2 \lambda_\alpha(t) \alpha_0}{S_{k}}
\Phi_\vk(t)^2.
\hspace{2em}
\label{sxySS}
\end{eqnarray}
We can see that a correction due to the dissipative coupling, which
originates in the current fluctuation incorporated in $\alpha(\vg)$,
arises to the well-known formula for the overdamped case \cite{FC2009}.
This is in contrast to the case of $\alpha(\vg) = \alpha_0$, where the
steady-state stress formula is coincident with the overdamped case
\cite{CK2009}.
Note that this correction parallels the ``cooling effect'' of the
kinetic temperature, which compensates the ``heat-up'' by the shearing
and keeps the kinetic temperature unchanged from its initial equilibrium
value.

\section{Numerical calculation}
\label{sec:Num}

To demonstrate the validity of our formulation, we show the results of
the numerical calculations in this section.
As is well known, it is ineffective at present to perform grid
calculations for a three-dimensional system, due to limitations of
computational resources.
In this work, we adopt the ``isotropic approximation'', which has been
formulated and implemented by FC \cite{FC2005, FC2009} and
CK~\cite{CK2009}.
The grid calculations in two-dimensional sheared Brownian systems
\cite{MRY2004, HWCF2009} show that the anisotropy is relatively small,
which assures the validation of this approximation, at least in two
dimensions
\footnote{At the quantitative level, it is known that the isotropic
approximation underestimates the effect of shearing. Hence, a more
accurate approximation scheme is desirable for three-dimensional
calculations.}.
This approximation also enables us to compare formally similar equations
for the underdamped and overdamped cases, as is discussed in Appendix
\ref{App:sec:Mori}.
The details of the formulation of the isotropic approximation is well
described by CK \cite{CK2009}, so we only show the results below and
give some additional remarks in Appendix \ref{App:sec:Num}.

Sheared systems are genuinely anisotropic, which can be seen from, e.g.
the existence of the anisotropic term $-\left[ \ve{\kappa} \cdot
\vH_\vq(t) \right]^\lambda$ in the Mori-type equation,
Eq.~(\ref{eom-H3}).
By the application of the isotropic approximation, the anisotropic terms
are neglected, which allows us to obtain a single second-order equation
for the density time-correlator by combining Eqs.~(\ref{eom-Phi}) and
(\ref{eom-H3}).
The resulting equation (MCT equation) is
\begin{eqnarray}
\frac{d^2}{dt^2}
\Phi_q(t)
&\simeq&
-
v_T^2
\frac{q(t)^2}{S_{q(t)}}
\Phi_q(t)
\nn \\
&&
-
\left[
\lambda_\alpha(t) \alpha_0
-
\dgamma
\frac{\frac{2}{3}\dgamma t}
{1 + \frac{1}{3}\left( \dgamma t \right)^2}
\right]
\frac{d}{dt}
\Phi_q(t)
\nn \\
&&
-
\int_0^t ds
\bar{M}_{\bq(s)}(t-s)
\frac{d}{ds}
\Phi_q(s),
\hspace{2em}
\label{iso_eq}
\end{eqnarray}
where the memory kernel is given by
\begin{eqnarray}
&&
\hspace{-1.5em}
\bar{M}_{\bq(s)}(t-s)
\nn \\
&=&
\frac{nv_T^2}{2q^2}
\left[
1 + \frac{1}{3} \left( \dgamma t \right)^2
\right]
\int \frac{d^3 \vk}{(2\pi)^3}
\left[
\left(\vq\cdot\vk\right) c_{\bk(t)}
+
\left(\vq\cdot\vp\right) c_{\bp(t)}
\right]
\nn \\
&&
\times
\left[
\left(\vq\cdot\vk\right) c_{\bk(s)}
+
\left(\vq\cdot\vp\right) c_{\bp(s)}
\right]
\Phi_{\bk(s)}(t-s)
\Phi_{\bp(s)}(t-s).
\label{iso_M}
\nn \\
\end{eqnarray}
Here, $\vp \equiv \vq - \vk$ is assumed.
The notation and the derivation of Eqs.~(\ref{iso_eq}) and (\ref{iso_M})
are shown in Appendix \ref{App:subsec:Iso}.
Since the multiplier $\lambda_\alpha(t)$ given by Eq.~(\ref{Eq:lambda_MCT})
is a functional of $\Phi_q(t)$, Eq.~(\ref{iso_eq}) is solved by
iteration between $\Phi_q(t)$ and $\lambda_\alpha(t)$.

\subsection{Time-correlators}
\label{subsec:Num_TC}

In the isotropic approximation, the MCT equation Eq.~(\ref{iso_eq}) is numerically solved
on a one-dimensional spatial and temporal grids.
The spatial grid is for the wavenumber (modulus of the wavevector).
The discretized form of the memory kernel Eq.~(\ref{iso_M}) on the
spatial grid is shown in Appendix \ref{App:subsec:Disc}.
For the time integration, we adopt the algorithm of
Ref.~\cite{FGHL1991}, which enables us to calculate robustly in long
time-scales by the gradual coarse-graining in the temporal grid.

For the units of non-dimensionalization,
we choose the diameter $d$ and the mass $m$ of the sphere for the length
and mass, respectively, and $\tau_0 \equiv d/v_0$, where $v_0 \equiv
v_0^{(+)} - v_0^{(-)} = \dgamma L$ is the relative shear velocity at the
two boundaries, for the time.

The conditions of the calculation are as follows.
The spatial grid is chosen as $q d = \hat{q} \Delta$, where $\Delta =
0.4$ is the grid spacing and $\hat{q} = (2m-1)/2$ ($m = 1, 2, \cdots,
100$) is the discretized index.
The cut-off of the wavenumber is $q_{\mathrm{max}} d =39.8$.
The number of the temporal grids is $N_t = 256$.
The time-step is initially $\Delta t_0 = 10^{-6} \tau_0$, which is
doubled in every $N_t/2$ steps.
There are three inputs; the volume fraction $\varphi \equiv \pi n d^3/6$,
the static structure factor $S_q$, and the shear rate $\dgamma$.
The volume fraction is expressed in terms of the ``distance'' from the
critical volume fraction of the MCT transition in the equilibrium MCT,
$\varphi_c = 0.51591213$ \cite{FFGMH1997}, which is denoted as $\ep \equiv
\left( \varphi - \varphi_c \right)/\varphi_c$.
This definition of the ``distance'' implies $\ep > 0$ for the glass
phase, while $\ep < 0$ for the liquid phase.
The value of $\ep$ is fixed at $\ep = +10^{-3}$ for the calculation of
the time-correlator, while it is varied for the shear stress, as is shown
in subsection \ref{subsec:Num_ShearStress}.
As for the static structure factor, the analytic solution of the
Percus-Yevick equation \cite{HM} for three-dimensional hard-sphere
systems is adopted, whose explicit expression in the Fourier space can
be found in e.g.  Ref.~\cite{A1973}.
The initial conditions are $\Phi_q(t=0) = S_q$ and $\left[\del
\Phi_q(t)/ \del t\right]_{t=0} = 0$.
\begin{figure}[htb]
\includegraphics[width=8.5cm]{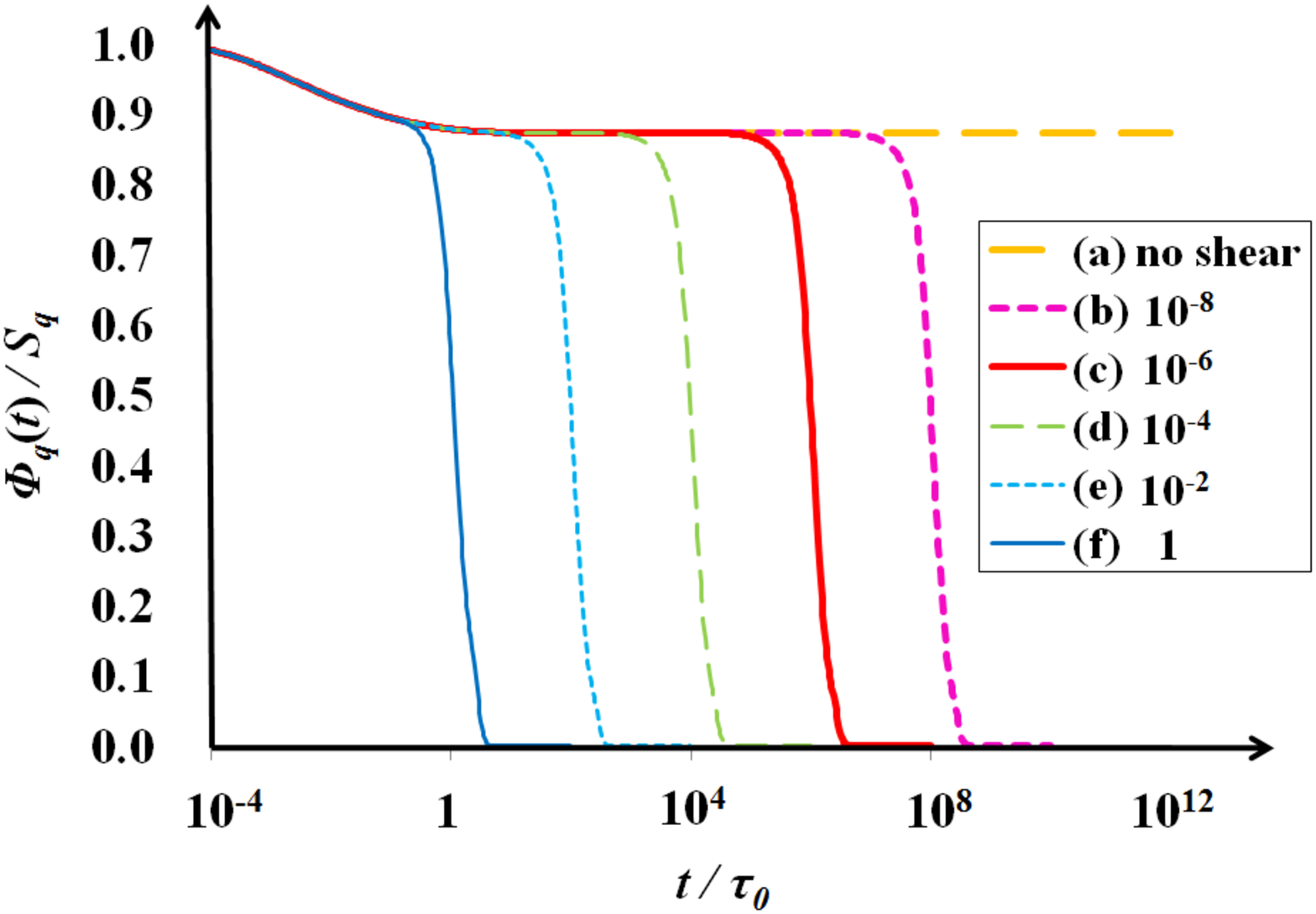}
\caption{(color online)
Numerical solution for the normalized density
 time-correlator. The wavenumber is fixed at $qd = 7.0$ (the first,
 highest peak of the static structure factor), while the shear rate
 $\dgamma$ is varied.  The lines correspond to $\dgamma \tau_0$ with (a)
 no shear (zero), (b) $10^{-8}$, (c) $10^{-6}$, (d) $10^{-4}$, (e)
 $10^{-2}$, and (f) $1$.}
\label{Fig_Phit_shear}
\includegraphics[width=8.5cm]{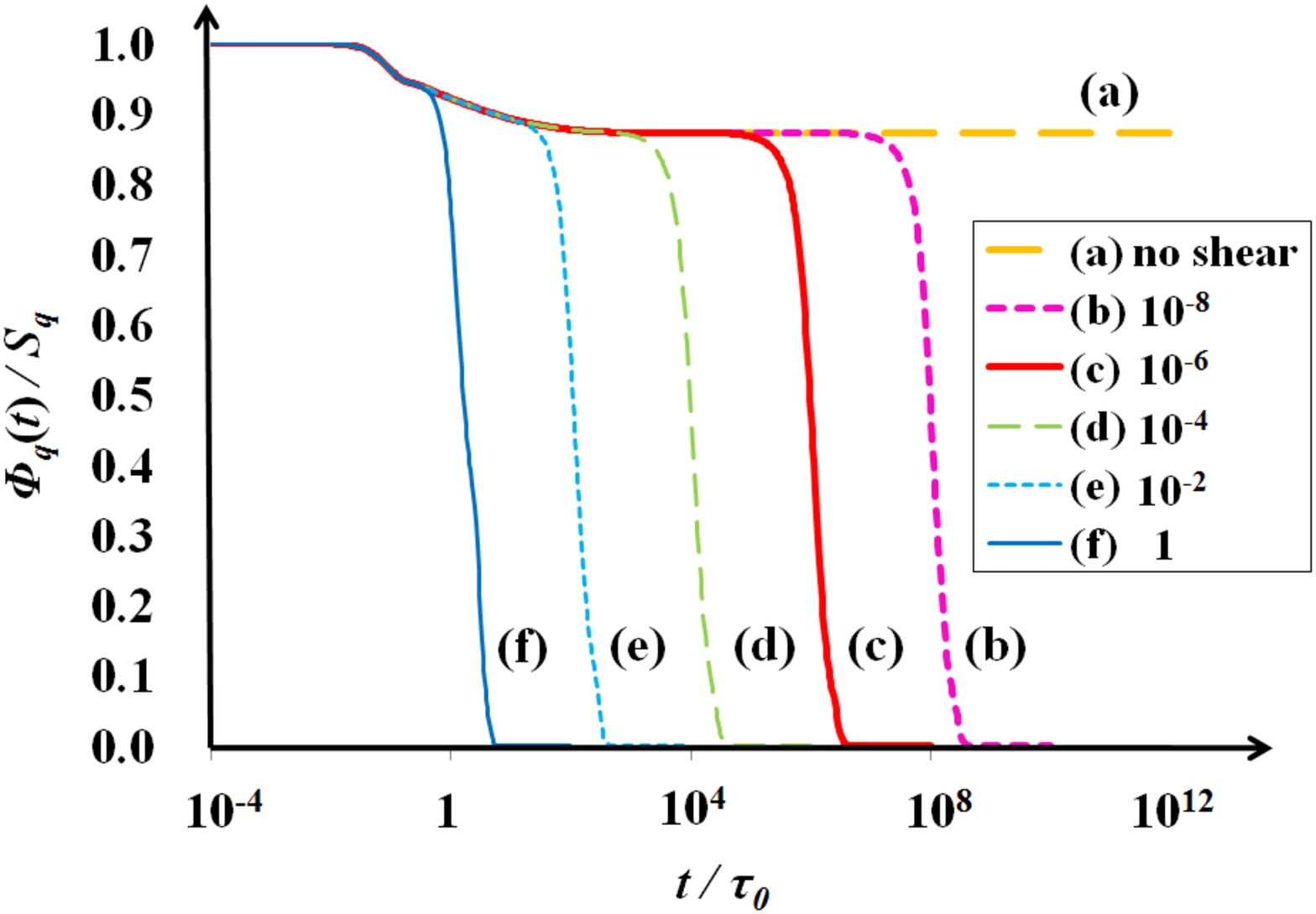}
\caption{(color online)
Numerical solution for the normalized density
time-correlator in the overdamped limit. The conditions and the captions are
the same as Fig.\ref{Fig_Phit_shear}.}
\label{Fig_Phit_shear_od}
\end{figure}

The result of the calculation is shown in Fig.~\ref{Fig_Phit_shear}.
Here, the wavenumber is fixed at $qd = 7.0$ (the first, highest peak of
the static struture factor; refer to Fig.~\ref{Fig_SSF}), while the shear
rate $\dgamma$ is varied.  The lines correspond to $\dgamma \tau_0 = 0$
(no shear), $10^{-8}$, $10^{-6}$, $10^{-4}$, $10^{-2}$, and $1$.

For comparison, we show the result for the overdamped case in
Fig.~\ref{Fig_Phit_shear_od}.
The MCT equation for this case is shown in Appendix \ref{App:sec:Mori},
Eq.~(\ref{App:Eq:mct-od}).
The effective friction coefficient which appears in
Eq.~(\ref{App:Eq:mct-od}), $\alpha_{\mathrm{od}}$, is fixed as
$\alpha_{\mathrm{od}} d^2 / (v_T^2 \tau_0) = 0.1$.
The result of Fig.~\ref{Fig_Phit_shear_od} is qualitatively in
accordance with the previous results \cite{MRY2004, HWCF2009}, and
almost coincident with them at the quantitative level as well.

From Figs.~\ref{Fig_Phit_shear} and \ref{Fig_Phit_shear_od}, the density
time-correlator decays due to shearing around the time-scale
$\tau_\alpha \simeq \dgamma^{-1}$ ($\alpha$-relaxation time), for both
the underdamped and the overdamped cases.
The resemblance between the two cases can be seen not only in the
$\alpha$-relaxation time $\tau_\alpha$, but also in the non-ergodic
parameter (NEP), which is almost coincident. 
These results are consistent with the observation that the long-time
dynamics after the early-$\beta$-relaxation time $\tau_\beta$ is
dominated by the memory kernel, and the instantaneous dynamics are
invalid at this time-scale \cite{SL1991, GKB1998, SF2004}.
The difference between the two cases can be seen at the early stage
before $\tau_\beta$ \cite{MRY2004}.
As for the underdamped case, the density time-correlator is held
constant at its initial value until $t \simeq 10^{-1} \tau_0$, since the
frequency of the sound wave, $\omega_q(t) \equiv \sqrt{ v_T^2 q(t)^2 /
S_{q(t)} }$, dominates the transient behavior at this stage
(for $q d = 7.0$, $\omega_q(t)^{-1} \simeq 0.27 \tau_0$ at $t \ll
\tau_\beta$).
On the other hand, for the overdamped case, the density time-correlator
is already decreasing at $t \simeq 10^{-4} \tau_0$, which can be seen
from its approximate solution at this stage, $\Phi_q(t) \simeq \exp
\left[ - t / \tau_{\mathrm{od}} \right]$, where $\tau_{\mathrm{od}}
\equiv \alpha_{\mathrm{od}} S_{q(t)}/ \left[ v_T^2 q(t)^2 \right] =
\alpha_{\mathrm{od}}/\omega_q(t)^2$ is the time-scale of this damping
(for $q d = 7.0$, $\tau_{\mathrm{od}} \simeq 7\times 10^{-3} \tau_0$ at
$t \ll \tau_\beta$).
The emergence of two time-scales is one of the significant features of
the underdamped systems.
In overdamped systems, there is only a single time-scale which is a
ratio of $\alpha_{\mathrm{od}}$ and $\omega_q(t)$, while
$\lambda_\alpha(t) \alpha_0$ and $\omega_q(t)$ settle independent time
scales in the underdamped case.
Due to this fact, overdamped systems are scaled by a single
non-dimensional parameter, the P\'{e}clet number $\mathrm{Pe} \equiv
\dgamma \tau_0$, while this is not the case for the underdamped case.
There are also effects of the difference on the steady-state shear
stress, which will be discussed in subsection
\ref{subsec:Num_ShearStress}.
\begin{figure}[thb]
\includegraphics[width=8.5cm]{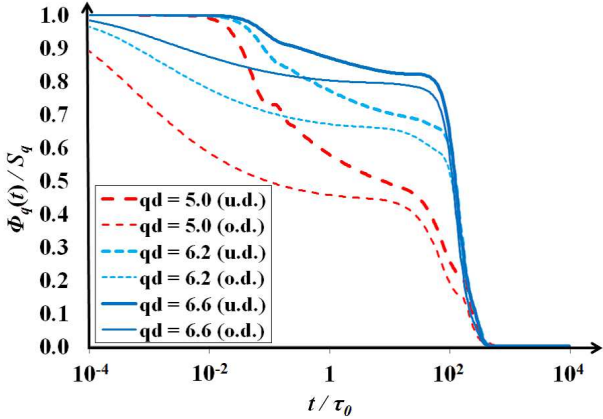}
\caption{(color online)
Numerical solution for the normalized density
 time-correlator. The shear rate is fixed at $\dgamma \tau_0 = 10^{-2}$,
 while results for several wavenumbers below the first, highest peak of
 the static structure factor are shown.  The lines correspond to
 wavenumbers $qd$ with 5.0, 6.2, and 6.6. ``u.d.'' and ``o.d.'' in the
 caption correspond to the underdamped and the overdamped cases,
 respectively.}
\label{Fig_Phit_k}
\end{figure}
%
\begin{figure}[htb]
\includegraphics[width=8.5cm]{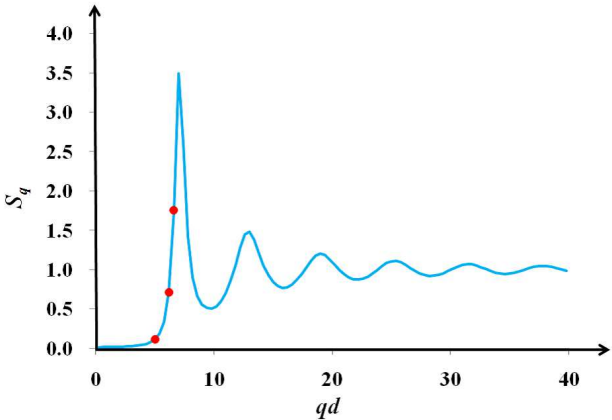}
\caption{(color online) 
The structure factor used as an input in the
calculation. The three wavenumbers whose density time-correlator is shown in
Fig.~\ref{Fig_Phit_k} are highlighted in (red) solid circles; $qd = 5.0$,
$6.2$, and $6.6$, from left to right. $qd = 7.0$ corresponds to the first, highest peak.}
\label{Fig_SSF}
\end{figure}
%
%
\begin{figure}[htb]
\includegraphics[width=8.5cm]{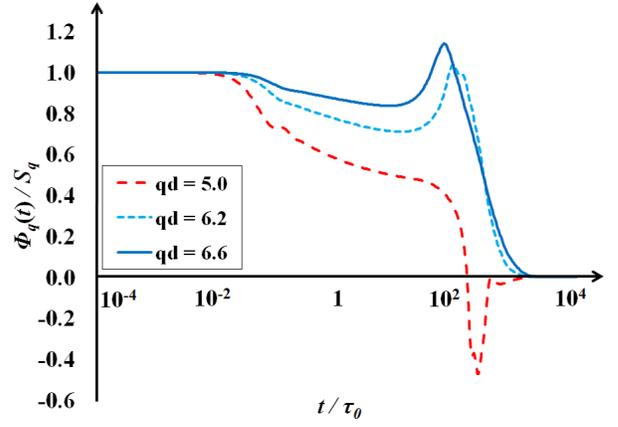}
\caption{(color online)
Numerical solution for the normalized density
 time-correlator of the theory by Chong and Kim \cite{CK2009}. The
 conditions are the same as Fig.\ref{Fig_Phit_k}.}
\label{Fig_Phit_k_CK}
\end{figure}
Next, we show the results for several wavenumbers in
Fig.~\ref{Fig_Phit_k}.
The shear rate is fixed at $\dgamma \tau_0 = 10^{-2}$, and other
conditions are the same as those in Fig.~\ref{Fig_Phit_shear}.
Three wavenumbers below the first, highest peak of the static structure
factor are chosen; $qd = 5.0$, $6.2$, and $6.6$.
They are depicted in Fig.~\ref{Fig_SSF} in (red) solid circles, where the
static structure factor we adopt is shown.
We can see that the density time-correlator is almost monotonically
decreasing, except for the spike around $t \simeq 10^{-1} \tau_0$, for
the underdamped case.
This spike is the vestige of the oscillation of the sound wave, which is
smeared out in longer time-scales.
In fact, there is no spike in the result for the overdamped case, which
is shown together in Fig.~\ref{Fig_Phit_k}.

Next, we show the result for the CK theory \cite{CK2009} in
Fig.~\ref{Fig_Phit_k_CK}, where the conditions are the same as those in
Fig.~\ref{Fig_Phit_k}.
The difference with the result of our formulation is obvious;
there are significant signals of overshoot/undershoot, i.e. the
normalized density time-correlator exceeds 1.0 or becomes negative.
For equilibrium systems, it is easy to prove that the absolute value of
the normalized density time-correlator is less than 1.0 \cite{HM}, and
that the density time-correlator monotonically decays in the overdamped
limit \cite{G}.
On the other hand, for general nonequilibrium systems, there seems to
be no rigorous proof of the bounded property or the monotonicity of the
density time-correlators.
However, it is natural to expect that these properties also hold as
well, at least for cases with small shear.
In addition, it is obvious that the overshoot/undershoot is not the
result of the oscillating nature of the underdamped system.
The overshoot/undershoot appears at the $\alpha$-relaxation regime,
where the instantaneous oscillation is sufficiently damped already.
From these considerations, we conclude that the overshoot/undershoot
found in CK theory \cite{CK2009} is an artifact of the
inappropriate definition of the density time-correlator.
The problem of overshoot/undershoot will be discussed further in section
\ref{sec:Disc} around Eq.~(\ref{M_CK}).

\subsection{Shear stress}
\label{subsec:Num_ShearStress}

\begin{figure}[htb]
\includegraphics[width=8.5cm]{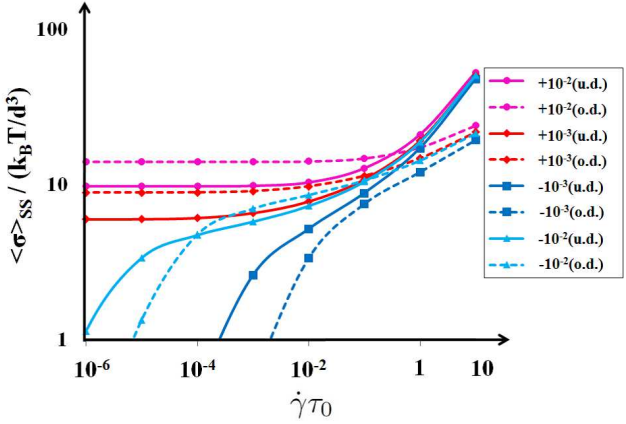}
\caption{(color online)
Numerical result for the steady-state shear
 stress, in unit of $k_B T / d^3$. The solid lines are for the
 underdamped case, while the dotted ones are for the overdamped
 case. The four lines for each case are for $\ep = \pm 10^{-2}$, $\pm
 10^{-3}$, where $\ep \equiv \left( \varphi - \varphi_c
 \right)/\varphi_c$ is the distance of the volume fraction $\varphi$
 from the MCT transition point $\varphi_c$. ``u.d.'' and ``o.d.'' in the
 caption correspond to the underdamped and the overdamped cases,
 respectively.}
\label{Fig_sigma}
\end{figure}

Now we present the result for the steady-state shear stress in unit of
$k_B T / d^3$, where $d$ is the diameter of the sphere, in
Fig.~\ref{Fig_sigma}, which is calculated from the solution of the
density time-correlator by Eq.~(\ref{sxySS}).
The conditions are the same as those in section \ref{subsec:Num_TC},
aside from two exceptions.
One is the strength of the thermostat in the overdamped case
$\alpha_{\mathrm{od}}$, which is fixed $\alpha_{\mathrm{od}} d^2 /
\left( v_T^2 \tau_0 \right) = 1$ here.
This value is chosen to conform with the previous work \cite{FC2002},
which is the direct reference of our calculation.
Another is the volume fraction, where four cases, $\ep = \pm 10^{-2}$,
$\pm 10^{-3}$, are considered for underdamped and overdamped cases,
respectively.
The results for the underdamped case are shown in solid lines, while
those for the overdamped case are shown in dotted lines.

As discussed in section \ref{subsec:Num_TC}, underdamped systems are not
scaled by a single parameter, the P\'{e}clet number $\mathrm{Pe}$, in contrast to
overdamped systems.
Hence, we choose as the horizontal axis the non-dimensionalized shear
rate, $\dgamma \tau_0$.
\begin{figure}[htb]
\includegraphics[width=8.5cm]{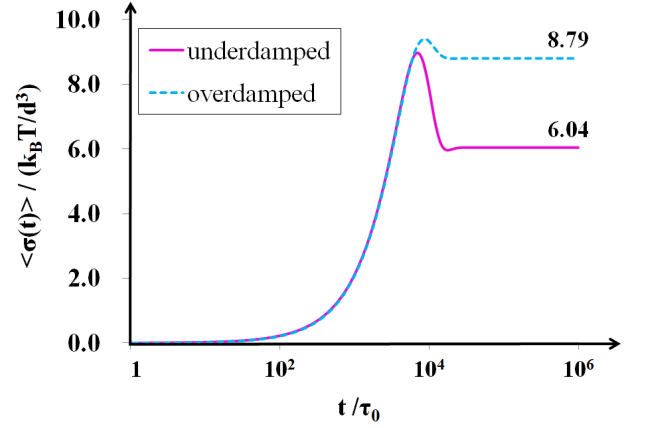}
\caption{(color online) Numerical result for the accumulated
steady-state shear stress $\eav{\sigma(t)}$, Eq.~(\ref{Eq:sxyt}),
in unit of $k_B T / d^3$.
The two lines are for the underdamped and overdamped cases,
respectively, for $\ep = +10^{-3}$ and $\dgamma \tau_0 = 10^{-4}$.
The steady-state value of the shear stress is attatched to each line.}
\label{Fig:sigma_t}
\end{figure}
\begin{figure}[htb]
\includegraphics[width=8.5cm]{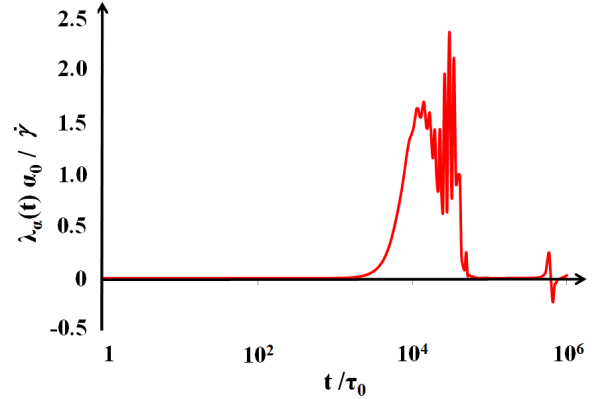}
\caption{(color online) 
Numerical result for the mutiplier
 $\lambda_\alpha(t)$.
We plot $\lambda_\alpha(t) \alpha_0$
in unit of $\dgamma$.
The conditions are the same as those for
Fig.~\ref{Fig:sigma_t}.}
\label{Fig:alpha0}
\end{figure}

The results for the overdamped case reproduces those found in
Ref.~\cite{FC2002}.
We can see from Fig.~\ref{Fig_sigma} that the underdamped case shows
similar tendencies with the overdamped case.
That is, in the liquid phase with $\ep < 0$, the shear stress shows the
Newtonian behavior, $\sigma_{xy} \propto \dgamma$, for small shear
rates.
For large shear rates, the above linearity is broken, which signals the
``shear-thinning''.
In the glass phase with $\ep > 0$, the shear stress remains finite in
the limit $\dgamma \ra 0$, which is nothing but the yield stress.

At the quantitative level, however, discrepancies can be observed.
For high shear rates, e.g. $\dgamma \tau_0 > 1.0$, the shear stress is
larger for the underdamped case.
This is since the density time-correlator is held constant in the
short-time regime $t < 0.1 \tau_0$ due to the inertia effect in the
underdamped case, while it is already decreasing in the overdamped case,
as previously discussed.

On the other hand, for low shear rates, the yield stress which emerges
in the glass phase ($\ep > 0$) is smaller for the underdamped case.
This is due to the analogue of the ``cooling effect'', which is
explained below Eq.~(\ref{sxySS}).
To convince this statement, the accumulated shear stress as a function
of time, which we define by
\begin{eqnarray}
\eav{\sigma(t)}
&\equiv&
\frac{k_B T}{2}
\int_0^t ds
\nn \\
&&
\times
\int \frac{d^3 \vk}{(2\pi)^3}
\frac{W_{\vk(s)}}{S_{k(s)}}
\frac{\dgamma W_{\vk} - 2 \lambda_\alpha(s) \alpha_0}{S_{k}}
\Phi_\vk(s)^2,
\hspace{2em}
\label{Eq:sxyt}
\end{eqnarray}
is shown in Fig.~\ref{Fig:sigma_t}, and the multiplier
$\lambda_\alpha(t)$, given by Eq.~(\ref{Eq:lambda_MCT}), is shown in
Fig.~\ref{Fig:alpha0}, for the case $\ep = +10^{-3}$ and $\dgamma \tau_0
= 10^{-4}$, respectively.
In Fig.~\ref{Fig:sigma_t}, the steady-state value of the shear stress is
attatched to each result.
We can see that the discrepancy between the underdamped and the
overdamped cases arises in the $\alpha$-relaxation regime, $t \sim
\dgamma^{-1} = 10^4 \tau_0$.
It is notable that the multiplier $\lambda_\alpha(t)$ correspondingly
magnifies in this regime, which suggests the growth of the current
fluctuation incorporated in $\alpha(\vg)$.
This implies that the growth of the current fluctuation in the
dissipative coupling relaxes the shear stress in the underdamped case.
It might also be worth noting that the density time-correlator are
almost coincident in the $\alpha$-relaxation regime for the underdamped
and the overdamped cases.
However, in the underdamped case, the time derivative of the density
time-correlator is nothing but the density-current cross
time-correlator, Eq.~(\ref{App:Eq:HL}), which also exhibits a growth in
the $\alpha$-relaxation regime.
This is another evidence that the current fluctuation incorporated in
$\alpha(\vg)$ is the origin of the stress relaxation.

Now it is clear why the correction in Eq.~(\ref{sxySS}) appears in the
vertex function; since it originates in the current fluctuation, it
cannot appear in the density time-correlator, which leaves the vertex
function as the only possibility.
Together with the fact that the correction is actually the multiplier
Eq.~(\ref{Eq:lambda_MCT}) which is written in terms of the density
time-correlator, Eq.~(\ref{Eq:lambda_MCT}) can be regarded as a current
fluctuation, which expectedly is highly nonlinear.

It is remarkable that the existence of a discrepancy between the
overdamped MCT and the molecular dynamics (MD) simulation, which has an
inertia effect in the $\alpha$-relaxation regime, has been reported
\cite{ZH2008}.
Our result presented here suggests that the discrepancy is quite
generic. 
Thus, we should be careful in comparing the results of MD
with overdamped dynamics such as Brownian dynamics or the overdamped
MCT.

\subsection{Response to a perturbation}
\label{sec:Resp}
As an application of the underdamped MCT so far constructed, we
calculate the response of the density time-correlator to a perturbation
of the shear rate, and demonstrate the significance of the underdamped
formulation.

We discuss a response to an instantaneous change in the shear rate
$\dot{\gamma}$ at the quasi-steady state which corresponds to the
plateau of the density time-correlator at $t=t_0$.
Specifically, we consider a pulse-like perturbation of the form
\begin{eqnarray}
\dot{\gamma}(t)
=
\dot{\gamma} + \Delta \dot{\gamma} \left[ \Theta(t-t_0) -
\Theta(t-(t_0+\Delta t_{\dot{\gamma}}))\right],
\hspace{1.5em}
\label{Eq:dgamma_pert}
\end{eqnarray}
where $\Delta t_{\dot{\gamma}}$ is the width of a rectangular pulse,
$\Theta(t)$ is the Heaviside's step function,
and $t_0$ is of the order of the early-$\beta$-relaxation time, $t_0
\sim \tau_\beta$.

For a time-dependent external field (the shear rate in our case), the
time-evolution operator is expressed in terms of a time-ordered
exponential \cite{EM}, and hence we cannot simply apply the MCT
formulated in this paper, which is valid for a constant shear
rate.
However, for a pulse-like perturbation, it can be shown that the
time-ordered exponential reduces to a {\it normal} exponential for the
case of a weak and instantaneous perturbation, i.e. $\Delta \dot{\gamma}
/ \dot{\gamma} \ll 1$ and $\Delta t_{\dot{\gamma}} / (t - t_0) \ll 1$.
In this case, the time-evolution operator $U_{\to}(t_0,t)$ can be
expanded as
\begin{eqnarray}
U_{\to}(t_0,t)
=e^{i\mathcal{L}(\Delta \dot{\gamma}) (t - t_0) }
+
\mathcal{O}( \Delta \dot{\gamma}^2 (t-t_0)^2 ),
\end{eqnarray}
where $i\mathcal{L}(\Delta \dot{\gamma})$ is the perturbed Liouvillian,
whose shear part $i\mathcal{L}_{\dot{\gamma}} \equiv \dot{\gamma}
i\tilde{\mathcal{L}}_{\dot{\gamma}}$, $i\tilde{\mathcal{L}}_{\dot{\gamma}}
\equiv \sum_{i=1}^N \left( y_i \partial / \partial x_i - p_i^y \partial
/ \partial p_i^x \right)$ is given by
\begin{eqnarray}
i\mathcal{L}_{\dot{\gamma}}(\Delta \dot{\gamma})
&=&
\tilde{\dot{\gamma}}(t)
i\tilde{\mathcal{L}}_{\dot{\gamma}},
\\
\tilde{\dot{\gamma}}(t)
&\equiv&
\dot{\gamma}
\left\{
1
+ \frac{\Delta \dot{\gamma}}{\dot{\gamma}}
\frac{\Delta t_{\dot{\gamma}}}{t - t_0}
\Theta( t - (t_0 + \Delta t_{\dot{\gamma}}) )
\right\}.
\hspace{2em}
\label{Eq:dgammat}
\end{eqnarray}
That is, we can apply the MCT formulated for a {\it normal} exponential,
together with a time-dependent shear rate, Eq.~(\ref{Eq:dgammat}).

The conditions of the calculation are as follows.
The time when the perturbation is switched on, i.e. $t_0$, is set to the
order of the early-$\beta$-relaxation time $\tau_\beta \sim 10^2
\tau_0$.
This choice is made by the observation that the onset of the plateau of
the density time-correlator is around $10^2 \tau_0$, which can be seen
from Fig.~\ref{Fig_Phit_shear}.
The shear rate and its magnitude of perturbation are set to
$\dot{\gamma} = 10^{-4} \tau_0$ and $\Delta \dot{\gamma} / \dot{\gamma}
= 0.1$, respectively.
Note that the $\alpha$-relaxation time is $\tau_\alpha \sim
\dot{\gamma}^{-1} = 10^4 \tau_0$, and hence the quasi-steady state
lasts for a time interval of $99 t_0$.
The initial time step is set to $\Delta t_0 = 10^{-6}\tau_0$, which is
doubled in every 4096 steps ($N_t = 8192$).
The width of the pulse is set to $\Delta t_{\dot{\gamma}} = \Delta t_0$,
which is small enough compared to the time-scale of the response.
Hence, the shear rate Eq.~(\ref{Eq:dgammat}) is pulse-like at this time
scale.

\begin{figure}[thb]
\includegraphics[width=8.5cm]{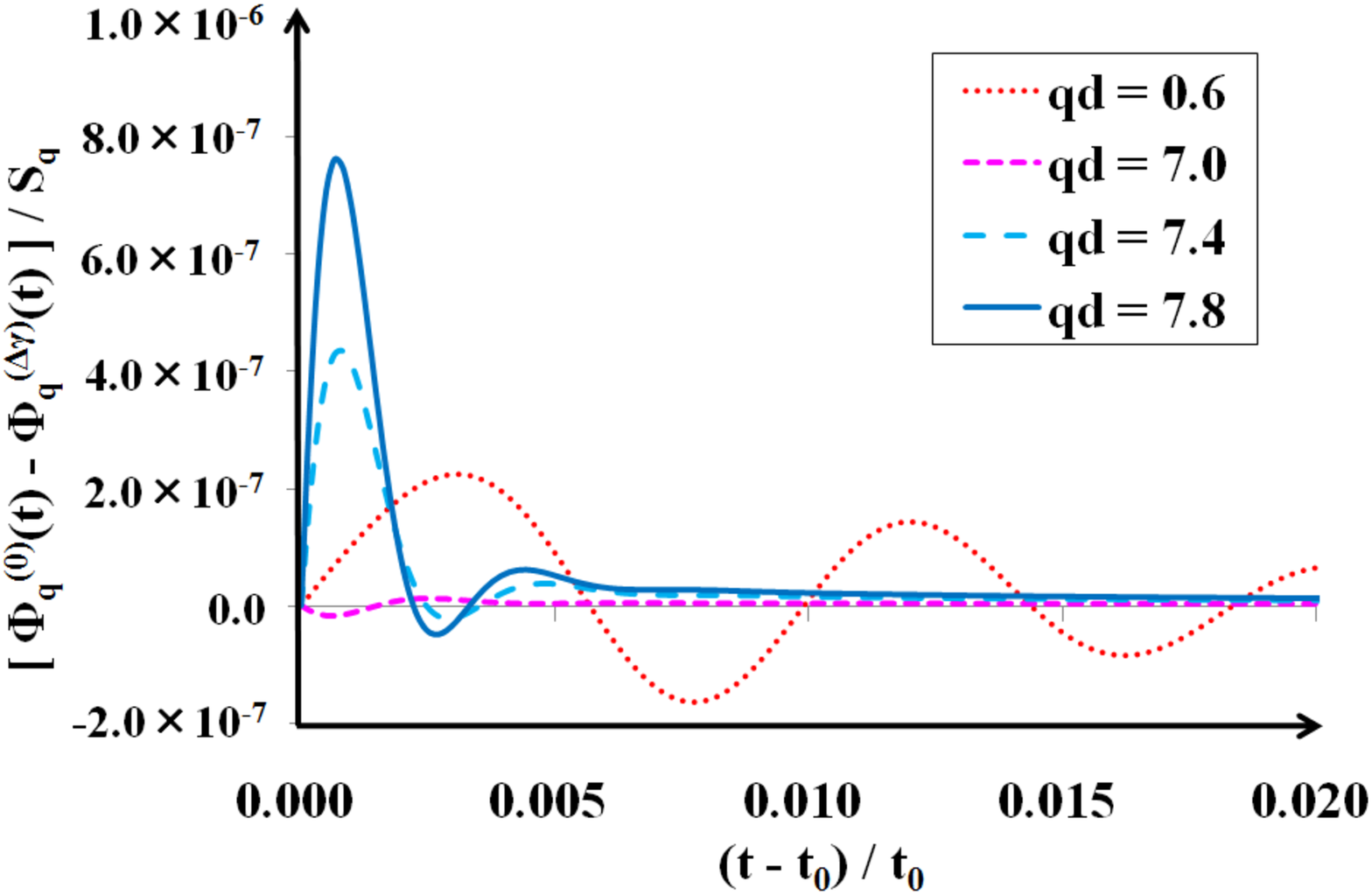}
\includegraphics[width=8.5cm]{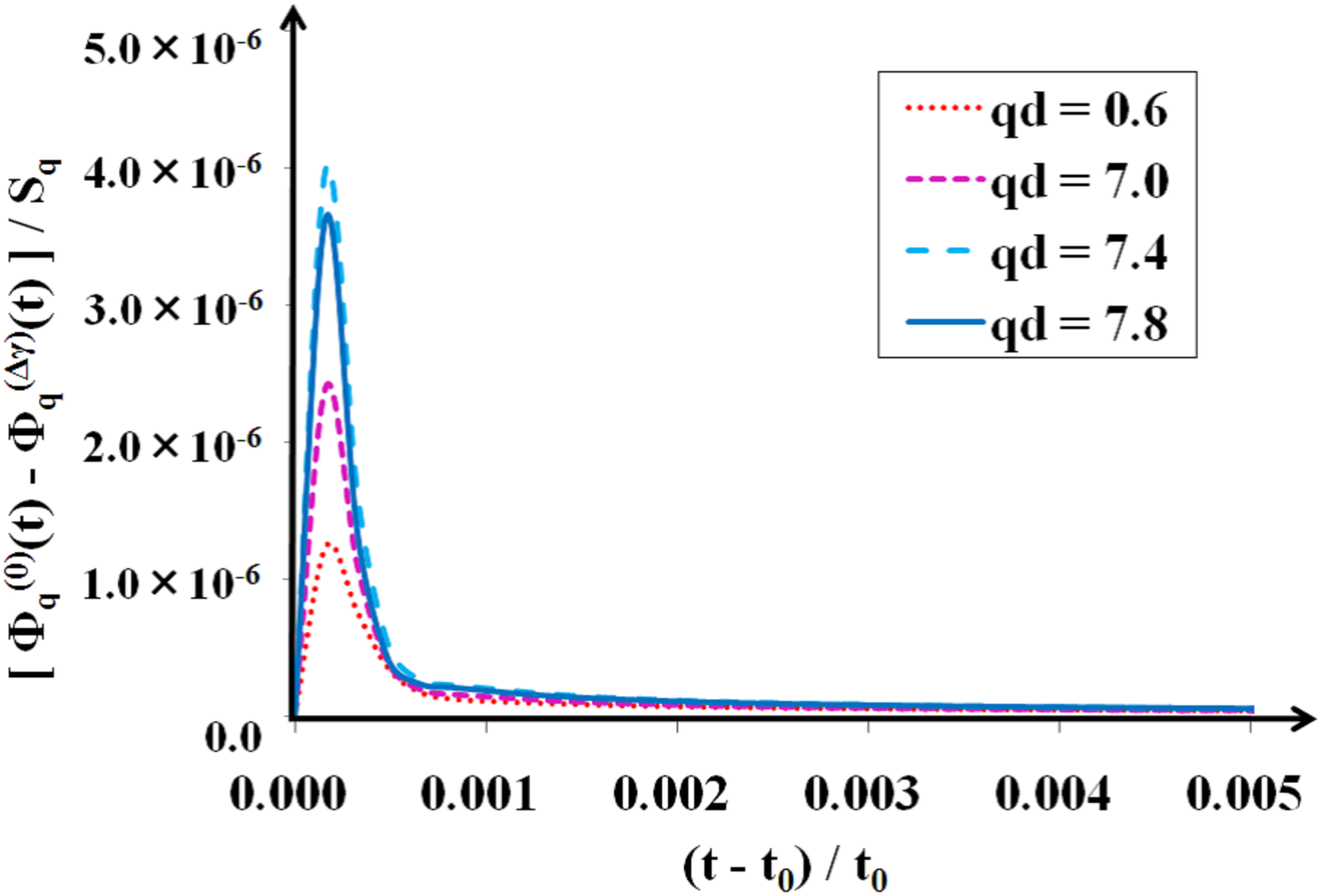}
\caption{(color online) Response of the density time-correlator. 
The upper (lower) figure is for the
underdamped (overdamped) case. The four lines correspond to
 non-dimensionalized wavenumbers $qd = 0.6, 7.0, 7.4, 7.8$}
\label{Fig:dPhi}
\end{figure}

The result for the response of the normalized density time-correlator,
$-\left[ \Phi_q^{(\Delta \dot{\gamma})}(t) - \Phi_q^{(0)}(t) \right] /
S_q$, is shown in Fig.~\ref{Fig:dPhi}, both for the underdamped and
overdamped cases.
Here, $\Phi_q^{(\Delta \dot{\gamma})}(t)$ and $\Phi_q^{(0)}(t)$ are the
perturbed and the unperturbed density time-correlators, respectively,
and the sign convention is chosen simply for convenience.
We can see that, in the underdamped case, there is a delay in the
emergence of the response, and an oscillation of period typically of the
order of $5\times10^{-3} t_0$ can be seen.
This feature is due to the inertia effect of the underdamped system.
On the other hand, in the overdamped case, the response emerges
instantaneously and then decays with the relaxation time $10^{-3} t_0$.
Hence, even though the density plateau coincides for the underdamped and
overdamped cases, a clear difference can be observed in the response
at this quasi-steady state.

The result of this subsection can be further applied to the response
theory.
A formulation and calculation of the response of the shear stress to a
perturbation of the shear rate is presented in Appendix
\ref{App:sec:response}.
%

\section{Discussion}
\label{sec:Disc}
In this section, we first compare our work with the previous works.
To the best of our knowledge, the major representative works in sheared
MCTs are Fuchs-Cates1 (FC1) \cite{FC2002, FC2005}, Fuchs-Cates2 (FC2)
\cite{FC2009}, Miyazaki {\it et al.}  \cite{MR2002, MRY2004}, Chong-Kim
(CK) \cite{CK2009}, and Hayakawa-Otsuki (HO) \cite{HO2008}.
Besides HO, which is formulated for inelastic granular systems, they are
for sheared Brownian or thermostatted systems.
The relations between the above theories might be confusing, so we
briefly review their basic setups, and then discuss about the resulting
formulations.

\begin{figure}[htb]
\includegraphics[width=8.5cm]{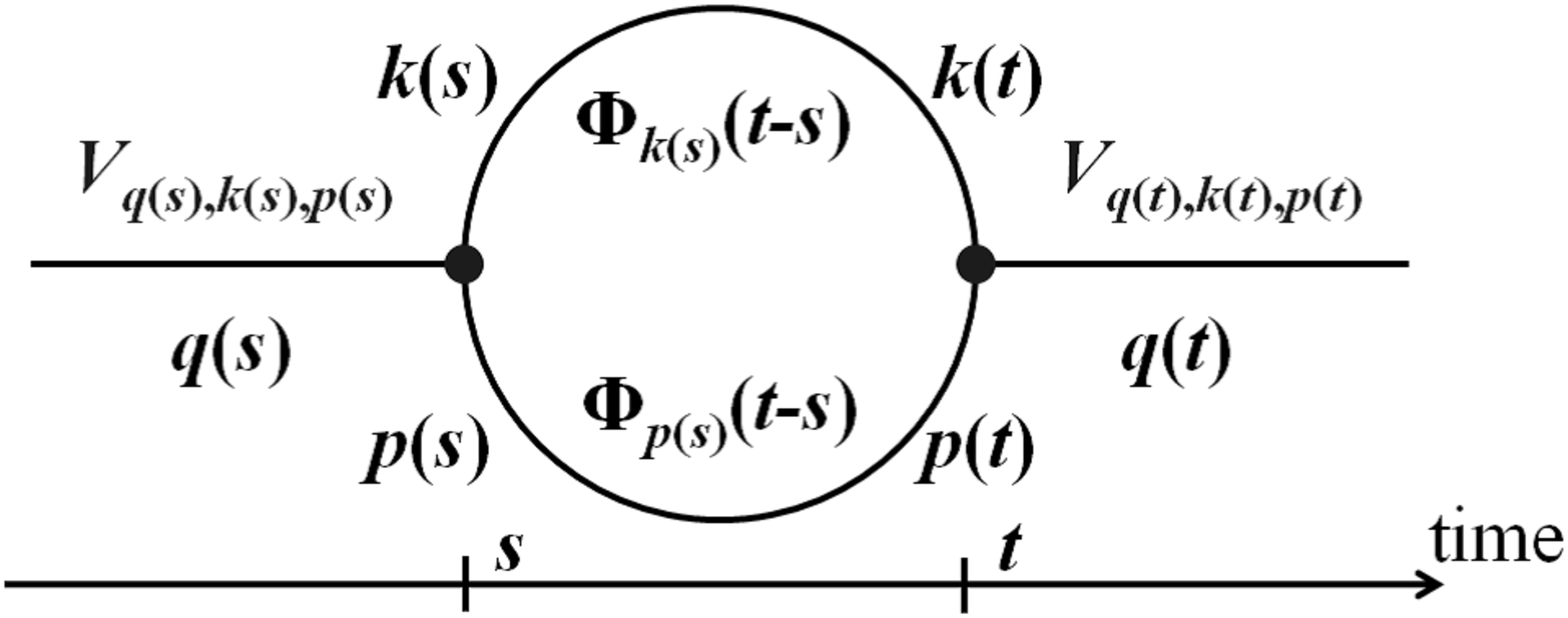}
\caption{A schematic diagram for the structure of the memory kernel in
our formulation. The kernels in the theories by Fuchs-Cates
\cite{FC2009} and Miyazaki {\it et al.} \cite{MR2002, MRY2004} also have
the same structure. }
\label{Fig_M_SH}
\includegraphics[width=8.5cm]{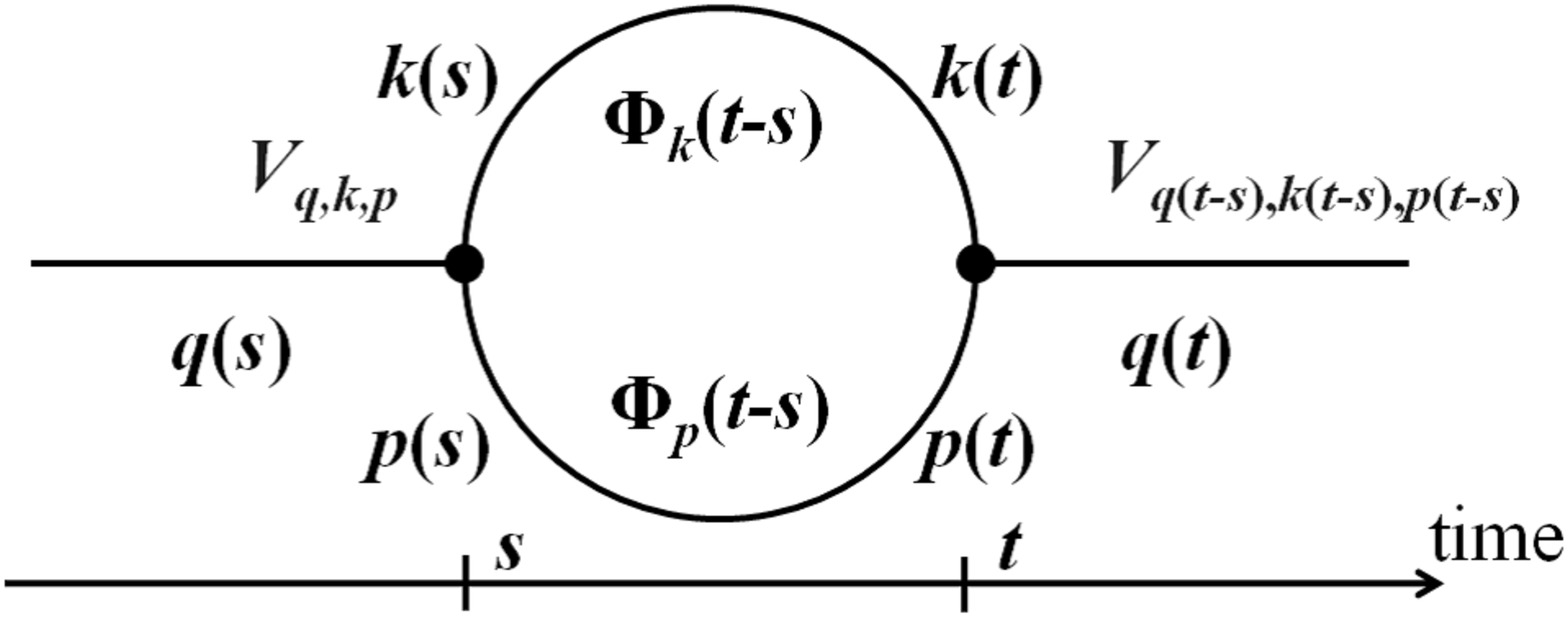}
\caption{A schematic diagram for the structure of the memory kernel in
 the theory by Chong-Kim \cite{CK2009}.}
\label{Fig_M_CK}
\end{figure}

The FC theories, FC1 and FC2, are for overdamped systems whose dynamics
is governed by the Smoluchowski operator.
In FC1, the density time-correlator is defined as
$\Phi_\vq^{(\mathrm{\mathrm{FC1}})}(t) \equiv \eav{ n_\vq(t)
n_{\vq(t)}^*(0)}/N$, with $\vq(t) \equiv \vq + \vq\cdot\ve{\kappa}t$.
It is discussed in Ref.~\cite{FC2009} that the application of MCT to the
above time-correlator leads to non-positive-definite ``initial decay
rates'', which causes numerical instabilities.
This has motivated the modification which leads to FC2.
In FC2, the density time-correlator is defined as
$\Phi_\vq^{(\mathrm{FC2})}(t) \equiv \eav{ n_{\vq(t)}(t) n_\vq^*(0)}/N$,
with $\ve{q}(t) \equiv \vq - \vq\cdot\ve{\kappa}t$, which is identical
to our definition.
Moreover, the resulting MCT equation is also formally correspondent if
we neglect the transverse mode (or adopt the isotropic approximation)
and take the overdamped limit (i.e. neglect the inertia term) in our
framework; refer to Appendix \ref{App:sec:Mori} for this issue.
Hence, our formulation can be regarded as an extension of that of FC2 to
underdamped systems.
To be more concrete, the initial decay rate $\Gamma_\vq(t) \equiv
\vq(t)^2 v_T^2 / S_{q(t)}$ is properly advected, and the memory kernel
possesses the structure of wavevector indices depicted schematically in
Fig.~\ref{Fig_M_SH};
i.e. the memory kernel consists of vertex functions with wavevector
indices advected to time $s$, $V_{\vq(s),\vk(s),\vk(s)}$, and time $t$,
$V_{\vq(t),\vk(t),\vp(t)}$, respectively, which are bridged by a
projection-free propagator starting from time $s$ with interval $t-s$,
e.g. $\Phi_{\vk(s)}(t-s)$.
These features seem physically sensible, which manifests the ``alignment
of the wavevectors'' which we adopt as a principle in sections
\ref{subsec:TC} and \ref{subsec:PO}.
As discussed in Ref.~\cite{FC2009}, note that the application of the
{\it time-dependent} projection operators enables the preservation of
this principle.
%

The CK theory starts with an almost identical microscopic framework with
ours; i.e. an underdamped SLLOD equation governed by a Liouvillian.
The crucial difference at this level is that the dissipative coupling to
the thermostat is a constant, while it contains current fluctuations in
order to satisfy the isothermal condition in our framework.
At the level of the Mori-type equations, another crucial difference is
that the time-correlators are defined as equivalent to FC1, e.g.
$\Phi_\vq^{(\mathrm{CK})}(t) = \Phi_\vq^{(\mathrm{FC1})}(t)$ for the
density time-correlator, which are inequivalent to our definition.
The conventional static projection operators are applied, and the
resulting MCT equation differs from ours in two respects:
(a) it breaks the alignment of the wavevectors, and (b) the memory
kernel $L_\vq^\lambda$ survives.
The feature (a) can be seen in (i) the initial decay rate
$\Gamma^{(\mathrm{CK})}_\vq \equiv q^2 v_T^2 / S_q$, which is not
advected, (ii) the memory kernel, which has the structure of wavevector
indices schematically depicted in Fig.\ref{Fig_M_CK},
and probably the most significantly in (iii) the wavevector-structure of
the time integration, which reads
\begin{eqnarray}
\int_0^t ds
M_q(t-s)
\frac{\del}{\del s}
\Phi_{\bq(t-s)}(s)
\label{M_CK}
\end{eqnarray}
in the isotropic approximation.
The advection of the wavenumber index appears in the {\it
time-derivative} of the density time-correlator, while it appears in the
memory kernel in our framework, cf. Eq.~(\ref{iso_eq}).
Numerically, the overshoot/undershoot found in the CK theory is a
consequence of the term $\del \Phi_{\bq(t-s)}(s) / \del s$, which shows
a singular behavior when the advected wavenumber passes the first peak
of the static structure factor.
On the other hand, since the memory kernel includes only the density
time-correlator, not its time-derivative, a singular behavior is not
found in our result.
As for the feature (b), the memory kernel $L_\vq^\lambda$ has been
confirmed to be numerically negligible \cite{CK2009}, at least for the
situation in concern, which is compatible with our result,
$L_\vq^\lambda = 0$.

In Miyazaki {\it et al.} \cite{MR2002, MRY2004}, an alternative approach
is adopted.
They start with a generalized fluctuating hydrodynamics for the density
and the velocity fields with a Gaussian noise.
Aside from the drawbacks of this approach, which are already discussed
(e.g. (i) the assumption of the fluctuation-dissipation theorem which is
known not to hold in sheared systems, (ii) the formulation based on the
steady-state fluctuations), the resulting MCT parallels that derived by
the projection operator formalism.
The time-correlators are defined in accordance with our work and FC2,
and it turns out that the MCT equation for the density time-correlator
in the overdamped limit coincides with that of FC2.

Next, we discuss the novelties which result from our framework.
First, the relaxation of the shear stress at the $\alpha$-relaxation
regime is a genuine feature of our underdamped formulation, which cannot
be derived from the overdamped framework.
This is since the above relaxation is due to the growth of current
fluctuations, which is not considered in the overdamped case.
In the overdamped case, there might be a possibility to incorporate
density fluctuations into the dissipative coupling
$\alpha_{\mathrm{od}}$, but this does not lead to a relaxation of the
shear stress discussed here.

Second, as demonstrated in section \ref{sec:Resp}, we consider an
application to the calculation of a response.
In addition, we show in Appendix \ref{App:sec:response} the
corresponding response of the stress tensor, which is formulated by the
response of the density time-correlator.
From these results, it can be seen that although there is no difference
in the plateau of the density time-correlator for the underdamped and
overdamped cases in the absense of a perturbation, there is a clear
difference in a short-time response to a perturbation for the two cases.
This is another clear evidence for the relevancy to use the underdamped
model even in glassy systems, where the dynamics is expected to be slow
and the effects of current fluctuations nor those of inertia are
believed to be irrelevant.

Another virtue of the present framework is that its extension to other
systems, such as an assembly of granular particles, is relatively
simple.
One just has to replace the coupling to the thermostat (the term
$-\alpha(\vg) \vp_i(t)$ in Eq.~(\ref{pdot})) with e.g.
the interparticle viscous interactions of the Stokes' type.
In fact, it is difficult to incorporate viscous interactions inherent in
granular materials into the formulation of FC \cite{FC2002, FC2005,
FC2009} nor Miyazaki {\it et al.} \cite{MR2002, MRY2004}.
Dissipation has been introduced in MCT by HO \cite{HO2008} and Kranz
{\it et al.} (KSZ) \cite{KSZ2010}, but they seem not to be satisfactory.
As for HO, which deals with sheared granular systems, dissipation is
introduced by the inelastic Boltzmann operator, i.e. the
pseudo-Liouvillian \cite{BP}.
The model considered in KSZ is for a driven granular system under a
white-noise thermostat, which does not have any advected
wavenumber. Moreover, the connection between the model by KSZ and
actual vibrating granular systems is unclear.
%
The application of MCT formulated here to sheared granular fluids is now
in progress \cite{SH2012p2, CSOH2012}.
We have already found some remarkable features in MCT for sheared
granular fluids, where the projection onto the density-current mode
plays an important role in destructing the plateau of the density
time-correlator \cite{SH2012p2}.
Details will be reported elsewhere \cite{CSOH2012}.

\section{Summary and concluding remarks}
\label{sec:Sum}

In this paper, we have constructed a nonequilibrium mode-coupling theory
for uniformly sheared {\it underdamped} systems.
For such systems, the theory by Chong and Kim (CK) \cite{CK2009} has
been known, but the discrepancies with the theory by Fuchs and Cates
(FC) \cite{FC2009} in the overdamped limit have been left unresolved.
We have figured out that the formulation of CK is physically not
sensible;
i.e. it does not satisfy the translational invariance in the sheared
frame, and leads to peculiar overshoot/undershoot of the
density time-correlator.
We have performed a reformulation, starting from the redefinition of the
time-correlators to satisfy the ``alignment of the wavevectors'', which
is a consequence of the translational invariance.
The resulting MCT equation preserves this alignment, coincides with that
of FC in the overdamped limit, and leads to a monotonically-decreasing
density time-correlator.
Furthermore, motivated by the observation that the constant dissipative
coupling to the thermostat does not lead to a steady-state kinetic
termperature, we have implemented the isothermal condition at the level
of the Mori-type equations.
It is essential to incorporate current fluctuations in the dissipative
coupling to satisfy the isothermal condition.
Hence, it may well be said that we have non-trivially extended the FC
theory to underdamped systems in a physically sensible way.

Although it has been believed that the equivalence of long-time dynamics
in underdamped and overdamped systems (e.g. the NEP and the scaling
properties of the density time-correlator at the $\alpha$-relaxation)
also holds for sheared systems, we have figured out two deviations of the
underdamed from the overdamped case.
These are the relaxation of the shear stress at the $\alpha$-relaxation
regime due to a growth of current fluctuations in the dissipative
coupling to the thermostat, and the short-time response to a
perturbation of the shear rate.
These findings are the genuine features of our underdamped formulation,
which cannot be derived from the conventional overdamped framework.
We also stress that our formulation provides a physical explanation to
the discrepancy between MD and the overdamped MCT reported in
Ref.~\cite{ZH2008}.

An attractive feature of our formulation is that it is relatively simple
to extend to granular systems.
In these strongly nonequilibrium, nonlinear systems, genuine features of
the underdamped dynamics are also expected to be observed.
The construction of the theory and its numerical analysis is under
investigation, and will be published elsewhere \cite{SH2012p2,
CSOH2012}.

\begin{acknowledgments}

Numerical calculations in this work were carried out at the computer
facilities at the Yukawa Institute and Canon Inc.
The authors are grateful to S.-H. Chong, M. Otsuki, and G. Szamel for
stimulating discussions and careful reading of the manuscript.
They also thank K. Miyazaki for his kind advice and comments, 
%
M. Fuchs for his useful comments and providing the authors with the
information on Ref.~\cite{ZH2008},
and Canon Inc. for hosting the collaboration.
K.S. is also grateful to Mr. Shinjo, Mr. Nagane, and the members of the
Analysis Technology Development Department 1 for their continuous
encouragement.
\end{acknowledgments}

\appendix
\section{Application to the response theory}
\label{App:sec:response}

In this Appendix we apply the MCT to the calculation of the response
formula for the stress tensor.
We discuss a response to an instantaneous change in the shear rate
$\dot{\gamma}$ at the quasi-steady state which corresponds to the
plateau of the density time-correlator at $t=t_0$, where $t_0$ is of the
order of the early-$\beta$-relaxation time, $\tau_\beta$.
We consider a pulse-like perturbation identical to the one specified
in Eq.~(\ref{Eq:dgamma_pert}) in section \ref{sec:Resp}.
For a time-reversible thermostatted system, $\dot{\gamma}^{-1}$ is of the order of the
$\alpha$-relaxation time, $\dot{\gamma} \tau_\alpha \sim 1$.
On the other hand, there is a large hierarchy between these two times in
the weakly-sheared case, $\tau_\beta \ll \tau_\alpha$, and hence
$\dot{\gamma} t_0 \ll 1$ holds.
In the remainder, we perform expansions with $\dgamma t_0$, and neglect
terms with order $\mathcal{O}(\dgamma^2 t_0^2)$.

\subsection{Formulation}

\vspace{0.5em}
We first formulate a response formula of the stress tensor.
We consider a response of the stress tensor at $t>t_0 + \Delta
t_{\dot{\gamma}}$,
\begin{eqnarray}
\delta \sigma_{xy}(\bm{\Gamma}(t))
\equiv
\sigma_{xy}^{(\Delta \dot{\gamma})}(\bm{\Gamma}(t))
-
\sigma_{xy}^{(0)}(\bm{\Gamma}(t)),
\label{Eq:resp0}
\end{eqnarray}
where
$\sigma_{xy}^{(\Delta \dot{\gamma})}(\bm{\Gamma}(t))$ and
$\sigma_{xy}^{(0)}(\bm{\Gamma}(t))$ are the perturbed and the
unperturbed stress tensor, respectively.
Our aim is to derive a formula for the statistical average of this
quantity at the quasi-steady state.
In the absence of perturbation, the average of a phase-space variable
$A(\bm{\Gamma}(t))$ at the quasi-steady state is given by
\begin{eqnarray}
\left\langle
A(\bm{\Gamma}(t))
\right\rangle_{\mathrm{SS}}
=
\int d\bm{\Gamma}(t_0)
\rho_{\mathrm{SS}}(\bm{\Gamma}, t_0)
A(\bm{\Gamma}(t))
\end{eqnarray}
for $t>t_0$, where \cite{EM, COH2010-2}
\begin{eqnarray}
\rho_{\mathrm{SS}}(\bm{\Gamma}, t_0)
\equiv
\rho_{\mathrm{eq}}(\bm{\Gamma})
\exp \left[
\int_{0}^{t_0} ds \Omega_0 (\bm{\Gamma}(-s))
\right]
\label{Eq:rho_SS}
\end{eqnarray}
is the distribution function of the quasi-steady state, expressed in
terms of the initial equilibrium distribution function
$\rho_{\mathrm{eq}}(\bm{\Gamma})$ and the unperturbed work function
$\Omega_0(\bm{\Gamma}(t))$.
The unperturbed work function is given by
\begin{eqnarray}
\Omega_0(\bm{\Gamma})
=
- \beta \dot{\gamma} \sigma_{xy}^{(0)}(\bm{\Gamma})
- 2 \beta \alpha(\bm{\Gamma}) \delta K(\bm{\Gamma}),
\end{eqnarray}
where we include only the conservative part, $-\beta \Delta \dot{\gamma}
\sigma_{xy}^{(0)}(\bm{\Gamma})$, to the the perturbation of the work
function.
While the conservative part of the work function is
$\mathcal{O}(\dot{\gamma})$, the dissipative part, $-
2 \beta \alpha(\bm{\Gamma}) \delta K(\bm{\Gamma})$, is
$\mathcal{O}(\dot{\gamma}^2 t)$.
This can be seen from the fact that $\alpha(\vg)$ is accompanied by the
time-evolution operator $U_R(t)$ given by Eq.~(\ref{Eq:multiplier}),
where $\lambda_\alpha(t)$ is given by Eq.~(\ref{Eq:lambda_MCT}), and the
fact that the isotropic part of $k^x k^y$ is of the order of
$\mathcal{O}(\dot{\gamma} t)$.
Hence, the following equality holds,
\begin{eqnarray}
\Omega_0(\bm{\Gamma})
=
- \beta \dot{\gamma}
\sigma_{xy}^{(0)}(\bm{\Gamma})
+
\mathcal{O}(\dot{\gamma}^2 t),
\label{Eq:Omega0_expand}
\end{eqnarray}
which validates the approximation mentioned above.
We should note that $\dgamma t \ll 1$ holds, since $t \sim t_0$.

Then, the average of the response of the stress tensor at the
quasi-steady state is given by
\begin{eqnarray}
\left\langle
\delta \sigma_{xy}(\bm{\Gamma}(t))
\right\rangle_{\mathrm{SS}}
&=&
\int d\bm{\Gamma}(t_0)
\rho^{(\Delta \dot{\gamma})}
(\bm{\Gamma}, t_0 + \Delta t_{\dot{\gamma}})
\delta \sigma_{xy}(\bm{\Gamma}(t))
\nonumber \\
&&
+
\mathcal{O}(\Delta \gamma^2)
\label{Eq:dsigma_SS}
\end{eqnarray}
for $t > t_0 + \Delta t_{\dot{\gamma}}$, where
\begin{eqnarray}
\rho^{(\Delta \dot{\gamma})}
(\bm{\Gamma}, t_0 + \Delta t_{\dot{\gamma}})
&=&
\rho_{\mathrm{SS}}(\bm{\Gamma}, t_0)
\hspace{9em}
\nonumber \\
&&
\hspace{-6em}
\times
\left\{
1 - \beta \frac{\Delta \gamma}{\Delta t_{\dot{\gamma}}}
\int_{0}^{\Delta t_{\dot{\gamma}}} ds
\sigma_{xy}^{(0)}(\bm{\Gamma}(t_0-s))
\right\}
\label{Eq:rho_pert}
\end{eqnarray}
is the perturbed distribution function at $t = t_0 + \Delta
t_{\dot{\gamma}}$.
Here, $\Delta \gamma$ is the strain generated by the perturbation,
\begin{eqnarray}
\Delta \gamma
\equiv
\Delta \dot{\gamma}
\Delta t_{\dot{\gamma}},
\end{eqnarray}
which we fix at a small value $\Delta \gamma \ll 1$,
and take the limit $\Delta t_{\dot{\gamma}} \to 0$ later.

The linear response of the stress tensor is given from
Eq.~(\ref{Eq:dsigma_SS}) by
\begin{eqnarray}
&&
\hspace{-5em}
\left[
\frac{\partial}{\partial \Delta \gamma}
\left\langle
\delta \sigma_{xy}(\bm{\Gamma}(t))
\right\rangle_{\mathrm{SS}}
\right]_{\Delta \gamma = 0}
\nonumber \\
\hspace{3em}
&=&
-
\frac{\beta }{\Delta t_{\dot{\gamma}}}
\int d\bm{\Gamma}(t_0)
\rho_{\mathrm{SS}}(\bm{\Gamma}, t_0)
\delta \sigma_{xy}(\bm{\Gamma}(t))
\nonumber \\
&&
\times
\int_{0}^{\Delta t_{\dot{\gamma}}} ds
\sigma_{xy}^{(0)}(\bm{\Gamma}(t_0-s))
\nonumber \\
&=&
-
\frac{\beta }{\Delta t_{\dot{\gamma}}}
\int d\bm{\Gamma}(t_0)
\int_{0}^{\Delta t_{\dot{\gamma}}} ds
\rho_{\mathrm{SS}}(\bm{\Gamma}, t_0-s)
\nonumber \\
&&
\times
\delta \sigma_{xy}(\bm{\Gamma}(t+s))
\sigma_{xy}^{(0)}(\bm{\Gamma}(t_0))
\nonumber \\
&\simeq&
-
\beta
\left\langle
\delta \sigma_{xy}(\bm{\Gamma}(t))
\sigma_{xy}^{(0)}(\bm{\Gamma}(t_0))
\right\rangle_{\mathrm{SS}},
\hspace{3em}
\label{Eq:resp}
\end{eqnarray}
where in the final equality we utilized the fact that the pulse width is
sufficiently small, $\Delta t_{\dot{\gamma}} \ll t_0$, which validates
the following approximations, $\rho_{\mathrm{SS}}(\bm{\Gamma}, t_0-s)
\simeq \rho_{\mathrm{SS}}(\bm{\Gamma}, t_0)$ and $\delta
\sigma_{xy}(\bm{\Gamma}(t+s)) \simeq \delta \sigma_{xy}(\bm{\Gamma}(t))$
($0 \leq s \leq \Delta t_{\dot{\gamma}}$).
Note that Eq.~(\ref{Eq:resp}) is equivalent to the conventional linear
response formula \cite{KTH} if the steady state is sufficiently close to
equilibrium.

\subsection{Steady-state distribution function}

The response formula Eq.~(\ref{Eq:resp}) is still merely a formal
expression, and additional approximations are necessary to perform a
concrete calculation.
In this section we derive an approximate expression for the distribution
function of the quasi-steady state, Eq.~(\ref{Eq:rho_SS}), under a
weakly-sheared condition.

It is notable that $\rho_{\mathrm{SS}}(\bm{\Gamma},t_0)$ can be
rewritten as \cite{KN2008,COH2010-2}
\begin{eqnarray}
\rho_{\mathrm{SS}}(\bm{\Gamma},t_0)
&=&
\rho_{\mathrm{eq}}(\bm{\Gamma})
\cdot
\exp
\left[
\frac{1}{2}
\left(
\left\langle
\Theta_{-}
\right\rangle_{\bm{\gamma};0}
+
\left\langle
\Theta_{+}
\right\rangle_{\bm{\gamma};t_0}
\right)
\right]
\nonumber \\
&&
+
\mathcal{O}\left( \dot{\gamma}^3 t_0^3 \right),
\label{Eq:rho_SS3}
\end{eqnarray}
where $\Theta_{\pm}$ is defined as
$\Theta_{\pm}
\equiv
\int_{0}^{t_0} dt
\Omega_0(\bm{\Gamma}(\pm t))$,
and $\left\langle \cdots \right\rangle_{\bm{\gamma};t_0}$,
$\left\langle \cdots \right\rangle_{\bm{\gamma};0}$ are the
``conditioned averages''.
Refer to Eqs.~(3.2), (3.3) of Ref.~\cite{COH2010-2} for their
definition.
By expanding the Liouvillian as $i\mathcal{L} = i\mathcal{L}_0 +
\mathcal{O}(\dot{\gamma})$, there holds $\bm{\Gamma}(t) =
\bm{\Gamma}_0(t) + \mathcal{O}(\dot{\gamma} t)$, where $\bm{\Gamma}_0(t)
\equiv e^{i\mathcal{L}_0 t}\bm{\Gamma}(0)$.
Then, the two conditioned averages which appear in
Eq.~(\ref{Eq:rho_SS3}) can be expanded in terms of $\dot{\gamma}$ as
\begin{eqnarray}
\int_0^{t_0} ds
\left\langle
\Omega_0(\bm{\Gamma}(-s))
\right\rangle_{\bm{\gamma};0}
&=&
\int_0^{t_0} ds
\left\langle
\Omega_0(\bm{\Gamma}_0(-s))
\right\rangle_{\bm{\gamma};0}
\nonumber \\
&&
+
\mathcal{O}( \dot{\gamma}^2 t_0^2 ),
\label{Eq:Approx_0}
\\
\int_0^{t_0} ds
\left\langle
\Omega_0(\bm{\Gamma}(s))
\right\rangle_{\bm{\gamma};t_0}
&=&
\int_0^{t_0} ds
\left\langle
\Omega_0(\bm{\Gamma}_0(s))
\right\rangle_{\bm{\gamma};t_0}
\nonumber \\
&&
+
\mathcal{O}( \dot{\gamma}^2 t_0^2 ),
\label{Eq:Approx_t0}
\end{eqnarray}
where $\Omega_0 = \mathcal{O}(\dot{\gamma})$ should be reminded.
Note that the phase-space point which appears in the conditioned
averages are $\bm{\Gamma}_0(0)$ and $\bm{\Gamma}_0(t_0)$ for
$\left\langle \cdots \right\rangle_{\bm{\gamma};0}$ and $\left\langle
\cdots \right\rangle_{\bm{\gamma};t_0}$,
respectively.
As in Ref.~\cite{COH2010-2}, we can show that
\begin{eqnarray}
\hspace{-0.5em}
\int_0^{t_0} ds
\left\langle
\Omega_0(\bm{\Gamma}_0(-s))
\right\rangle_{\bm{\gamma};0}
=
\int_0^{t_0} ds
\left\langle
\Omega_0(\bm{\Gamma}_0(s))
\right\rangle_{\bm{\gamma};t_0}.
\hspace{2em}
\label{Eq:reflection}
\end{eqnarray}
From Eqs.~(\ref{Eq:Approx_0})--(\ref{Eq:reflection}), the following
approximation holds,
\begin{eqnarray}
\int_0^{t_0} ds
\left\langle
\Omega_0(\bm{\Gamma}(s))
\right\rangle_{\bm{\gamma};t_0}
&=&
\int_0^{t_0} ds
\left\langle
\Omega_0(\bm{\Gamma}(-s))
\right\rangle_{\bm{\gamma};0}
\nonumber \\
&&
+
\mathcal{O}(\dot{\gamma}^2 t_0^2),
\end{eqnarray}
and hence
\begin{eqnarray}
\left\langle
\Theta_{-}
\right\rangle_{\bm{\gamma};0}
+
\left\langle
\Theta_{+}
\right\rangle_{\bm{\gamma};t_0}
&=&
2
\int_0^{t_0} ds
\Omega_0(\bm{\gamma}(-s))
\nn \\
&&
+
\mathcal{O}(\dot{\gamma}^2 t_0^2),
\label{Eq:exponent2}
\end{eqnarray}
where $\left\langle
\Omega_0(\bm{\Gamma}(-s))
\right\rangle_{\bm{\gamma};0} =
\Omega_0(\bm{\gamma}(-s))$ should be noted.
From Eqs.~(\ref{Eq:rho_SS3}) and (\ref{Eq:exponent2}), we obtain
\begin{eqnarray}
\rho_{\mathrm{SS}}(\bm{\Gamma},t_0)
&=&
\rho_{\mathrm{eq}}(\bm{\Gamma})
\cdot
\exp
\left[
\int_0^{t_0} ds
\Omega_0(\bm{\gamma}(-s))
\right]
\nonumber \\
&&
+
\mathcal{O}\left( \dot{\gamma}^2 t_0^2 \right),
\label{Eq:rho_SS5}
\end{eqnarray}
where it should be reminded that $\int_0^{t_0} ds
\Omega_0(\bm{\gamma}(-s))$ depends on the choice of the phase-space
point $\bm{\gamma}$.
Eq.~(\ref{Eq:rho_SS5}) might seem similar to Eq.~(\ref{Eq:rho_SS}), but
actually it is not.
The factor $\exp \left[ \int_0^{t_0} ds
\Omega_0(\bm{\gamma}(-s)) \right]$ in Eq.~(\ref{Eq:rho_SS5}) is merely a
number and factors out from the phase-space integral, which is not the
case for the factor $\exp \left[ \int_0^{t_0} ds
\Omega_0(\bm{\Gamma}(-s)) \right]$ in Eq.~(\ref{Eq:rho_SS}).

The remaining task is to evaluate the factor $\exp \left[ \int_0^{t_0}
ds \Omega_0(\bm{\gamma}(-s)) \right]$.
One way to perform this is to expand $\Omega_0(\bm{\gamma}(-t))$ in terms
of its {\it equilibrium} ensemble average and the correction terms.
By introducing the Kawasaki transform $\bm{\gamma}^K$ \cite{ME1989}
of $\bm{\gamma}$, and from the Kawasaki-transform property $\left\langle
\Omega_0(\bm{\Gamma}(-t)) \right\rangle_{\bm{\gamma};0} = - \left\langle
\Omega_0 (\bm{\Gamma}(t)) \right\rangle_{\bm{\gamma}^K;0}$,
$\Omega_0(\bm{\gamma}(-t))$ can be approximated as
\begin{eqnarray}
\Omega_0(\bm{\gamma}(-t))
&\simeq&
-
\left\langle
\Omega_0(\bm{\Gamma}(t))
\right\rangle_{\mathrm{eq}}
\nonumber \\
&&
-
\left(
\left\langle
\Omega_0(\bm{\Gamma}(t))
\right\rangle^2_{\bm{\gamma}^K;0}
-
\left\langle
\Omega_0(\bm{\Gamma}(t))^2
\right\rangle_{\mathrm{eq}}
\right)^{1/2}
\nonumber \\
&\simeq&
-
\left\langle
\Omega_0(\bm{\Gamma}(t))
\right\rangle_{\mathrm{eq}},
\label{Eq:OmegaExpansion}
\end{eqnarray}
where the correction term is expected to be negligible compared to the
equilibrium average.
Furthermore, by only retaining the conservative contribution to the work
function as in Eq.~(\ref{Eq:Omega0_expand}), $\Omega_0(\bm{\Gamma})
\simeq - \beta \dot{\gamma} \sigma_{xy}^{(0)}(\bm{\Gamma})$, the
exponent of the factor is approximated as
\begin{eqnarray}
\int_0^{t_0} ds
\Omega_0(\bm{\gamma}(-s))
&=&
\beta \dot{\gamma}
\int_0^{t_0} ds
\left\langle
\sigma_{xy}^{(0)}(\bm{\Gamma}(s))
\right\rangle_{\mathrm{eq}}
\nonumber \\
&&
+
\mathcal{O}( \dot{\gamma}^2 t_0^2).
\label{Eq:exponent3}
\end{eqnarray}
From Eqs.~(\ref{Eq:rho_SS5}) and (\ref{Eq:exponent3}), the distribution
function for the quasi-steady state can be approximated as
\begin{eqnarray}
\rho_{\mathrm{SS}}(\bm{\Gamma},t_0)
&=&
\rho_{\mathrm{eq}}(\bm{\Gamma})
\cdot
\exp
\left[
\beta \dot{\gamma}
\int_0^{t_0} ds
\left\langle
\sigma_{xy}^{(0)}(\bm{\Gamma}(s))
\right\rangle_{\mathrm{eq}}
\right]
\nonumber \\
&&
+
\mathcal{O}\left( \dot{\gamma}^2 t_0^2 \right).
\label{Eq:rho_SS6}
\end{eqnarray}
Finally, from Eqs.~(\ref{Eq:resp}) and (\ref{Eq:rho_SS6}), the
approximate formula for the linear response is given by
\begin{eqnarray}
&&
\hspace{-3.5em}
\left[
\frac{\partial}{\partial \Delta \gamma}
\left\langle
\delta \sigma_{xy}(\bm{\Gamma}(t))
\right\rangle_{\mathrm{SS}}
\right]_{\Delta \gamma = 0}
\nonumber \\
\hspace{1.5em}
&=&
-
\beta
\exp
\left[
\beta \dot{\gamma}
\int_0^{t_0} ds
\left\langle
\sigma_{xy}^{(0)}(\bm{\Gamma}(s))
\right\rangle_{\mathrm{eq}}
\right]
\nonumber \\
&&
\times
\left\langle
\delta \sigma_{xy}(\bm{\Gamma}(t))
\sigma_{xy}^{(0)}(\bm{\Gamma}(t_0))
\right\rangle_{\mathrm{eq}}
+
\mathcal{O}\left( \dot{\gamma}^2 t_0^2 \right).
\hspace{2em}
\label{Eq:resp2}
\end{eqnarray}

\subsection{Mode-Coupling Approximation}

\hspace{0.5em}
So far we have derived a response formula in terms of an
{\it equilibrium} two-point function.
One way to calculate this function is to apply the mode-coupling
approximation, which we perform in this section.

From the definition of the response, Eq.~(\ref{Eq:resp0}), the two-point
function in Eq.~(\ref{Eq:resp2}) consists of the following two terms,
\begin{eqnarray}
\left\langle
\sigma_{xy}^{(\Delta \dot{\gamma})}(\bm{\Gamma}(t))
\sigma_{xy}^{(0)}(\bm{\Gamma}(t_0))
\right\rangle_{\mathrm{eq}}
\label{Eq:2point_eq2}
\end{eqnarray}
and
\begin{eqnarray}
\left\langle
\sigma_{xy}^{(0)}(\bm{\Gamma}(t))
\sigma_{xy}^{(0)}(\bm{\Gamma}(t_0))
\right\rangle_{\mathrm{eq}}.
\label{Eq:2point_eq_unpert}
\end{eqnarray}
By inserting the ``zero-mode'' projection operator $\mathcal{P}_2^0(t)$
onto pair-density modes, the perturbed two-point function in
Eq.~(\ref{Eq:2point_eq2}), for instance, can be approximated as
\begin{eqnarray}
&&
\hspace{-2em}
\left\langle
\sigma_{xy}^{(\Delta \dot{\gamma})}(\bm{\Gamma}(t))
\sigma_{xy}^{(0)}(\bm{\Gamma}(t_0))
\right\rangle_{\mathrm{eq}}
\nonumber \\
&\simeq&
\left\langle
\left[
\tilde{U}_0(t,t_0)
\mathcal{P}_2^0(t)
\sigma_{xy}^{(\Delta \dot{\gamma})}(\bm{\Gamma}(t_0))
\right]
\mathcal{P}_2^0(t_0)
\sigma_{xy}^{(0)}(\bm{\Gamma}(t_0))
\right\rangle_{\mathrm{eq}}.
\label{Eq:2point-fact}
\nonumber \\
\hspace{-9em}
&&
\hspace{-9em}
\end{eqnarray}
Applying the factorization approximation, Eq.~(\ref{Eq:2point-fact}) can
be expressed in terms of the density time-correlator $\Phi_{\bm{q}}(t)$
as
\begin{eqnarray}
&&
\hspace{-2.5em}
\left\langle
\sigma_{xy}^{(\Delta \dot{\gamma})}(\bm{\Gamma}(t))
\sigma_{xy}^{(0)}(\bm{\Gamma}(t_0))
\right\rangle_{\mathrm{eq}}
\nonumber \\
&\simeq&
\frac{1}{2}
\frac{(k_B T_{\mathrm{eq}})^2 }{V}
\int \frac{d^3 \bm{k}}{(2\pi)^3}
\frac{W_{\bm{k}(t)}}{S_{k(t)}}
\frac{W_{\bm{k}(t_0)}}{S_{k(t_0)}}
\Phi_{\bm{k}(t_0)}^{(\Delta \dot{\gamma})}(t)^2,
\hspace{1em}
\end{eqnarray}
where $\Phi_{\bm{k}(t_0)}^{(\Delta \dot{\gamma})}(t)$ is the density
time-correlator in the presence of the perturbation.
A similar approximation holds for the unperturbed two-point function
Eq.~(\ref{Eq:2point_eq_unpert}) as well, with $\Phi_{\bm{q}}^{(\Delta
\dot{\gamma})}(t)$ replaced by its unperturbed counterpart,
$\Phi_{\bm{q}}^{(0)}(t)$.
The resulting expression reads
\begin{eqnarray}
&&
\hspace{-1.5em}
\left[
\frac{\partial}{\partial \Delta \gamma}
\left\langle
\delta \sigma_{xy} (\bm{\Gamma}(t))
\right\rangle_{\mathrm{SS}}
\right]_{\Delta \gamma = 0}
\nonumber \\
&\simeq&
- \frac{k_B T_{\mathrm{eq}}}{2}
\exp \left[
- \frac{\dot{\gamma}^2 V}{2}
\int_0^{t_0} ds
\int_0^{s} ds'
\right.
\nonumber \\
&&
\times
\left.
\int \frac{d^3 \bm{k}}{(2\pi)^3}
\frac{W_{\bm{k}(s')}}{S_{k(s')}}
\frac{W_{\bm{k}}}{S_{k}}
\Phi_{\bm{k}}^{(0)}(s')^2
\right]
\nonumber \\
&&
\times
\int \frac{d^3 \bm{k}}{(2\pi)^3}
\frac{W_{\bm{k}(t)}}{S_{k(t)}}
\frac{W_{\bm{k}(t_0)}}{S_{k(t_0)}}
\left\{
\Phi_{\bm{k}(t_0)}^{(\Delta \dot{\gamma})}(t)^2
-
\Phi_{\bm{k}(t_0)}^{(0)}(t)^2
\right\}.
\label{Eq:response-3}
\nonumber \\
\end{eqnarray}
Equation~(\ref{Eq:response-3}) is our stress formula for the isothermal
sheared thermostat system.
%
\begin{figure}[thb]
\includegraphics[width=8.0cm]{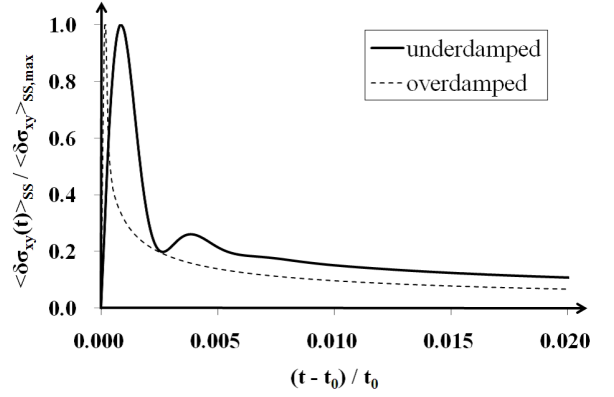}
%
\caption{Response of the stress tensor calculated from
 Eq.~(\ref{Eq:response-3}), normalized by their maximum values.}
\label{Fig:resp}
\end{figure}

The result for the response formula Eq.~(\ref{Eq:response-3}) is
shown in Fig.~\ref{Fig:resp}.
In this calculation, the result for the response of the density
time-correlator, which is shown in Fig.~\ref{Fig:dPhi}, is utilized.
As already seen in the result of Fig.~\ref{Fig:dPhi}, 
an oscillation can be observed for the underdamped case, while a
monotonically decreasing behavior is seen for the overdamped case.

\section{Mori-type equations for the underdamped and overdamped cases}
\label{App:sec:Mori}

In this Appendix, we discuss the relation between the Mori-type
equations for the underdamped and overdamped \cite{FC2009} cases.

In the underdamped case, the Mori-type equation consists of two
equations, Eqs.~(\ref{eom-Phi}) and (\ref{eom-H3}).
If we decompose $\ve{H}_\vq(t)$ into the longitudinal and transverse
components with respect to $\ve{q}(t)$,
\begin{eqnarray}
\ve{H}_\vq(t)
&=&
\ve{H}_\vq^L(t) + \ve{H}_\vq^T(t),
\end{eqnarray}
the longitudinal component can be explicitly given from
Eq.~(\ref{eom-Phi}) as
\begin{eqnarray}
\ve{H}_\vq^L(t)
&=&
\frac{\ve{q}(t)}{q(t)^2}
\frac{d}{dt} \Phi_\vq(t).
\label{App:Eq:HL}
\end{eqnarray}
By differentiating Eq.~(\ref{eom-Phi}) with time, and combining it with
Eq.~(\ref{eom-H3}), we obtain
\begin{eqnarray}
\frac{d^2}{dt^2}
\Phi_\vq(t)
&=&
\dot{\ve{q}}(t) \cdot \ve{H}_{\ve{q}}(t)
+
\ve{q}(t) \cdot \frac{d}{dt} \ve{H}_{\ve{q}}(t)
\nn \\
&=&
- v_T^2 \frac{q(t)^2}{S_{q(t)}} \Phi_\vq(t)
- \alpha_\vq(t) \frac{d}{dt} \Phi_\vq(t)
\nn \\
&&
- 2 \ve{q} \cdot \ve{\kappa} \cdot \ve{H}_{\vq}^T(t)
\nn \\
&&
- \int_0^t ds q(t)^\lambda L_\vq^\lambda(t,s) \Phi_\vq(s)
\nn \\
&&
- \int_0^t ds M_\vq(t,s)
\frac{d}{ds} \Phi_\vq(s)
\nn \\
&&
- \int_0^t ds q(t)^\lambda M_\vq^{\lambda \mu}(t,s) H_\vq^{T\mu}(s),
\label{App:Eq:mct-eq}
\end{eqnarray}
where we have utilized  Eq.~(\ref{App:Eq:HL}).
Here, the effective friction coefficient $\alpha_\vq(t)$ stands for
\begin{eqnarray}
\alpha_\vq(t)
&\equiv&
\alpha_0(t)
+
2 \frac{\ve{q} \cdot \ve{\kappa} \cdot \ve{q}(t)}{q(t)^2},
\end{eqnarray}
and the scalar memory kernel $M_\vq(t, s)$ is defined as
\begin{eqnarray}
M_\vq(t, s)
&\equiv&
q(t)^\lambda M_\vq^{\lambda \mu}(t,s) \frac{q(s)^\mu}{q(s)^2}.
\label{App:Eq:M}
\end{eqnarray}

On the other hand, the Mori-type equation for the overdamped case reads
\begin{eqnarray}
\frac{d}{dt}
\Phi_\vq(t)
&=&
- \Omega_\vq^2(t)
\Phi_\vq(t)
-
\int_0^t ds
m_\vq(t,s) \frac{d}{ds} \Phi_\vq(s),
\hspace{2em}
\label{App:Eq:mct-od}
\end{eqnarray}
where $\Omega_\vq^2(t) \equiv v_T^2 q(t)^2 / [\alpha_{\mathrm{od}}
S_{q(t)}]$ and $m_\vq(t,s) \equiv M_\vq(t,s)/\alpha_{\mathrm{od}}$
\cite{MR2002, MRY2004, FC2009}.
Here, $\alpha_{\mathrm{od}}$ is the effective friction coefficient which
is related to the bare diffusion coefficient $D_0$ by
$\alpha_{\mathrm{od}} = k_B T / (m D_0)$.

Aside from the inertia term, Eq.~(\ref{App:Eq:mct-eq}) is not coincident
with Eq.~(\ref{App:Eq:mct-od}) in three aspects: (i) the transverse
component $\ve{H}_\vq^{T}(t)$ does not decouple, (ii) there exists an
additional memory kernel $L_\vq^\lambda(t)$, (iii) the effective
friction coefficient $\alpha_\vq(t)$ depends on $\vq$ and $t$.

In situations where we can neglect $\ve{H}_\vq^T(t)$ and
$L_\vq^\lambda(t)$, a concrete example of which is shown below,
Eq.~(\ref{App:Eq:mct-eq}) is reduced to
\begin{eqnarray}
\frac{d^2}{dt^2}
\Phi_\vq(t)
&=&
- v_T^2 \frac{q(t)^2}{S_{q(t)}} \Phi_\vq(t)
-
\alpha_\vq(t)
\frac{d}{dt} \Phi_\vq(t)
\nn \\
&&
- \int_0^t ds M_\vq(t,s) \frac{d}{ds} \Phi_\vq(s).
\label{App:Eq:mct-eq2}
\end{eqnarray}
Although Eq.~(\ref{App:Eq:mct-eq2}) is formally coincident with
Eq.~(\ref{App:Eq:mct-od}) in the overdamped limit (i.e. upon neglection
of the inertia term), an exact correspondence is lacking due to the
discrepancy between $\alpha_\vq(t)$ and $\alpha_{\mathrm{od}}$.
Still, however, it is tempting and meaningful to compare the solution of
Eqs.~(\ref{App:Eq:mct-eq2}) and (\ref{App:Eq:mct-od}), which is performed
in this work.

One manifestation to neglect $\ve{H}_\vq^T(t)$ and $L_\vq^\lambda(t)$ is
to resort to the isotropic approximation, which we adopt in this work
(refer to Appendix \ref{App:subsec:Iso} for details).
In this approximation, $\ve{H}_\vq^T(t)$ is ignored by definition, and
it is shown in section \ref{sec:MCA} that the memory kernel
$L_\vq^\lambda(t)$ exactly vanishes in the mode-coupling approximation.
Furthermore, the scalar memory kernel defined in Eq.~(\ref{App:Eq:M})
coincides with that for the overdamped case in this approximation.
%

\section{Details of the derivations}
\label{App:sec:Details}

Large portion of the details of the calculations can be found in
previous papers, e.g. CK \cite{CK2009} and FC \cite{FC2009}.
In this Appendix, we show some details which is specific in this work.

\subsection{Steady-state formula}
\label{App:subsec:SS}

One of the goals of this study is to derive a formula for the
statistical ensemble average of a phase-space variable at the
nonequilibrium steady state.
For this purpose, we derive an analogue of the Green-Kubo formula which
is valid for sheared thermostatted systems.

To begin with, we adopt the ``Heisenberg picture'' for the statistical
ensemble average of a phase-space variable $A(\vg)$, defined as
\begin{eqnarray}
\eav{A(\vg(t))}
&=&
\int d\vg
\rho_{\mathrm{ini}}(\vg) A(\vg(t)).
\end{eqnarray}
Here, $A(\vg)$ is time-evolved while the distribution function remains
at its initial value, $\rho_{\mathrm{ini}}(\vg)$ \cite{EM}.
Since the system is at equilibrium with temperature $T$ at $t=0$,
$\rho_{\mathrm{ini}}(\vg)$ is the Maxwell-Boltzmann distribution
\begin{eqnarray}
\rho_{\mathrm{ini}}(\vg)
&\equiv&
\frac{e^{-\beta H_0(\vg)} }{\int d\vg e^{-\beta H_0(\vg)} },
\label{rho_ini}
\end{eqnarray}
where $H_0(\vg) \equiv K(\vg) + U(\vg)$ is the Hamiltonian, $K(\vg)
\equiv \sum_i p_i^2/2m$ is the total kinetic energy, and $\beta \equiv
1/\left(k_B T\right)$.

The starting point to derive the analogue of the Green-Kubo formula is
the integral expression for the nonequilibrium distribution function
$\rho(\vg, t)$,
\begin{eqnarray}
\rho(\vg,t)
&=&
\rho_{\mathrm{ini}}(\vg)
-
\beta \dgamma
\int_0^t ds
e^{-i\Liouv^\dagger s}
\left[
\rho_{\mathrm{ini}}(\vg) \sigma_{xy}(\vg)
\right]
\nn \\
&&
-
2 \beta
\int_0^t ds
e^{-i\Liouv^\dagger s}
\left[
\rho_{\mathrm{ini}}(\vg) \alpha(\vg) \delta K(\vg)
\right].
\label{rho2}
\end{eqnarray}
Here,
\begin{eqnarray}
\sigma_{xy}(\vg)
&\equiv&
\sum_i
\left(
\frac{p_i^x p_i^y}{m}
+
y_i F_i^x
\right),
\label{sxy}
\\
\delta K(\vg)
&\equiv&
K(\vg)
-
\frac{3}{2} N k_B T
\label{dK}
\end{eqnarray}
are the zero-wavevector limit of the shear stress and the fluctuation of
the kinetic energy, respectively, which together constitute the ``work
function'',
\begin{eqnarray}
\Omega(\vg)
&\equiv&
-
\beta \dgamma \sigma_{xy}(\vg)
-
2 \beta \alpha(\vg) \delta K(\vg).
\label{Omega}
\end{eqnarray}
As explained in section \ref{subsec:IsoTh}, note that the parameter
$\alpha(\vg)$ depends on $\vg$, whose specific choice in our work is
shown in Eq.~(\ref{Eq:alpha}).
From Eq.~(\ref{rho2}), we obtain the following expression for the
nonequilibrium ensemble average for a phase-space variable
$A(\vg(t))$,
\begin{eqnarray}
\eav{ A(\vg(t))}
&=&
\int d\vg \rho_{\mathrm{ini}}(\vg) A(\vg(t))
=
\int d\vg \rho(\vg, t) A(\vg(0))
\nn \\
&=&
\eav{ A(\vg(0))}
-
\beta \dgamma
\int_0^t ds
\eav{ A(\vg(t)) \sigma_{xy}(\vg(0)) }
\nn \\
&&
-
2 \beta
\int_0^t ds
\eav{ A(\vg(t)) \alpha(\vg) \delta K(\vg(0)) },
\label{vevA}
\end{eqnarray}
where the adjoint relation Eq.~(\ref{adj2}) is utilized, and the
integrations are assumed to converge uniformly.
Differentiation of Eq.~(\ref{vevA}) with time and the assumption of
``mixing'' results in the existence of a steady state in the limit $t
\ra \infty$, which leads to the following formula for the steady-state
ensemble average:
\begin{eqnarray}
\eav{ A}_{\mathrm{SS}}
&=&
\eav{ A(\vg)}
-
\beta \dgamma
\int_0^\infty ds
\eav{ A(\vg(t)) \sigma_{xy}(\vg) }
\nn \\
&&
-
2 \beta
\int_0^\infty ds
\eav{ A(\vg(t)) \alpha(\vg) \delta K(\vg) }.
\label{A_Ass}
\end{eqnarray}
Here, abbreviated notation $A(\vg) \equiv A(\vg(0))$ is adpoted.
This completes the derivation of Eq.~(\ref{Ass}).

\subsection{Fourier transform in the sheared frame}
\label{App:subsec:FT}

The derivation of Eqs.~(\ref{FT}), (\ref{advk}), and
(\ref{eomq})--(\ref{Liouv_dgammap}) is shown here.
Since the phase-space variables are defined in terms of the phase-space
coordinates, $\vg = \left\{ \ve{r}_i, \vp_i \right\}_{i=1}^N$, we need
to define ``field variables'' to consider the properties with respect to
spatial transformations, e.g. translational invariance.
A field variable for a phase-space variable $A(\vg)$ in the {\it
experimental frame} is introduced as
\begin{eqnarray}
A(\ve{r}, t)
&=&
\sum_i
A_i(\vg(t)) \delta( \ve{r} - \ve{r}_i(t) ),
\label{A_Art}
\end{eqnarray}
where $A_i(\vg(t))$ is a coefficient which depends only on the
phase-space variables.
Here, $\{\ve{r}\}$ is a coordinate fixed in space.
Its equation of motion is
\begin{eqnarray}
\frac{\del}{\del t}
A(\ve{r}, t)
&=&
\sum_i
\frac{\del}{\del t}
A_i(\vg(t)) \delta( \ve{r} - \ve{r}_i(t) )
\nn \\
&=&
\sum_i
i\Liouv
A_i(\vg(t)) \delta( \ve{r} - \ve{r}_i(t) )
=
i\Liouv
A(\ve{r}, t)
\nn \\
&=&
\left[
i\Liouv_0
+
i\Liouv_\alpha
+
i\Liouv_{\dgamma_r}
+
i\Liouv_{\dgamma_p}
\right]
A(\ve{r}, t),
\label{A_eom}
\end{eqnarray}
where the Liouvillians $i\Liouv_0$, $i\Liouv_\alpha$, $i\Liouv_{\dgamma
r}$ and $i\Liouv_{\dgamma_p}$ are given in Eqs.~(\ref{Liouv_0}),
(\ref{Liouv_alpha}), (\ref{Liouv_dgammar}) and (\ref{Liouv_dgammap}),
respectively.
Simple observation leads to
\begin{eqnarray}
i\Liouv_{\dgamma_r}
A(\ve{r}, t)
&=&
-
\ve{r} \cdot \vka^T \cdot \frac{\del}{\del \ve{r}}
A(\ve{r},t).
\label{A_LgammarA}
\end{eqnarray}
From Eqs.~(\ref{A_eom}) and (\ref{A_LgammarA}), we obtain
\begin{eqnarray}
\left[
\frac{\del}{\del t}
+
\ve{r} \cdot \vka^T \cdot \frac{\del}{\del \ve{r}}
\right]
A(\ve{r},t)
&=&
i\tLiouv
A(\ve{r},t),
\label{A_eom2}
\end{eqnarray}
where
\begin{eqnarray}
i\tLiouv
&\equiv&
i\Liouv_0
+
i\Liouv_\alpha
+
i\Liouv_{\dgamma_p}
\label{A_tLiouv}
\end{eqnarray}
is the Liouvillian which is obtained by subtracting $i\Liouv_{\dgamma
r}$ from $i\Liouv$.

Next we move to the {\it sheared frame} $\left( \tilde{\ve{r}},
\tilde{t} \right)$, which is defined by the coordinate transformation,
\begin{eqnarray}
\tilde{\ve{r}}
&\equiv&
\ve{r} - (\vka\cdot\ve{r})t
=
\left[
\ve{1} - \vka t
\right]
\cdot
\ve{r}
=
\left[
\begin{array}{c}
x - \left(\dgamma t\right) y \\
y \\
z
\end{array}
\right],
\nn \\
\tilde{t}
&\equiv&
t.
\label{A_trans}
\end{eqnarray}
This is a comoving frame with the stretching of the wavelengths due to
shearing.
The value of a field variable is unaltered by the transformation
Eq.~(\ref{A_trans}), so
\begin{eqnarray}
A(\ve{r}, t)
&=&
A(\tvr, \tit).
\label{A_id}
\end{eqnarray}
Simple calculation leads to
\begin{eqnarray}
\left[
\frac{\del}{\del t}
+
\ve{r} \cdot \vka^T \cdot \frac{\del}{\del \ve{r}}
\right]
A(\ve{r},t)
=
\frac{\del}{\del \tilde{t}}
A(\tilde{\ve{r}},\tilde{t}),
\end{eqnarray}
which implies, together with Eq.~(\ref{A_eom2}),
\begin{eqnarray}
\frac{\del}{\del \tilde{t}}
A(\tilde{\ve{r}},\tilde{t})
&=&
i\tLiouv
A(\tilde{\ve{r}}, \tilde{t}).
\label{A_eomq}
\end{eqnarray}
Hence, the time-evolution generator in the sheared frame is $i\tLiouv$,
Eq.~(\ref{A_tLiouv}).

Now we move on to the Fourier space.
Fourier transform in the experimental frame is
\begin{eqnarray}
A_\vq(t)
&\equiv&
\int d^3 \ve{r}
A(\ve{r}, t)
e^{i\vq\cdot\ve{r}}
=
\sum_i
A_i(\vg(t))
e^{i\vq\cdot\ve{r}_i(t)},
\nn \\
\label{A_FTex}
\end{eqnarray}
where $\vq$ is some wavevector.
It can easily be seen that $A_\vq(t)$ satisfies Eq.~(\ref{A_eom}) as
well,
\begin{eqnarray}
\frac{\del}{\del t}
A_\vq(t)
&=&
i\Liouv
A_\vq(t).
\end{eqnarray}
Fourier transform in the sheared frame is
\begin{eqnarray}
A_{\tilde{\vq}}(\tilde{t})
&\equiv&
\int d^3 \tilde{\ve{r}}
A(\tilde{\ve{r}}, \tilde{t})
e^{i\tilde{\vq}\cdot\tilde{\ve{r}}}
\nn \\
&=&
\int d^3 \ve{r}
\det
\left(
\frac{\del \tilde{\ve{r}}}{\del \ve{r}}
\right)
A(\ve{r}, t)
e^{i\tilde{\vq}\cdot(\ve{1}-\vka t)\cdot \ve{r}}
\nn \\
&=&
\int d^3 \ve{r}
A(\ve{r}, t)
e^{i\tilde{\vq}\cdot(\ve{1}-\vka t)\cdot \ve{r}},
\label{A_FTsh}
\end{eqnarray}
where Eqs.~(\ref{A_trans}), (\ref{A_id}) are applied.
Eq.~(\ref{A_id}) holds in the Fourier frame as well, so
comparing Eqs.~(\ref{A_FTex}) and (\ref{A_FTsh}), we obtain a relation
between the wavevectors in both frames,
\begin{eqnarray}
\vq
&=&
\tilde{\vq}\cdot(\ve{1}-\vka t).
\label{A_wv}
\end{eqnarray}
Inverting Eq.~(\ref{A_wv}),
\begin{eqnarray}
\tilde{\vq}
&=&
\vq\cdot
(\ve{1}+\vka t)
=
\vq(-t),
\label{A_tilq}
\\
\vq(t)
&\equiv&
\vq - \vq\cdot\vka t.
\label{A_fadv}
\end{eqnarray}
We verify that the wavevector in the sheared frame is Affine-deformed
by the shear-rate tensor $\ve{\kappa}$.
We choose the signature convention which is referred to as the ``forward
advection'' in FC \cite{FC2009}.
It can also easily be seen that $A_{\vq(-t)}(\tilde{t})$ satisfies
Eq.~(\ref{A_eomq}),
\begin{eqnarray}
\frac{\del}{\del \tilde{t}}
A_{\vq(-t)}(\tilde{t})
&=&
i \tLiouv
A_{\vq(-t)}(\tilde{t}).
\label{A_eomqt}
\end{eqnarray}
%

\subsection{The Mori-type equations}
\label{App:subsec:MoriEq}

The derivation of Eqs.~(\ref{eom-H2})--(\ref{tU0}), i.e. the Mori-type
equation without isothermal condition, is shown here.
Let us start with the basic properties, i.e. the action of the
Liouvillians on the density and the current-density fluctuations,
$n_\vq$ and $j_\vq^\lambda$.
The following equalities can be verified by straightforward
manipulations.
First, for the density fluctuation:
\begin{eqnarray}
i\tLiouv n_\vq
&=&
\left[
i\Liouv_0
+
i\Liouv_\alpha
+
i\Liouv_{\dgamma_p}
\right]
\sum_i e^{i\vq\cdot \ve{r}_i},
\end{eqnarray}
%
%
\begin{eqnarray}
i\Liouv_0 n_\vq
&=&
\sum_i
\left(
\frac{\ve{p}_i}{m}
\cdot
\frac{\del}{\del \ve{r}_i}
+
\ve{F}_i
\cdot
\frac{\del}{\del \ve{p}_i}
\right)
\sum_j e^{i\vq\cdot \ve{r}_j}
\nn \\
&=&
\sum_i i\vq \cdot \frac{\vp_i}{m} e^{i\vq\cdot \ve{r}_i}
=
i\vq \cdot \vj_\vq,
\label{L0n}
\end{eqnarray}
%
%
\begin{eqnarray}
i\Liouv_\alpha n_\vq
&=&
-
\alpha(\vg)
\sum_i
\vp_i
\cdot
\frac{\del}{\del \ve{p}_i}
\sum_j e^{i\vq\cdot \ve{r}_j}
=
0,
\label{Lalphan}
\end{eqnarray}
%
%
\begin{eqnarray}
i\Liouv_{\dgamma_p} n_\vq
&=&
-
\sum_i
\ve{p}_i
\cdot
\ve{\kappa}^T
\cdot \frac{\del}{\del \ve{p}_i}
\sum_j e^{i\vq\cdot \ve{r}_j}
=
0.
\label{Ldgpn}
\end{eqnarray}
From Eqs.~(\ref{L0n})--(\ref{Ldgpn}), the action of $i\tLiouv$ is
obtained as
\begin{eqnarray}
i\tLiouv n_\vq
&=&
i \vq\cdot\vj_\vq.
\label{tLn}
\end{eqnarray}

Next, for the current-density fluctuation:
\begin{eqnarray}
i\tLiouv j_\vq^\lambda
&=&
\left[
i\Liouv_0
+
i\Liouv_\alpha
+
i\Liouv_{\dgamma_p}
\right]
\sum_i \frac{p_i^\lambda}{m}e^{i\vq\cdot\ve{r}_i},
\end{eqnarray}
%
%
\begin{eqnarray}
i\Liouv_0 j_\vq^\lambda
&=&
\sum_i
\left(
\frac{\ve{p}_i}{m}
\cdot
\frac{\del}{\del \ve{r}_i}
+
\ve{F}_i
\cdot
\frac{\del}{\del \ve{p}_i}
\right)
\sum_j \frac{p_j^\lambda}{m}e^{i\vq\cdot \ve{r}_j}
\nn \\
&=&
\sum_i
\left(
iq^\mu
\frac{p_i^\mu p_i^\lambda}{m^2}
+
\frac{1}{m}
\ve{F}_i^\lambda
\right)
e^{i\vq\cdot \ve{r}_i},
\label{L0j}
\end{eqnarray}
%
%
\begin{eqnarray}
i\Liouv_\alpha j_\vq^\lambda
&=&
-
\alpha(\vg)
\sum_i
\vp_i
\cdot
\frac{\del}{\del \ve{p}_i}
\sum_j
\frac{p_j^\lambda}{m}
e^{i\vq\cdot \ve{r}_j}
=
- \alpha(\vg)
j_\vq^\lambda,
\label{Lalphaj}
\nn \\
\end{eqnarray}
%
%
\begin{eqnarray}
i\Liouv_{\dgamma_p}
j_\vq^\lambda
&=&
-
\sum_i
\ve{p}_i
\cdot
\ve{\kappa}^T
\cdot \frac{\del}{\del \ve{p}_i}
\sum_j \frac{p_j^\lambda}{m}e^{i\vq\cdot \ve{r}_j}
\nn \\
&=&
-
\sum_i
\frac{1}{m} (\vka\cdot\vp_i)^\lambda
e^{i\vq\cdot \ve{r}_i}
=
-
\left[ \vka \cdot \vj_\vq \right]^\lambda.
\nn \\
\label{Ldgpj}
\end{eqnarray}

Now we calculate the ``correlated part'', which is the first term on the
r.h.s. of Eq.~(\ref{dUj}).
It is projected by the rescaled projection operator Eq.~(\ref{bP}) as
\begin{widetext}
\begin{eqnarray}
U(t)
\bar{\mP}_t
e^{i\Liouv_{\dgamma_r}^\dagger t}
i\tLiouv
j_{\vq(t)}^\lambda
&=&
U(t)
\sum_\vk
\left\{
\eav{
\left[
e^{i\Liouv_{\dgamma_r}^\dagger t}
i\tLiouv
j_{\vq(t)}^\lambda
\right]
n_\vk^*
}
\frac{1}{NS_{k(t)}}
n_\vk
+
\eav{
\left[
e^{i\Liouv_{\dgamma_r}^\dagger t}
i\tLiouv
j_{\vq(t)}^\lambda
\right]
j_\vk^{\mu *}
}
\frac{1}{N v_T^2}
j_\vk^\mu
\right\}
\nn \\
&=&
\sum_\vk
\left\{
\eav{
i\tLiouv
j_{\vq(t)}^\lambda
\left[
e^{-i\Liouv_{\dgamma_r}t}
n_\vk
\right]^*
}
\frac{1}{NS_{k(t)}}
n_{\vk(t)}(t)
+
\eav{
i\tLiouv
j_{\vq(t)}^\lambda
\left[
e^{-i\Liouv_{\dgamma_r}t}
j_\vk^{\mu}
\right]^*
}
\frac{1}{N v_T^2}
j_{\vk(t)}^\mu(t)
\right\}
\nn \\
&=&
\sum_\vk
\left\{
\eav{
\left[
i\tLiouv
j_{\vq(t)}^\lambda
\right]
n_{\vk(t)}^*
}
\frac{1}{NS_{k(t)}}
n_{\vk(t)}(t)
+
\eav{
\left[
i\tLiouv
j_{\vq(t)}^\lambda
\right]
j_{\vk(t)}^{\mu *}
}
\frac{1}{N v_T^2}
j_{\vk(t)}^\mu(t)
\right\}.
\label{A_corr}
\end{eqnarray}
The two-point functions which appear in Eq.~(\ref{A_corr}) are obtained
by explicit manipulations.

First,
\begin{eqnarray}
\eav{\left[ i \tLiouv j_{\vq(t)}^\lambda \right] n_{\vk(t)}^*}
&=&
-\eav{j_{\vq(t)}^\lambda \left[ i\tLiouv n_{\vk(t)} \right]^*}
+\eav{j_{\vq(t)}^\lambda \left[ n_{\vk(t)} \tOmega \right]^*}
\nn \\
&=&
-\eav{j_{\vq(t)}^\lambda \left[ i \vk(t)\cdot \vj_{\vk(t)} \right]^*}
=
i k(t)^\mu \eav{j_{\vq(t)}^\lambda j_{\vk(t)}^{\mu *}}
=
i N v_T^2 q(t)^\lambda
\delta_{\vq,\vk},
\label{A_corrn}
\end{eqnarray}
where Eq.~(\ref{tLn}) and the fact that terms with odd number of momentum
variables vanish, are applied.

Next,
\begin{eqnarray}
\eav{\left[ i \tLiouv j_{\vq(t)}^\lambda \right]
j_{\vk(t)}^{\mu *}}
&=&
\eav{\left[ i \Liouv_\alpha j_{\vq(t)}^\lambda \right]
j_{\vk(t)}^{\mu *}}
+
\eav{\left[ i \Liouv_{\dgamma_p} j_{\vq(t)}^\lambda \right]
j_{\vk(t)}^{\mu *}},
\label{Eq:iLjj}
\end{eqnarray}
where the term with $i\Liouv_0$ vanishes due to odd number of momentum
variables.
In order to calculate the first term of Eq.~(\ref{Eq:iLjj}), it is
necessary to specify the explicit form of $\alpha(\vg)$.
As explained in section \ref{sec:IsoTh_MCT}, we adopt the form of
Eq.~(\ref{Eq:alpha}), which leads to
\begin{eqnarray}
\eav{\left[ i \tLiouv j_{\vq(t)}^\lambda \right]
j_{\vk(t)}^{\mu *}}
&=&
-
N v_T^2
\left\{
\alpha_0
\left( 1 + \frac{2}{3N} \right)
\delta^{\lambda \mu}
+
\ve{\kappa}^{\lambda \mu}
\right\}
\delta_{\vq,\vk}
\nn \\
&\simeq&
-
N v_T^2
\left\{
\alpha_0 \delta^{\lambda \mu}
+
\ve{\kappa}^{\lambda \mu}
\right\}
\delta_{\vq,\vk}
\label{A_corrj}
\end{eqnarray}
in the thermodynamic limit.
From Eqs.~(\ref{A_corr}), (\ref{A_corrn}), and (\ref{A_corrj}), the
result for the correlated part is
\begin{eqnarray}
U(t)
\bar{\mP}_t
e^{i\Liouv_{\dgamma_r}^\dagger t}
i\tLiouv
j_{\vq(t)}^\lambda
&=&
i v_T^2 \frac{q(t)^\lambda}{S_{q(t)}} n_{\vq(t)}(t)
-
\alpha_0 j_{\vq(t)}^\lambda(t)
-
\left[
\ve{\kappa} \cdot \vj_{\vq(t)}(t)
\right]^\lambda.
\end{eqnarray}
This leads to the first three terms in the r.h.s. of Eq.~(\ref{eom-H2}).

\vspace{1em}
Next, we calculate the ``random part''.
The term without time integration is defined as the ``random force'' in
Eqs.~(\ref{Randt}), (\ref{Rand}).
This is nothing but the fourth term in the r.h.s. of Eq.~(\ref{eom-H2}).
The part which requires some calculation is the one with memory kernels.
Projection with the rescaled projection operator Eq.~(\ref{bP}) is
\begin{eqnarray}
\left[
\frac{d}{dt} U(t) j_{\vq}^\lambda
\right]^{(\mathrm{mem})}
&\equiv&
\int_0^t ds
U(s)
\bar{\mP}_s
e^{i\Liouv_{\dgamma_r}^\dagger s}
i\tLiouv
e^{-i\Liouv_{\dgamma_r}s}
U_0(t,s)
e^{i\Liouv_{\dgamma_r}t}
R_{\vq(t)}^\lambda
\nn \\
&=&
\int_0^t ds
U(s)
\sum_\vk
\left\{
\eav{
\left[
e^{i\Liouv_{\dgamma_r}^\dagger s}
i\tLiouv
\tilde{U}_0(t,s)
R_{\vq(t)}^\lambda
\right]
n_\vk^*
}
\frac{1}{NS_{k(t)}}
n_\vk
+
\eav{
\left[
e^{i\Liouv_{\dgamma_r}^\dagger s}
i\tLiouv
\tilde{U}_0(t,s)
R_{\vq(t)}^\lambda
\right]
j_\vk^{\mu *}
}
\frac{1}{N v_T^2}
j_\vk^\mu
\right\}
\nn \\
&=&
\int_0^t ds
\sum_\vk
\left\{
\eav{
\left[
i\tLiouv
\tilde{U}_0(t,s)
R_{\vq(t)}^\lambda
\right]
n_{\vk(s)}^*
}
\frac{1}{NS_{k(t)}}
n_{\vk(s)}(s)
+
\eav{
\left[
i\tLiouv
\tilde{U}_0(t,s)
R_{\vq(t)}^\lambda
\right]
j_{\vk(s)}^{\mu *}
}
\frac{1}{N v_T^2}
j_{\vk(s)}^\mu(s)
\right\},
\label{A_dUj_mem}
\nn \\
\end{eqnarray}
where we introduced the abbreviated notation,
\begin{eqnarray}
\tilde{U}_0(t,s)
&\equiv&
e^{-i\Liouv_{\dgamma_r}s}
U_0(t,s)
e^{i\Liouv_{\dgamma_r}t}.
\end{eqnarray}
Substitution of Eq.~(\ref{A_dUj_mem}) into Eq.~(\ref{eom-H}) yields
\begin{eqnarray}
\frac{i}{N}
\eav{
\left[
\frac{d}{dt} U(t) j_{\vq}^\lambda
\right]^{(\mathrm{mem})}
n_\vq^*
}
&=&
\int_0^t ds
\left\{
\eav{
\left[
i\tLiouv
\tilde{U}_0(t,s)
R_{\vq(t)}^\lambda
\right]
n_{\vq(s)}^*
}
\frac{i}{NS_{q(t)}}
\Phi_\vq(s)
+
\eav{
\left[
i\tLiouv
\tilde{U}_0(t,s)
R_{\vq(t)}^\lambda
\right]
j_{\vq(s)}^{\mu *}
}
\frac{1}{N v_T^2}
H_\vq^\mu(s)
\right\}
\nn \\
&\equiv&
-
\int_0^t ds
L_\vq^\lambda (t,s) \Phi_\vq(s)
-
\int_0^t ds
M_\vq^{\lambda \mu}(t,s) H_\vq^\mu(s),
\end{eqnarray}
where the memory kernels $L_\vq^\lambda(t,s)$ and $M_\vq^{\lambda
\mu}(t,s)$ are the ones defined in Eqs.~(\ref{L}) and (\ref{M}).
These are the last two terms in the r.h.s. of Eq.~(\ref{eom-H2}).

\subsection{Insertion of the projection operator $\mQ$}
\label{App:subsec:InsertionQ}
The derivation of Eq.~(\ref{insertQ}) is shown here.
Let $X$ be an arbitrary phase-space variable, and consider the following
expression,
\begin{eqnarray}
U_0^\dagger (t,t')
e^{i\Liouv_{\dgamma_r} t'}
X^*
&=&
U_0^\dagger (t,t')
e^{i\Liouv_{\dgamma_r} t'}
\left[
\mP(t')
+
\mQ(t')
\right]
X^*,
\label{U0dagX}
\nn \\
\end{eqnarray}
where $U_0^\dagger (t,t')$ is the adjoint of $U_0(t,t')$,
\begin{eqnarray}
U_0^\dagger (t,t')
&\equiv&
\exp_\la
\left[
-
\int_{t'}^t ds
e^{i\Liouv_{\dgamma_r}^\dagger s}
i\tLiouv^{\dagger}
\mQ(s)
e^{-i\Liouv_{\dgamma_r}^\dagger s}
\right].
\end{eqnarray}
Here, the adjoint Liouvillians $i\tLiouv^\dagger$ and $i\Liouv_{\dgamma
r}^\dagger$ are defined in Eqs.~(\ref{tLiouv_dagger}) and
(\ref{Ldgammar_dag}), respectively.
The correlated part of Eq.~(\ref{U0dagX}) is
\begin{eqnarray}
U_0^\dagger (t,t')
e^{i\Liouv_{\dgamma_r} t'}
\mP(t')
X^*
&=&
U_0^\dagger (t,t')
e^{i\Liouv_{\dgamma_r} t'}
\sum_\vk
\left\{
\frac{
\eav{X^* n_{\vk(t')}}
}
{N S_{k(t')}}
n_{\vk(t')}^*
+
\frac{
\eav{X^* j_{\vk(t')}^{\mu}}
}
{N v_T^2}
j_{\vk(t')}^{\mu *}
\right\}
\nn \\
&=&
\sum_\vk
\left\{
\frac{
\eav{X^* n_{\vk(t')}}
}
{N S_{k(t')}}
U_0^\dagger (t,t')
n_{\vk}^*
+
\frac{
\eav{X^* j_{\vk(t')}^{\mu}}
}
{N v_T^2}
U_0^\dagger (t,t')
j_{\vk}^{\mu *}
\right\}.
\label{UdagPX}
\end{eqnarray}
The action of $U_0^\dagger (t,t')$ on $\xi = n$ and $j$ is
\begin{eqnarray}
U_0^\dagger (t,t')
\xi_{\vk}^*
&=&
\exp_\la
\left[
-
\int_{t'}^t ds
e^{i\Liouv_{\dgamma_r}^\dagger s}
i\tLiouv^{\dagger}
\mQ(s)
e^{-i\Liouv_{\dgamma_r}^\dagger s}
\right]
\xi_\vk^*
\nn \\
&=&
\exp_\la
\left[
-
\int_{t'}^t ds
e^{i\Liouv_{\dgamma_r}^\dagger s}
i\tLiouv^{\dagger}
\mQ(s)
\left[
1 + \Sigma(s)
\right]
e^{-i\Liouv_{\dgamma_r} s}
\right]
\xi_\vk^*
\nn \\
&\simeq&
\exp_\la
\left[
-
\int_{t'}^t ds
e^{i\Liouv_{\dgamma_r}^\dagger s}
i\tLiouv^{\dagger}
\mQ(s)
\xi_{\vk(s)}^*
\right]
=
0,
\label{Udagxi}
\end{eqnarray}
where we neglected $\Sigma(s)$ in the last step, as explained below
Eq.~(\ref{id}) in section \ref{sec:MCA}.
\end{widetext}
From Eqs.~(\ref{UdagPX}) and (\ref{Udagxi}), the correlated
part vanishes, and hence
\begin{eqnarray}
U_0^\dagger (t,t')
e^{i\Liouv_{\dgamma_r} t'}
&=&
U_0^\dagger (t,t')
e^{i\Liouv_{\dgamma_r} t'}
\mQ(t'),
\end{eqnarray}
or, taking the adjoint,
\begin{eqnarray}
e^{-i\Liouv_{\dgamma_r}^\dagger t'}
U_0(t,t')
&=&
\mQ(t')
e^{-i\Liouv_{\dgamma_r}^\dagger t'}
U_0(t,t').
\end{eqnarray}
This is the desired equality.

\subsection{The memory kernels}
\label{App:subsec:MemKer}

The derivation of Eqs.~(\ref{L2})--(\ref{dRand}) is shown here.
We start with the ensemble average of Eq.~(\ref{L}).
From the adjoint relation Eq.~(\ref{adj4}),
%
\begin{eqnarray}
&&
\hspace{-5em}
\eav{
\left[
i\tLiouv
\tilde{U}_0(t,s)
R_{\vq(t)}^\lambda
\right]
n_{\vq(s)}^*
}
\nn \\
&=&
-
\eav{
\left[
\tilde{U}_0(t,s)
R_{\vq(t)}^\lambda
\right]
\left(
i\tLiouv
n_{\vq(s)}
\right)^*
}
\nn \\
&&
+
\eav{
\left[
\tilde{U}_0(t,s)
R_{\vq(t)}^\lambda
\right]
n_{\vq(s)}^*
\tOmega
}.
\label{L5}
\end{eqnarray}
Substituting Eq.~(\ref{approx2}) into the two terms of Eq.~(\ref{L5}),
\begin{eqnarray}
&&
\hspace{-4em}
\eav{
\left[
\tilde{U}_0(t,s)
R_{\vq(t)}^\lambda
\right]
X^*
}
\nn \\
&\simeq&
\eav{
\left[
\mQ(s)
e^{-i\Liouv_{\dgamma_r}^\dagger s}
U_0(t,s)
e^{i\Liouv_{\dgamma_r}t}
R_{\vq(t)}^\lambda
\right]
X^*
}
\nn \\
&=&
\eav{
\left[
\mQ(s)
\tilde{U}_0'(t,s)
\mQ(t)
R_{\vq(t)}^\lambda
\right]
\mQ(s)
X^*
},
\end{eqnarray}
where $X = i\tLiouv n_{\vq(s)}, n_{\vq(s)}\tOmega$.
Here, the idempotency and the Hermiticity of $\mQ(s)$, and the
abbreviated notation of Eq.~(\ref{U0prime}) is used.
From Eq.~(\ref{tLn}), the first term of Eq.~(\ref{L5}) is projected out by
$\mQ(s)$, which leaves
\begin{eqnarray}
\eav{
\left[
\mQ(s)
\tilde{U}_0'(t,s)
\mQ(t)
R_{\vq(t)}^\lambda
\right]
\mQ(s)
\left[
n_{\vq(s)} \tOmega
\right]^*
}.
\end{eqnarray}
Similar manipulation for Eq.~(\ref{M}) leads to
\begin{eqnarray}
&&
\hspace{-4em}
\eav{
\left[
i\tLiouv
\tilde{U}_0(t,s)
R_{\vq(t)}^\lambda
\right]
j_{\vq(s)}^{\mu *}
}
\nn \\
&=&
-
\eav{
\left[
\mQ(s)
\tilde{U}_0'(t,s)
\mQ(t)
R_{\vq(t)}^\lambda
\right]
\Delta
R_{\vq(s)}^{\mu *}
},
\end{eqnarray}
where $\Delta R_{\vq(s)}^{\mu *}$ is the modified random force defined
in Eq.~(\ref{dRand}).

\subsection{The vertex functions}
\label{App:subsec:VF}

The derivation of Eqs.~(\ref{vf1})--(\ref{vf3}) is shown here.
We start with Eq.~(\ref{vf1}).
From the definition of Eq.~(\ref{Rand}),
\begin{eqnarray}
&&
\hspace{-1.5em}
\eav{
R_{\vq(t)}^\lambda
n_{\vk(t)}^*
n_{\vp(t)}^*
}
\nn \\
&=&
\eav{
\left[
\mQ(t)
i\tLiouv
j_{\vq(t)}^\lambda
\right]
n_{\vk(t)}^*
n_{\vp(t)}^*
}
\nn \\
&=&
\eav{
\left[
i\tLiouv
j_{\vq(t)}^\lambda
\right]
n_{\vk(t)}^*
n_{\vp(t)}^*
}
-
\eav{
\left[
\mP(t)
i\tLiouv
j_{\vq(t)}^\lambda
\right]
n_{\vk(t)}^*
n_{\vp(t)}^*
}.
\nn \\
\label{A_Rnn}
\end{eqnarray}
The first term of Eq.~(\ref{A_Rnn}) is evaluated, by the use of the adjoint
relation Eq.~(\ref{adj4}) and Eq.~(\ref{tLn}), as
\begin{eqnarray}
&&
\hspace{-1.5em}
\eav{
\left[i\tLiouv j_{\vq(t)}^\lambda \right] n_{\vk(t)}^* n_{\vp(t)}^*
}
\nn \\
&=&
-\eav{
j_{\vq(t)}^\lambda \left[ i\tLiouv n_{\vk(t)} \right]^* n_{\vp(t)}^*
}
-\eav{
j_{\vq(t)}^\lambda n_{\vk(t)}^* \left[ i\tLiouv n_{\vp(t)} \right]^*
}
\nn \\
&=&
ik(t)^\mu \eav{j_{\vq(t)}^\lambda j_{\vk(t)}^{\mu *} n_{\vp(t)}^*}
+
ip(t)^\mu \eav{j_{\vq(t)}^\lambda j_{\vp(t)}^{\mu *} n_{\vk(t)}^*}.
\nn \\
\label{A_tLjnn}
\end{eqnarray}
The three-point function in Eq.~(\ref{A_tLjnn}) can be calculated explicitly
as
\begin{eqnarray}
\eav{j_\vq^\lambda j_{\vk}^{\mu *} n_\vp^*}
&=&
\sum_{i,j}
\eav{\frac{p_i^\lambda}{m}e^{i\vq\cdot\ve{r}_i}
\frac{p_j^\mu}{m}e^{-i\vk\cdot\ve{r}_j}n_\vp^*}
\nn \\
&=&
\delta^{\lambda \mu} \delta_{\vq-\vk,\vp} N v_T^2 S_p.
\label{A_jnn}
\end{eqnarray}
From Eqs.~(\ref{A_tLjnn}) and (\ref{A_jnn}),
\begin{eqnarray}
&&
\hspace{-4em}
\eav{
\left[i\tLiouv j_{\vq(t)}^\lambda \right] n_{\vk(t)}^* n_{\vp(t)}^*
}
\nn \\
&=&
i\delta_{\vq,\vk+\vp}
N v_T^2
\left[ k(t)^\lambda S_{p(t)} + p(t)^\lambda S_{k(t)} \right].
\end{eqnarray}
An explicit manipulation of the projection operator in the second term
of Eq.~(\ref{A_Rnn}) leads to
\begin{eqnarray}
&&
\hspace{-1.5em}
\eav{
\left[
\mP(t)
i\tLiouv
j_{\vq(t)}^\lambda
\right]
n_{\vk(t)}^*
n_{\vp(t)}^*
}
\nn \\
&=&
\sum_{\vq'}
\left\{
\eav{
\left[
i\tLiouv
j_{\vq(t)}^\lambda
\right]
n_{\vq'(t)}^*
}
\frac{1}{N S_{q'(t)}}
\eav{
n_{\vq'(t)}
n_{\vk(t)}^*
n_{\vp(t)}^*
}
\right.
\nn \\
&&
\left.
+
\eav{
\left[
i\tLiouv
j_{\vq(t)}^\lambda
\right]
j_{\vq'(t)}^{\mu *}
}
\frac{1}{N v_T^2}
\eav{
j_{\vq'(t)}^\mu
n_{\vk(t)}^*
n_{\vp(t)}^*
}
\right\}
\nn \\
&=&
- \frac{1}{N S_{q'(t)}}
\sum_{\vq'}
\eav{
j_{\vq(t)}^\lambda
\left[
i\tLiouv
n_{\vq'(t)}
\right]^*
}
\eav{
n_{\vq'(t)}
n_{\vk(t)}^*
n_{\vp(t)}^*
}
\nn \\
&=&
\frac{i q'(t)^\mu}{N S_{q'(t)}}
\sum_{\vq'}
\eav{
j_{\vq(t)}^\lambda
j_{\vq'(t)}^{\mu *}
}
\eav{
n_{\vq'(t)}
n_{\vk(t)}^*
n_{\vp(t)}^*
}
\nn \\
&=&
i v_T^2 \frac{q(t)^\mu}{S_{q(t)}}
\eav{
n_{\vq(t)}
n_{\vk(t)}^*
n_{\vp(t)}^*
}.
\end{eqnarray}
The convolution approximation \cite{HM},
\begin{eqnarray}
\eav{n_\vq n_\vk^* n_\vp^*}
&\simeq &
\delta_{\vq,\vk+\vp} N S_q S_k S_p,
\end{eqnarray}
is applied for the three-point function, which finally leads to the
following expression,
\begin{eqnarray}
&&
\hspace{-1.5em}
\eav{
R_{\vq(t)}^\lambda
n_{\vk(t)}^*
n_{\vp(t)}^*
}
\nn \\
&=&
i\delta_{\vq,\vk+\vp}
N v_T^2
\left[
k(t)^\lambda S_{p(t)} + p(t)^\lambda S_{k(t)}
\right]
\nn \\
&&
-
i v_T^2 \frac{q(t)^\lambda}{S_{q(t)}}
\delta_{\vq,\vk+\vp}
N S_{q(t)} S_{k(t)} S_{p(t)}
\nn \\
&=&
i\delta_{\vq,\vk+\vp}
N v_T^2
\left[
k(t)^\lambda S_{p(t)} + p(t)^\lambda S_{k(t)} - q(t)^\lambda S_{k(t)} S_{p(t)}
\right]
\nn \\
&=&
-
i\delta_{\vq,\vk+\vp}
N v_T^2
S_{k(t)} S_{p(t)}
\nn \\
&&
\times
\left(
k(t)^\lambda + p(t)^\lambda
- \frac{k(t)^\lambda}{S_{k(t)}} - \frac{p(t)^\lambda}{S_{p(t)}}
\right)
\nn \\
&=&
-
i\delta_{\vq,\vk+\vp}
N v_T^2
S_{k(t)} S_{p(t)} n
\left[
k(t)^\lambda c_{k(t)} + p(t)^\lambda c_{p(t)}
\right].
\label{A_Rnn2}
\nn \\
\end{eqnarray}
Here, $c_q$ is the direct correlation function \cite{HM} which is
related to the static structure factor $S_q$ as
\begin{eqnarray}
n c_q
&=&
1 - \frac{1}{S_q}
\end{eqnarray}
by the Ornstein-Zernike relation \cite{HM}.
From Eq.~(\ref{A_Rnn2}), it is straightforward to verify Eqs.~(\ref{vf1}) and
(\ref{V}).

Next we deal with Eq.~(\ref{vf2}).
From the definition of $\Delta R_\vq^\lambda$,
\begin{eqnarray}
&&
\hspace{-2em}
\frac{
\eav{n_{\vk(t)} n_{\vp(t)} \Delta R_{\vq(t)}^{\lambda *} }
}
{N^2 S_{k(t)} S_{p(t)}}
\nn \\
&=&
\frac{
\eav{n_{\vk(t)} n_{\vp(t)} R_{\vq(t)}^{\lambda *} }
}
{N^2 S_{k(t)} S_{p(t)}}
-
\frac{
\eav{n_{\vk(t)} n_{\vp(t)}
\mQ(t)
\left[
j_{\vq(t)}^\lambda \tOmega
\right]^* }
}
{N^2 S_{k(t)} S_{p(t)}},
\nn \\
\label{A_nndR}
\end{eqnarray}
where the first term of Eq.~(\ref{A_nndR}) is the complex conjugate of
Eq.~(\ref{vf1}).
Hence, we only need to handle the second term.
There are two terms in the modified work function, $\tOmega(\vg) =
-\beta \dgamma \sigma_{xy}^{(\mathrm{kin})}(\vg) - 2\beta \alpha(\vg)
\delta K(\vg)$.
For the first term,
\begin{eqnarray}
\mQ(t)
\left[
j_{\vq(t)}^\lambda
\sigma_{xy}^{(\mathrm{kin})}
\right]
&=&
j_{\vq(t)}^\lambda
\sigma_{xy}^{(\mathrm{kin})}
-
\mP(t)
\left[
j_{\vq(t)}^\lambda
\sigma_{xy}^{(\mathrm{kin})}
\right],
\label{A_Qjsxy}
\nn \\
\end{eqnarray}
where $\sigma_{xy}^{(\mathrm{kin})} = \sum_i p_i^x p_i^y / 2m$ is the kinetic
part of the shear stress.
The projected term is
\begin{eqnarray}
\mP(t)
\left[
j_{\vq(t)}^\lambda
\sigma_{xy}^{(\mathrm{kin})}
\right]
&=&
\sum_\vk
\frac{
\eav{\left[ j_{\vq(t)}^{\lambda} \sigma_{xy}^{(\mathrm{kin})}\right] n_{\vk(t)}^*}
}{N S_{k(t)}}
n_{\vk(t)}
\nn \\
&&
+
\sum_\vk
\frac{
\eav{\left[ j_{\vq(t)}^{\lambda} \sigma_{xy}^{(\mathrm{kin})}\right] j_{\vk(t)}^{\mu *}}
}{Nv_T^2}
j_{\vk(t)}^{\mu}
\nn \\
&=&
m v_T^2
\left(
\delta^{\lambda x}
j_{\vq(t)}^{y}
+
\delta^{\lambda y}
j_{\vq(t)}^{x}
\right).
\label{A_Pjsxy}
\end{eqnarray}
From Eqs.~(\ref{A_Qjsxy}) and (\ref{A_Pjsxy}), $\mQ(t) \left[
j_{\vq(t)}^\lambda \sigma_{xy}^{(\mathrm{kin})} \right]$ is an odd function of
the momentum variables, and hence vanishes.
For the second term,
\begin{eqnarray}
&&
\hspace{-3em}
\mQ(t)
\left[
j_{\vq(t)}^\lambda
\alpha(\vg)
\delta K(\vg)
\right]
\nn \\
&=&
j_{\vq(t)}^\lambda
\alpha(\vg)
\delta K(\vg)
-
\mP(t)
\left[
j_{\vq(t)}^\lambda
\alpha(\vg)
\delta K(\vg)
\right],
\hspace{0.5em}
\label{A_QjdK}
\end{eqnarray}
where $\delta K(\vg) = \sum_i \vp_i^2 / 2m - 3Nk_B T/2$ is the fluctuation of
the kinetic energy, and $\alpha(\vg)$ is given in Eq.~(\ref{Eq:alpha}).
The projected term is
\begin{eqnarray}
&&
\hspace{-4em}
\mP(t)
\left[
j_{\vq(t)}^\lambda
\alpha(\vg)
\delta K(\vg)
\right]
\nn \\
&=&
\sum_\vk
\frac{
\eav{\left[ j_{\vq(t)}^{\lambda} \alpha(\vg) \delta K(\vg) \right] n_{\vk(t)}^*}
}{N S_{k(t)}}
n_{\vk(t)}
\nn \\
&&
+
\sum_\vk
\frac{
\eav{\left[ j_{\vq(t)}^{\lambda} \alpha(\vg) \delta K(\vg) \right] j_{\vk(t)}^{\mu *}}
}{Nv_T^2}
j_{\vk(t)}^{\mu}
\nn \\
&\propto&
j_{\vq(t)}^\lambda.
\label{A_PjdK}
\end{eqnarray}
From Eqs.~(\ref{A_QjdK}) and (\ref{A_PjdK}), $\mQ(t)
\left[j_{\vq(t)}^\lambda \alpha(\vg) \delta K(\vg) \right]$ is an odd
function of the momentum variables, and hence vanishes as well.
This completes the derivation of Eq.~(\ref{vf2}).

Finally we show Eq.~(\ref{vf3}).
The part which corresponds to the first term of the modified work
function $\tOmega(\vg)$ is
\begin{eqnarray}
\mQ(t)
\left[
n_{\vq(t)}
\sigma_{xy}^{(\mathrm{kin})}
\right]
&=&
n_{\vq(t)}
\sigma_{xy}^{(\mathrm{kin})}
-
\mP(t)
\left[
n_{\vq(t)}
\sigma_{xy}^{(\mathrm{kin})}
\right],
\label{A_Qnsxy}
\nn \\
\end{eqnarray}
where the projected term vanishes,
\begin{eqnarray}
\mP(t)
\left[
n_{\vq(t)}
\sigma_{xy}^{(\mathrm{kin})}
\right]
&=&
\sum_\vk
\frac{
\eav{\left[ n_{\vq(t)} \sigma_{xy}^{(\mathrm{kin})}\right] n_{\vk(t)}^*}
}{N S_{k(t)}}
n_{\vk(t)}
\nn \\
&&
+
\sum_\vk
\frac{
\eav{\left[ n_{\vq(t)} \sigma_{xy}^{(\mathrm{kin})}\right] j_{\vk(t)}^{\mu *}}
}{Nv_T^2}
j_{\vk(t)}^{\mu}
\nn \\
&=&
0,
\label{A_Pnsxy}
\end{eqnarray}
since $\eav{p_i^x p_i^y}_\vp = 0$.
Here, $\eav{\cdots}_\vp$ is the ensemble average with only the momentum
variables integrated.
From Eqs.~(\ref{A_Qnsxy}) and (\ref{A_Pnsxy}),
\begin{eqnarray}
\eav{n_{\vk(t)} n_{\vp(t)}
\mQ(t) \left[ n_{\vq(t)}^* \sigma_{xy}^{(\mathrm{kin})} \right]}
&=&
\eav{n_{\vk(t)}
n_{\vp(t)} n_{\vq(t)}^* \sigma_{xy}^{(\mathrm{kin})} },
\nn \\
\end{eqnarray}
which again vanishes due to $\eav{p_i^x p_i^y}_\vp =0$.
The second term is
\begin{eqnarray}
&&
\hspace{-2.5em}
\mQ(t)
\left[
n_{\vq(t)}
\alpha(\vg)
\delta K(\vg)
\right]
\nn \\
&=&
n_{\vq(t)}
\alpha(\vg)
\delta K(\vg)
-
\mP(t)
\left[
n_{\vq(t)}
\alpha(\vg)
\delta K(\vg)
\right].
\hspace{1em}
\label{A_QndK}
\end{eqnarray}
Here, the projected term is
\begin{eqnarray}
&&
\hspace{-4em}
\mP(t)
\left[
n_{\vq(t)}
\alpha(\vg)
\delta K(\vg)
\right]
\nn \\
&=&
\sum_\vk
\frac{
\eav{\left[ n_{\vq(t)} \alpha(\vg) \delta K(\vg) \right] n_{\vk(t)}^*}
}{N S_{k(t)}}
n_{\vk(t)}
\nn \\
&&
+
\sum_\vk
\frac{
\eav{\left[ n_{\vq(t)} \alpha(\vg) \delta K(\vg) \right] j_{\vk(t)}^{\mu *}}
}{Nv_T^2}
j_{\vk(t)}^{\mu}
\nn \\
&=&
\alpha_0 k_B T
n_{\vq(t)},
\label{A_PndK}
\end{eqnarray}
where
\begin{eqnarray}
\eav{
\alpha(\vg) \delta K(\vg)
}_{\vp}
&=&
\alpha_0  k_B T
\label{App:Eq:adK}
\end{eqnarray}
is utilized.
From Eqs.~(\ref{A_QndK}) and (\ref{A_PndK}),
\begin{eqnarray}
&&
\hspace{-5em}
\eav{n_{\vk(t)} n_{\vp(t)}
\mQ(t) \left[ n_{\vq(t)}^* \alpha(\vg)\delta K(\vg) \right]}
\nn \\
&=&
\eav{n_{\vk(t)}
n_{\vp(t)} n_{\vq(t)}^* \alpha(\vg)\delta K(\vg) }
\nn \\
&&
-
\alpha_0 k_B T
\eav{n_{\vk(t)}
n_{\vp(t)} n_{\vq(t)}^*}
,
\end{eqnarray}
which again vanishes due to Eq.~(\ref{App:Eq:adK}). 

\vspace{1.5em}
\subsection{Factorization approximations}
\label{App:subsec:FA}

The derivation of Eqs.~(\ref{FA}) and (\ref{4point}) is shown here.
From the definition Eq.~(\ref{U0prime}), the left-hand side (l.h.s.) of
Eq.~(\ref{FA}) can be expanded as
\begin{eqnarray}
&&
\hspace{-1.5em}
\frac{1}{N^2}
\eav{
\left[
\tilde{U}_0'(t,s)
n_{\vk(t)}
n_{\vp(t)}
\right]
n_{\vk'(s)}^*
n_{\vp'(s)}^*
}
\nn \\
&=&
\frac{1}{N^2}
\eav{
\left[
e^{-i\Liouv_{\dgamma_r}^\dagger s}
U_0(t,s)
e^{i\Liouv_{\dgamma_r}t}
n_{\vk(t)}
n_{\vp(t)}
\right]
n_{\vk'(s)}^*
n_{\vp'(s)}^*
}
\nn \\
&=&
\frac{1}{N^2}
\eav{
\left[
U_0(t,s)
n_{\vk}
n_{\vp}
\right]
n_{\vk'}^*
n_{\vp'}^*
},
\label{App:lhs}
\end{eqnarray}
where the property $e^{i\Liouv_{\dgamma_r}t} n_{\vk(t)} n_{\vp(t)} =
n_\vk n_\vp$ is utilized.
It is clear that this is the propagator for the pair-density
fluctuations with respect to the projected time-evolution operator
$U_0(t,s)$.
As is familiar in conventional MCT, Eq.~(\ref{App:lhs}) is approximated
by factorizing into a product of propagators for the density fluctuation
with respect to the projection-free time-evolution operator $e^{i\Liouv
(t-s)}$,
\begin{widetext}
\begin{eqnarray}
\frac{1}{N^2}
\eav{
\left[
U_0(t,s)
n_{\vk}
n_{\vp}
\right]
n_{\vk'}^*
n_{\vp'}^*
}
&\simeq&
\frac{1}{N^2}
\eav{
\left[
e^{i\Liouv_{\dgamma_r}^\dagger s}
e^{i\Liouv (t-s)}
e^{-i\Liouv_{\dgamma_r}t}
n_{\vk}
\right]
n_{\vk'}^*
}
\eav{
\left[
e^{i\Liouv_{\dgamma_r}^\dagger s}
e^{i\Liouv (t-s)}
e^{-i\Liouv_{\dgamma_r}t}
n_{\vp}
\right]
n_{\vp'}^*
}
\nn \\
&=&
\frac{1}{N^2}
\eav{
\left[
e^{i\Liouv (t-s)}
e^{-i\Liouv_{\dgamma_r}(t-s)}
n_{\vk(s)}
\right]
n_{\vk'(s)}^*
}
\eav{
\left[
e^{i\Liouv (t-s)}
e^{-i\Liouv_{\dgamma_r}(t-s)}
n_{\vp(s)}
\right]
n_{\vp'(s)}^*
}
\nn \\
&=&
\frac{1}{N^2}
\eav{
\left[
U(t-s) n_{\vk(s)}
\right]
n_{\vk'(s)}^*
}
\eav{
\left[
 U(t-s) n_{\vp(s)}
\right]
n_{\vp'(s)}^*
}
\nn \\
&=&
\delta_{\vk', \vk}
\delta_{\vp', \vp}
\Phi_{\vk(s)}(t-s)
\Phi_{\vp(s)}(t-s).
\label{App:rhs}
\end{eqnarray}
\end{widetext}
In the first step of the r.h.s. of Eq.~(\ref{App:rhs}), the advection
generators $e^{i\Liouv_{\dgamma_r}^\dagger s}$ and $e^{-i\Liouv_{\dgamma
r}t}$ are inserted to account for the time when the propagation starts,
$s$, and ends, $t$.
Hence Eq.~(\ref{FA}) is shown.

Similarly, the l.h.s. of Eq.~(\ref{4point}) is
\begin{eqnarray}
&&
\hspace{-6em}
\frac{1}{N^2}
\eav{\left[
\tilde{U}_0(t,0)
n_{\vk(t)} n_{\vk(t)}^*
\right]
n_{\vk'} n_{\vk'}^*
}
\nn \\
&=&
\frac{1}{N^2}
\eav{\left[
U_0(t,0)
n_{\vk} n_{\vk}^*
\right]
n_{\vk'} n_{\vk'}^*
},
\end{eqnarray}
which is approximated as
\begin{eqnarray}
&&
\hspace{-1.5em}
\frac{1}{N^2}
\eav{\left[
U_0(t,0)
n_{\vk} n_{\vk}^*
\right]
n_{\vk'} n_{\vk'}^*
}
\nn \\
&\simeq &
\frac{1}{N}
\eav{\left[e^{i\Liouv t} e^{-i\Liouv_{\dgamma_r}t} n_{\vk} \right]n_{\vk'}^*}
\cdot
\frac{1}{N}
\eav{\left[e^{i\Liouv t}e^{-i\Liouv_{\dgamma_r}t} n_{\vk}^* \right](n_{\vk'}^*)^*}
\nn \\
&&
+
\frac{1}{N}
\eav{\left[e^{i\Liouv t} e^{-i\Liouv_{\dgamma_r}t} n_{\vk} \right](n_{\vk'}^*)^*}
\cdot
\frac{1}{N}
\eav{\left[e^{i\Liouv t} e^{-i\Liouv_{\dgamma_r}t} n_{\vk}^*
     \right]n_{\vk'}^*}
\nn \\
&=&
\frac{1}{N}
\eav{\left[e^{i\Liouv t} n_{\vk(t)} \right]n_{\vk'}^*}
\cdot
\frac{1}{N}
\eav{\left[e^{i\Liouv t} n_{\vk(t)}^* \right](n_{\vk'}^*)^*}
\nn \\
&&
+
\frac{1}{N}
\eav{\left[e^{i\Liouv t} n_{\vk(t)} \right](n_{\vk'}^*)^*}
\cdot
\frac{1}{N}
\eav{\left[e^{i\Liouv t} n_{\vk(t)}^* \right]n_{\vk'}^*}
\nn \\
&=&
\delta_{\vk',\vk} \Phi_\vk(t)
\cdot
\delta_{\vk',\vk} \Phi_{-\vk}(t)
+
\delta_{-\vk',\vk} \Phi_\vk(t)
\cdot
\delta_{\vk',-\vk} \Phi_{-\vk}(t)
\nn \\
&=&
\left[
\delta_{\vk',\vk}
+
\delta_{\vk',-\vk}
\right]
\Phi_\vk(t)^2.
\label{App:4point}
\end{eqnarray}
Hence Eq.~(\ref{4point}) is shown.
Note that Eq.~(\ref{App:4point}) is a special case of
Eq.~(\ref{App:rhs}) with $s=0$, aside from the possible pairings of the
density fluctuations.

\subsection{The projected time-correlator}
\label{App:subsec:PTC}

The derivation of Eq.~(\ref{GA2}) is shown here.
From Eq.~(\ref{U}), there holds
\begin{eqnarray}
U(t) A(\vg)
&=&
U_0(t,0) A(\vg)
\nn \\
&&
+
\int_0^t ds
U(s)
\bar{\mP}_s
e^{i\Liouv_{\dgamma_r}^\dagger s}
i\tLiouv
e^{-i\Liouv_{\dgamma_r}s}
U_0(t,s) A(\vg).
\nn \\
\end{eqnarray}
For the second term, the application of the rescaled projection operator
Eq.~(\ref{bP}) leads to the form
\begin{eqnarray}
&&
\int_0^t ds
\sum_\vk
\left\{
\frac{C_\vk^{(n)}}{S_{k(t)}}
n_{\vk(s)}(s)
+
\frac{C_\vk^{(j) \lambda}}{S_{k(t)}}
j_{\vk(s)}^\lambda(s)
\right\},
\end{eqnarray}
where $C_\vk^{(\xi)}$ $(\xi = n,j)$ is a correlator whose detailed
expression is not important for our purpose.
The time-correlators of the density and the current-density fluctuations
with zero-wavevector variables $B(\vg)$ vanish,
\begin{eqnarray}
\eav{
n_{\vk(s)}(s)
B(\vg)
}
&=&
\delta_{\vk,0}
\eav{
n_{\vk=0}(s)
B(\vg)
}
=
0,
\nn \\
\eav{
j_{\vk(s)}^\lambda(s)
B(\vg)
}
&=&
\delta_{\vk,0}
\eav{
j_{\vk=0}^\lambda(s)
B(\vg)
}
=
0,
\end{eqnarray}
since $n_{\vk=0}=0$ holds by definition, and
\begin{eqnarray}
j_{\vk=0}^\lambda(s)
&=&
e^{i\Liouv s} \sum_i \frac{p_i^\lambda}{m} = 0
\end{eqnarray}
from the definition of the peculiar momentum, $\vp_i \equiv m\left(
\dot{\ve{r}}_i - \ve{\kappa}\cdot\ve{r}_i \right).$
This completes the derivation.

\subsection{The projected shear stress}
\label{App:subsec:PSS}
The derivation of Eq.~(\ref{Psigma}) is shown here.
The shear stress is projected by the second projection operator as
\begin{eqnarray}
\mP_{2}^0(t) \sigma_{xy}
&=&
\sum_{\vk >0}
\frac{
\eav{\sigma_{xy}
n_{\vk(t)}^* n_{\vk(t)}}
}{N^2 S_{k(t)}^2}
n_{\vk(t)} n_{\vk(t)}^*.
\label{A_P2sxy}
\end{eqnarray}
Only the potential part of the shear stress $\sigma_{xy}^{(\mathrm{pot})}$
survives in the correlator of Eq.~(\ref{A_P2sxy}), since the kinetic part
vanishes due to $\eav{p_i^x p_i^y}_\vp =0$.
The remaining part is
\begin{eqnarray}
\hspace{-1.5em}
\eav{
\sigma_{xy}^{(\mathrm{pot})}
n_{\vk(t)}^* n_{\vk(t)}}
&=&
\sum_i
\eav{x_i F_i^{(el)y} n_{\vk(t)} n_{\vk(t)}^*}.
\label{A_}
\end{eqnarray}
We utilize the relation well known in equilibrium statistical mechanics
\cite{HM},
\begin{eqnarray}
\eav{A F_i^\lambda}
&=&
-
\eav{A \frac{\del U}{\del r_i^\lambda}}
=
-
k_B T
\eav{\frac{\del A}{\del r_i^\lambda}},
\end{eqnarray}
which also holds in our formulation since the ensemble average
$\eav{\cdots}$ is defined by an averaging with the
Maxwell-Boltzmann distribution, Eq.~(\ref{rho_ini}).
Then, it is straightforward to show
\begin{eqnarray}
&&
\hspace{-1.5em}
\eav{
\sigma_{xy}^{(\mathrm{pot})}
n_{\vk(t)}^* n_{\vk(t)}}
\nn \\
&=&
-k_B T
\sum_i
\eav{x_i \frac{\del}{\del y_i}
\left[ n_{\vk(t)} n_{\vk(t)}^* \right]}
\nn \\
&=&
-k_B T
k(t)^y
\sum_i
\eav{
i x_i e^{i\vk(t)\cdot\ve{r}_i} n_{\vk(t)}^*
-
i x_i e^{-i\vk(t)\cdot\ve{r}_i} n_{\vk(t)}
}
\nn \\
&=&
-k_B T
k(t)^y
\eav{
\frac{\del n_{\vk(t)}}{\del k(t)^x} n_{\vk(t)}^*
+
\frac{\del n_{\vk(t)}^*}{\del k(t)^x} n_{\vk(t)}
}
\nn \\
&=&
-k_B T
k(t)^y \frac{\del}{\del k(t)^x}
\eav{
n_{\vk(t)} n_{\vk(t)}^*
}
\nn \\
&=&
-N k_B T k(t)^y \frac{\del S_{k(t)}}{\del k(t)^x}
\nn \\
&=&
-N k_B T \frac{k(t)^x k(t)^y}{k(t)} \frac{\del S_{k(t)}}{\del k(t)}.
\end{eqnarray}
This, together with Eq.~(\ref{A_P2sxy}), proves Eq.~(\ref{Psigma}).

\section{Miscellaneouses of the numerical analysis}
\label{App:sec:Num}
%

\subsection{The isotropic approximation}
\label{App:subsec:Iso}
The basics of the isotropic approximation and the derivation of its
resulting equations, Eqs.~(\ref{iso_eq}) and (\ref{iso_M}), is shown here.
Refer to CK \cite{CK2009} for further details.

The fundamental idea of the isotropic approximation is to reduce to
dependence of the three-dimensional wavevector to its modulus.
This is accomplished for the time-correlators by the assumption of
\begin{eqnarray}
\Phi_\vq(t)
&\simeq&
\Phi_q(t),
\label{A_iso_Phi}
\\
\vH_\vq(t)
&\simeq&
\frac{\vq(t)}{q(t)^2}
\frac{d}{dt}
\Phi_q(t).
\label{A_iso_H}
\end{eqnarray}
In addition, there are polynomials of the wavevectors to be handled.
They are approximated by their mean values with respect to the solid
angles.
For instance,
\begin{eqnarray}
\vq(t) \cdot \vk(t)
&=&
\vq \cdot \vk
-
(\dgamma t)
\left(
q_x k_y + q_y k_x
\right)
+
(\dgamma t)^2
q_x k_x
\nn \\
&\simeq&
(\vq \cdot \vk)
\left[
1
+
\frac{1}{3}
(\dgamma t)^2
\right],
\label{A_iso_qk}
\\
k_x(t) k_y(t)
&=&
k_x
\left[
k_y - (\dgamma t) k_x
\right]
\simeq
-
\frac{1}{3}
(\dgamma t)
k^2.
\label{A_iso_kk}
\end{eqnarray}
Note that the anisotropic terms are neglected in the above
approximations, which are the leading terms in the shear-rate for the
case $\dgamma t \ll 1$.

The modulus of the advected wavevector is approximated as
\begin{eqnarray}
q(t)^2
&=&
q^2
-
2(\dgamma t) q_x q_y
+ (\dgamma t)^2 q_x^2
\simeq
\bq(t)^2,
\end{eqnarray}
where $\bq(t)$ is defined as
\begin{eqnarray}
\bq(t)
&\equiv&
q
\sqrt{
1
+
\frac{1}{3}
(\dgamma t)^2
}.
\end{eqnarray}
Combining Eqs.~(\ref{eom-Phi}) and (\ref{eom-H3}), with the application
of Eqs.~(\ref{A_iso_Phi}) and (\ref{A_iso_H}), leads to
\begin{eqnarray}
\hspace{-1em}
\frac{d^2}{dt^2}
\Phi_q(t)
&=&
-
\dgamma
\frac{q_x q_y(t)}{q(t)^2}
\frac{d}{dt}
\Phi_q(t)
-
v_T^2
\frac{q(t)^2}{S_{q(t)}}
\Phi_q(t)
\nn \\
&&
-
\lambda_\alpha(t) \alpha_0
\frac{d}{dt}
\Phi_q(t)
-
\dgamma
\frac{q_x(t)q_y(t)}{q(t)^2}
\frac{d}{dt}
\Phi_q(t)
\nn \\
&&
-
\int_0^t ds
M_{\vq}(t,s)
\frac{d}{ds}
\Phi_q(s).
\hspace{1.5em}
\label{A_iso_eq}
\end{eqnarray}
The scalar memory kernel $M_{\vq}(t,s)$, which is defined in
Eq.~(\ref{App:Eq:M}), reads
\begin{widetext}
\begin{eqnarray}
M_{\vq}(t,s)
&=&
\frac{nv_T^2}{2q(s)^2}
\int \frac{d^3 \vk}{(2\pi)^3}
\left[
\left(\vq(t)\cdot\vk(t)\right) c_{k(t)}
+
\left(\vq(t)\cdot\vp(t)\right) c_{p(t)}
\right]
\left[
\left(\vq(s)\cdot\vk(s)\right) c_{k(s)}
+
\left(\vq(s)\cdot\vp(s)\right) c_{p(s)}
\right]
\nn \\
&&
\times
\Phi_{k(s)}(t-s)
\Phi_{p(s)}(t-s),
\label{A_M}
\end{eqnarray}
where Eqs.~(\ref{M4}) and (\ref{V}) are utilized.
It is convenient to shift the integration variable in Eq.~(\ref{A_M}) as
$\vk \ra \vk' \equiv \vk(s)$, which leads to
\begin{eqnarray}
M_\vq(t,s)
&=&
\bar{M}_{\vq(s)}(t-s),
\end{eqnarray}
where
\begin{eqnarray}
\bar{M}_{\vq}(\tau)
&\equiv&
\frac{nv_T^2}{2q^2}
\int \frac{d^3 \vk}{(2\pi)^3}
\left[
\left(\vq\cdot\vk\right) c_{k}
+
\left(\vq\cdot\vp\right) c_{p}
\right]
\left[
\left(\vq(\tau)\cdot\vk(\tau)\right) c_{k(\tau)}
+
\left(\vq(\tau)\cdot\vp(\tau)\right) c_{p(\tau)}
\right]
\Phi_{k}(\tau)
\Phi_{p}(\tau).
\label{A_bM}
\end{eqnarray}
Application of Eqs.~(\ref{A_iso_qk}) and (\ref{A_iso_kk}) to Eq.~(\ref{A_iso_eq})
leads to Eq.~(\ref{iso_eq}), and to Eq.~(\ref{A_bM}) leads to Eq.~(\ref{iso_M}).

\subsection{Discretization}
\label{App:subsec:Disc}
The discretized form of the memory kernel on the one-dimensional spatial
grid is given by
\begin{eqnarray}
\bar{M}_q(\tau)
&=&
\frac{1}{32\pi^2}
\frac{n v_T^2}{q^3}
\left[
1 + \frac{1}{3} \left( \dgamma \tau \right)^2
\right]
\int_0^\infty dk
\int_{|q-k|}^{q+k} dp
\left[
(q^2+k^2-p^2) c_{\bk(\tau)} + (q^2-k^2+p^2) c_{\bp(\tau)}
\right]
\nn \\
&&
\times
\left[
(q^2+k^2-p^2) c_k + (q^2-k^2+p^2) c_p
\right]
kp
\Phi_k(\tau) \Phi_p(\tau)
\nn \\
&\simeq &
\frac{n v_T^2}{32\pi^2}
\frac{\Delta^5}{d^5}
\frac{1}{\hat{q}^3}
\left[
1 + \frac{1}{3} \left( \dgamma \tau \right)^2
\right]
\sum_{\hat{k}}
\hat{k} \Phi_k(\tau)
\sum_{\hat{p}} {}^{'}
\left[
(\hat{q}^2+\hat{k}^2-\hat{p}^2) c_{\bk(\tau)} + (\hat{q}^2-\hat{k}^2+\hat{p}^2) c_{\bp(\tau)}
\right]
\nn \\
&&
\times
\left[
(\hat{q}^2+\hat{k}^2-\hat{p}^2) c_k + (\hat{q}^2-\hat{k}^2+\hat{p}^2) c_p
\right]
\hat{p} \Phi_p(\tau).
\end{eqnarray}
In the last step, the wavenumber is discretized as $q d = \Delta \h{q}$,
where $\Delta$ is the grid spacing, and $\h{q}$ is the discretized index
of the wavenumber, $\h{q} = (2m-1)/2$ ($m=1,2,\cdots,M$), $M$ being the
number of grids.
The summation with respect to $\h{p}$ is restricted to those which
satisfy the triangle inequality, i.e.,
\begin{eqnarray}
\sum_{\hat{p}} {}^{'}
&= &
\sum_{\hat{p}=|\hat{q}-\hat{k}|+1/2}^{\hat{p}=\hat{q}+\hat{k}-1/2}.
\nn
\end{eqnarray}
\end{widetext}

\bibliography{Suzuki-Hayakawa}
\end{document}